%% file: main.tex
\documentclass[11pt,letterpaper]{article}
\usepackage[T1]{fontenc}
\usepackage[utf8]{inputenc}
\usepackage{amsmath,amssymb,amsthm}
\usepackage[margin=1in]{geometry}
\usepackage{setspace}
\usepackage{microtype}
\usepackage{graphicx}
\usepackage{booktabs}
\usepackage{array}
\usepackage{longtable}
\usepackage{subcaption}
\usepackage{enumitem}
\usepackage{float}
\usepackage{placeins}
\usepackage{tikz}
\usetikzlibrary{arrows.meta, positioning, patterns}
\usepackage{inconsolata}
\usepackage{palatino}
\usepackage{natbib}
\bibpunct[, ]{(}{)}{,}{a}{}{,}

\usepackage[hidelinks,hyperfootnotes=false,
  pdftitle={Paying Creators for Training Data Can Increase AI Displacement}]{hyperref}

\let\amseqref\eqref
\renewcommand{\eqref}[1]{Eq.~\amseqref{#1}}

\newtheorem{proposition}{Proposition}
\newtheorem{lemma}{Lemma}
\newtheorem{assumption}{Assumption}
\newtheorem{corollary}{Corollary}

\newtheorem{applemma}{Lemma}[section]

\title{\Large\textrm{Paying Creators for Training Data Can Increase AI Displacement}}

\newenvironment{sciabstract}{\begin{quote}\small\onehalfspacing}
{\end{quote}}

\author
{
  Lijuan Luo\footnote{Lijuan Luo (\underline{\smash{\href{mailto:ll954@cornell.edu}{ll954@cornell.edu}}}) is the corresponding author. She is a doctoral student at Cornell University. Emaad Manzoor is an Assistant Professor of Marketing at Cornell University. Nathan Yang is an Associate Professor and C.W. Park Faculty Fellow at the University of Illinois Urbana-Champaign. Fan Yao is an Assistant Professor in the Department of Statistics and Operations Research at UNC Chapel Hill.
The authors thank Tommaso Bondi, Andrey Fradkin, Sylvia Hristakeva, and Justin Johnson for their feedback. The authors also thank participants at 2024 ISMS Marketing Science Conference, 2024 Conference on Artificial Intelligence, Machine Learning, and Business Analytics (AIML), the Artificial Intelligence in Management (AIM) 2025 Conference, the 2025 PhD student workshop at Cornell University, the Brownbag at UIUC, and AI \& Marketing: New Methods and New Risks Conference at Stanford for helpful discussions. Previously circulated under the title \textit{Platform Design When Creators Train Their AI Substitutes} (November 7, 2025; \url{http://dx.doi.org/10.2139/ssrn.5719862}).}\\
  Cornell University \\
  \and
  Emaad Manzoor\\
  Cornell University\\
  \and
  Nathan Yang\\
  UIUC\\
  \and 
  Fan Yao \\
  UNC Chapel Hill
}

\date{November 15, 2026}

\usepackage{xcolor}
\begin{document}

\maketitle
\thispagestyle{empty}

\begin{center}
\textbf{Abstract}
\end{center}

\begin{sciabstract}
\noindent Generative AI increasingly competes with creators while using their content as training data. In response, policymakers have called for platforms to compensate creators for training data use and allow opting out. We study how such policies affect creator displacement in a model of a platform where creators choose whether to contribute training data and consumers choose between purchasing from creators and prompting generative AI trained on their content. 
When compensation is required, the low-quality creator chooses to remain solely as a training data supplier while AI fully displaces her demand, despite AI quality and the price the platform charges for AI-generated content being identical with and without compensation.
Without compensation, the risk of losing creators and their training data constrains the platform to set the price of creators' content to preserve their demand. This constraint also prevents further AI displacement as AI improves, with the platform raising AI prices while keeping creators' prices and demand unchanged to prevent exit. 
Compensation increases creators' payoffs and total surplus but reduces consumer surplus. Our findings highlight how compensation policies for training data can shape creators' economic roles in a future with more capable generative AI.
\end{sciabstract}

\begin{quote}\small\textit{Keywords:} Generative AI; Platform Design; Creator Economy; Future of Work; Ethical AI\end{quote}

\clearpage

\onehalfspacing

\input{sections/intro}

\input{sections/related}
\input{sections/model}
\input{sections/analysis}

\input{sections/extension}
\input{sections/conclusion}

\bibliographystyle{informs2014}
\bibliography{references}

\clearpage
\appendix
\input{sections/appendix}

\end{document}

%% file: sections/intro.tex
\section{Introduction}
\label{sec:introduction}

Creative platforms increasingly offer consumers a choice between purchasing creators' content and prompting an integrated generative AI system trained on their content.\footnote{For example, consumers on the Shutterstock and Adobe Stock photography platforms can prompt generative AI trained on creators' content to generate a customized image rather than purchase one from a creator.} While consumers benefit from tailoring content to their preferences by prompting AI, creators risk being displaced. In response, policymakers have called for compensating creators and allowing them to opt out of AI training \citep{europeanparliament2026copyright}. However, compensation can change the platform's pricing incentives. Its equilibrium effects on creators' demand, participation, and payoffs are thus unclear.

We show that compensating creators can increase AI displacement and change creators' economic role from selling content to supplying training data. The result arises because compensating creators for training data weakens the platform's incentive to preserve demand for their content. Without compensation, the platform must preserve content sales to prevent creators from exiting and withdrawing their training data. Compensation relaxes this constraint: the platform can retain creators, and their training data, without preserving demand for their content. 

To formalize this, we extend a Hotelling-style model of a two-sided platform to include an integrated generative AI system that competes with creators. Consumers with heterogeneous tastes are distributed between two creators that differ in quality. Consumers choose between purchasing creators' content and exerting prompting effort to generate content with AI that matches their tastes exactly. The platform sets a common price for creators' content and a separate price for AI-generated content. The quality of AI-generated content and the required prompting effort depend on the creators' content used for training.

We use this model to compare two regimes. In the regime without compensation, providing training data is a condition of platform participation, and the platform retains all revenue from AI-generated content sales. In the regime with compensation and opt-out, creators choose whether to contribute training data, and the platform shares revenue from AI-generated content sales among contributing creators.\footnote{This revenue sharing scheme is inspired by Shutterstock's Contributor Fund \citep{shutterstock2023} and resembles the pro-rata revenue sharing principle used by music streaming platforms \citep{moreau2024alternative}.} In both regimes, creators receive a share of revenue from their own content sales. Creators incur a participation cost and choose whether to remain on the platform, withdrawing their training data upon exit.

The platform preserves demand for both creators' content without compensation but fully displaces the low-quality creator's demand with compensation. Without compensation, the platform sets the price of creators' content to retain the low-quality creator, whose exit would deprive the platform of her training data. With compensation, the platform sets a higher price for creators' content while retaining the low-quality creator solely as a training data supplier. In both regimes, both creators contribute training data (by choice under compensation), and the quality and price of AI-generated content are identical. Greater AI displacement thus arises not from a better or cheaper AI, but from the requirement to compensate creators changing the platform's pricing incentives.

Requiring compensation also changes how the platform responds to improvements in AI capability. In both regimes, the platform raises the price of AI-generated content identically as AI improves. Without compensation, the platform keeps the price of creators' content unchanged to preserve their demand, prevent exit, and retain their training data. With compensation, the platform instead raises the price of creators' content, progressively displacing the remaining demand for creators until AI serves the entire market. Both creators nevertheless remain as training data suppliers, earning more from their data as AI improves. Better AI thus increases displacement only when compensation allows the platform to retain training data without preserving creators' sales.

Requiring compensation increases both creators' payoffs and total surplus but reduces consumer surplus. The total surplus increase comes from reallocating demand toward AI-generated content that better matches consumers' tastes, rather than expanding market coverage or improving AI quality. Consumers lose surplus as the platform raises the price of creators' content and induces more consumers to purchase AI-generated content, from which the platform captures their full willingness to pay. As AI improves, the differences in both creators' payoffs and total surplus between compensation and no compensation increase. The platform may prefer not to compensate creators when preserving demand for their content is cheaper than sharing AI revenue. A requirement to compensate creators can thus increase total surplus when the platform would otherwise choose not to compensate, but reduces consumer surplus and platform profit.

We extend our model to examine imperfect AI ``steerability'' (the prompting effort required to generate content that matches consumers' tastes) \citep{vafa2025whats}, heterogeneous transfer rates of creators’ training data to improve AI quality, and alternative compensation rules. Compensation-induced displacement persists with moderate prompting effort, although sufficiently high prompting effort can restore demand for creators' content. It also persists when creators’ data differ in their contributions to AI quality, as equal revenue sharing allows displaced creators to remain as training data suppliers even when their contributions are small. When compensation is weighted by contributions to AI quality, displacement persists under equal transfer rates, but revenue from AI-generated content sales is redistributed toward the high-quality creator. With heterogeneous transfer rates, however, quality-weighted compensation may fail to sustain the participation of displaced creators whose data contribute little to AI quality.

Our findings highlight a distinct feature of AI displacement. While automation typically reduces demand for the labor it replaces, generative AI can create a new source of demand for displaced creators: their content as training data. Compensation allows creators to earn revenue from AI-generated content sales as training data suppliers even after losing demand for their content, leaving them better off as AI improves and further displaces their sales. Compensation also increases total surplus by shifting demand toward AI-generated content, although consumers lose surplus. Compensation policies may thus shape creators' economic roles and payoffs in a future with increasingly capable generative AI.

%% file: sections/related.tex
\subsection{Related Work}
\label{sec:literature}

Our paper contributes to research on technological change and labor displacement. This literature examines how automation displaces labor from existing tasks \citep{autor2003skill,acemoglu2018race,acemoglu2020robots,acemoglu2022tasks}, how new tasks create labor demand \citep{acemoglu2019automation,autor2024new}, and how displaced workers adjust \citep{autor2013growth,feigenbaum2024answering}. The displaced creators in our model continue supplying the content they produced, but now as training data for the AI that replaces their sales. Compensation for training data allows them to benefit from AI improvements without undertaking new tasks.

We also contribute to research on AI copyright. This literature examines how copyright protection, licensing, and opt-out rights affect access to AI training data \citep{gans2026copyright,derassenfosse2025intellectual,shcherbakov2025ai,peukert2025copyright,cooper2025law,DelaDurantaye2025Control,samuelson2026collective}, how training on creators' content affects their subsequent contributions \citep{peukert2025dynamic}, and how to compensate creators for AI training data \citep{yang2024generative,peukert2025economics,gao2025fair,deng2024economic,senftleben2025win,orstavik2026copyright,bostoen2026putting}.

We contribute to this literature in three ways. First, compensation changes the platform's price for creators' content, and hence displacement, even when training data access and AI capability are unchanged. Second, creators can remain as training data suppliers and receive higher payoffs than without compensation, even when AI displaces demand for their content. Third, we show that compensation can increase total surplus while reducing consumer surplus by inducing the platform to reallocate demand toward AI, in addition to redistributing revenue to creators.

Our paper also relates to theoretical research on platform encroachment. This literature examines how platforms' competing offerings affect third-party sellers' participation and investment incentives \citep{jiang2011firm,hagiu2013first,hagiu2022should}, and how platform pricing, channel structure, and information policies shape these incentives \citep{abhishek2016agency,choi2025sherlocking}. Relatedly, \citet{bhargava2026strategic} show that data sharing with competitors can create strategic interdependence that softens competition. \citet{yao2024human} model competition between creators and AI trained on their content, taking data sharing as given. \citet{taitler2025data} and \citet{keinan2026strategic} examine strategic data sharing with competing AI and the role of data pricing and revenue sharing. Unlike these studies, which focus on content production and data sharing, we examine competition through the platform's joint pricing of creators' and AI-generated content.

Creators in our model can earn revenue in two economic roles, by selling content to consumers or supplying training data to the platform. The platform trains AI on their content and competes with them through AI-generated content. Compensation changes which of these roles must generate revenue to keep a creator on the platform. Without compensation, the creator has to sell content; with compensation, her share of revenue from AI-generated content sales can cover her participation cost. Dependence on a creator's training data can thus sustain her participation without protecting her content from competition.

%% file: sections/model.tex
\section{Model Setup}
\label{sec:model}

We model a platform where consumers choose between purchasing content from two quality-differentiated creators and generating customized content with AI that the platform trains on the creators' content.\footnote{We abstract away from creator-facing AI tools that lower creators' production costs, such as AI editing assistants.} The platform sets a common price for creators' content and a separate price for AI-generated content, while creators make participation and, under compensation, data sharing decisions. In this section, we begin by describing consumer preferences and the effects of training data on AI quality and prompting effort. We then describe creators' payoffs and decisions, followed by the platform's pricing problem. We conclude by describing the timing of the game before proceeding to the main analysis in Section~\ref{sec:analysis}.

\subsection{Consumers}
\label{subsec:consumers}

A unit mass of consumers, each with unit demand, is uniformly distributed on the Hotelling line $[0,1]$, with creator $L$ located at $0$ and creator $H$ at $1$.
Creator $j$ produces content of quality $q_j$, where $0<q_L<q_H<1$. The platform sets a common price $P$ for creators' content and a separate price $P_{\mathrm{AI}}$ for AI-generated content of quality $q_{\mathrm{AI}}$. A consumer at location $x$ incurs mismatch costs of $x$ and $1-x$ when purchasing from $L$ and $H$, respectively, with the transportation cost normalized to one. A consumer can use AI to generate content that perfectly matches her taste but incurs prompting effort $\kappa(x)$, denoted by $\kappa$ henceforth. Each consumer chooses the option that yields the highest utility, including an outside option with utility normalized to zero, such as general-purpose AI.\footnote{\label{fn:consumer-tie-breaking}We adopt two tie-breaking rules for indifferent consumers: consumers favor a platform product over the outside option and creators' content over AI-generated content. When $\kappa_{\textrm{data}}=1$, the second rule affects aggregate demand only when the platform trains AI on one creator's content.}

The resulting consumer utilities are:
\begin{align}
&U_{L}=q_{L}-P-x, && \text{if the consumer purchases from $L$,}\nonumber\\
&U_{H}=q_{H}-P-(1-x), && \text{if the consumer purchases from $H$,}\nonumber\\
&U_{AI}=q_{AI}-P_{AI}-\kappa, && \text{if the consumer generates a product using AI.}
\label{eq:consumer-utility}
\end{align}

\begin{figure}[t]
\caption{Prompting Effort with Different Training Data}
\label{fig:prompting-effort}
\centering
\begin{subfigure}[b]{0.24\textwidth}
\centering
\begin{tikzpicture}[scale=0.62, thick]
\draw[thick] (-2,0) -- (2,0);
\draw[thick] (-2,-0.2) -- (-2,0.2);
\draw[thick] (2,-0.2) -- (2,0.2);
\node[left] at (-2,0) {$L$};
\node[right] at (2,0) {$H$};
\draw[thick] (-2,1.5) -- (2,1.5);
\node at (0,1.95) {AI};
\node at (-1,0) [circle,fill,inner sep=1.5pt]{};
\node at (-1,-0.5) {$x$};
\draw[<->, thick, dashed] (-1,0.075) -- (-1,1.5);
\node[right, font=\scriptsize] at (-1,0.75) {$\kappa(x)=\kappa_0+\kappa_{\textrm{data}}$};
\end{tikzpicture}
\caption{Before training}
\label{fig:before-fine-tuning}
\end{subfigure}\hfill
\begin{subfigure}[b]{0.24\textwidth}
\centering
\begin{tikzpicture}[scale=0.62, thick]
\draw[thick] (-2,0) -- (2,0);
\draw[thick] (-2,-0.2) -- (-2,0.2);
\draw[thick] (2,-0.2) -- (2,0.2);
\node[left] at (-2,0) {$L$};
\node[right] at (2,0) {$H$};
\draw[thick] (-2,1.5) -- (2,0.5);
\node[rotate=-14] at (0,1.45) {AI};
\node at (-1,0) [circle,fill,inner sep=1.5pt]{};
\node at (-1,-0.5) {$x$};
\draw[<->, thick, dashed] (-1,0.075) -- (-1,1.25);
\node[right, font=\scriptsize] at (-1,0.6) {$\kappa(x)$};
\end{tikzpicture}
\caption{Trained on $H$}
\label{fig:after-fine-tuning-H}
\end{subfigure}\hfill
\begin{subfigure}[b]{0.24\textwidth}
\centering
\begin{tikzpicture}[scale=0.62, thick]
\draw[thick] (-2,0) -- (2,0);
\draw[thick] (-2,-0.2) -- (-2,0.2);
\draw[thick] (2,-0.2) -- (2,0.2);
\node[left] at (-2,0) {$L$};
\node[right] at (2,0) {$H$};
\draw[thick] (-2,0.5) -- (2,1.5);
\node[rotate=14] at (0,1.45) {AI};
\node at (-1,0) [circle,fill,inner sep=1.5pt]{};
\node at (-1,-0.5) {$x$};
\draw[<->, thick, dashed] (-1,0.075) -- (-1,0.75);
\node[right, font=\scriptsize] at (-1,0.38) {$\kappa(x)$};
\end{tikzpicture}
\caption{Trained on $L$}
\label{fig:after-fine-tuning-L}
\end{subfigure}\hfill
\begin{subfigure}[b]{0.24\textwidth}
\centering
\begin{tikzpicture}[scale=0.62, thick]
\draw[thick] (-2,0) -- (2,0);
\draw[thick] (-2,-0.2) -- (-2,0.2);
\draw[thick] (2,-0.2) -- (2,0.2);
\node[left] at (-2,0) {$L$};
\node[right] at (2,0) {$H$};
\draw[thick] (-2,0.5) -- (2,0.5);
\node at (0,1.15) {AI};
\node at (-1,0) [circle,fill,inner sep=1.5pt]{};
\node at (-1,-0.5) {$x$};
\draw[<->, thick, dashed] (-1,0.075) -- (-1,0.5);
\node[right, font=\scriptsize] at (-1,0.28) {$\kappa(x)=\kappa_0$};
\path (0,1.95) node {\phantom{AI}};
\end{tikzpicture}
\caption{Trained on both}
\label{fig:after-fine-tuning-both}
\end{subfigure}
\vspace{2pt}

\parbox{\textwidth}{\small
\textit{Note.} The prompting effort $\kappa(x)$ is the vertical distance from a consumer at $x$ to the line representing the AI generated content, as in \eqref{eq:kappa}. Training on one creator's content reduces prompting effort for consumers with tastes closer to that creator's style. Training on both creators' content eliminates the taste-dependent component of prompting effort. The panels illustrate $\kappa_0>0$; the main analysis sets $\kappa_0=0$.
}
\end{figure}
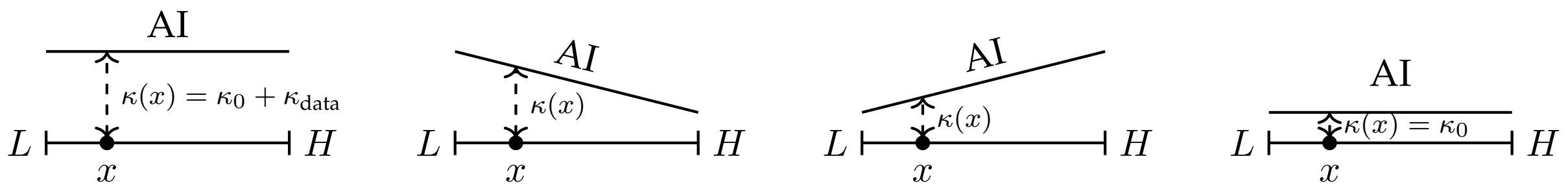

\paragraph{Prompting effort $\kappa$.}
 Prompting effort consists of two components. The base component $\kappa_0\geq0$ captures the generative AI model's inherent steerability \citep{vafa2025whats}, determined by its architecture, pretraining, and post-training rather than by creators' training data. A larger $\kappa_0$ corresponds to a less steerable model. The data component $\kappa_{\textrm{data}}\geq0$ captures the additional effort required when the model lacks training data close to the consumer's taste. Letting $s_j\in\{0,1\}$ indicate whether the platform trains AI on creator $j$'s content, we model prompting effort as:
\begin{equation}
\kappa(x;s_H,s_L)\;=\;\begin{cases}
\kappa_0+\kappa_{\textrm{data}},& s_H=0\text{ and }s_L=0,\\
\kappa_0+\kappa_{\textrm{data}}\,x,& s_H=0\text{ and }s_L=1,\\
\kappa_0+\kappa_{\textrm{data}}(1-x),& s_H=1\text{ and }s_L=0,\\
\kappa_0,& s_H=1\text{ and }s_L=1.
\end{cases}
\label{eq:kappa}
\end{equation}
Training on one creator's content reduces prompting effort most for consumers whose tastes are closer to that creator's style. Figure~\ref{fig:prompting-effort} illustrates \eqref{eq:kappa} visually.

For the main analysis, we assume perfect steerability ($\kappa_0=0$) when the platform trains AI on both creators' content and consider imperfect steerability ($\kappa_0>0$) in Section~\ref{subsec:imperfect-steerability}. We also set $\kappa_{\textrm{data}}=1$ to treat prompting effort and content mismatch symmetrically.\footnote{$\kappa_{\textrm{data}}>1$ may arise if, for example, the consumer's cost of eliminating the mismatch between the creator's content and her own taste by prompting the platform's AI is higher than that using her own tools (e.g., Adobe Photoshop).} When the platform trains AI on only one creator's content (as illustrated in Figures~\ref{fig:after-fine-tuning-H} and \ref{fig:after-fine-tuning-L}), a consumer located $z$ units from that creator incurs $z$ units of mismatch cost when purchasing her content and $z$ units of prompting effort when using AI to generate content that perfectly matches her taste.

\paragraph{Quality of AI generated content $q_{AI}$.}
Scaling studies show that AI quality improves with training data size \citep{kaplan2020scaling,hoffmann2022training} and with training on specific high-quality examples (e.g., professional writing and photographs) \citep{zhou2023lima,dai2023emu}.

Hence, we model the quality of AI-generated content as increasing with the number of creators whose content the platform trains AI on and the quality of their content:
\begin{equation}
q_{AI}\;=\;q^{\textrm{base}}\;+\;\tau_{H}\,q_{H}\,s_{H}\;+\;\tau_{L}\,q_{L}\,s_{L}.
\label{eq:q-ai}
\end{equation}
Here, $q^{\textrm{base}}\geq0$ is the generative AI model's quality before training on creators' content, and $\tau_j\in(0,1)$ measures how much creator $j$'s content contributes to AI quality.

When creator $j$ exits, the platform loses access to her data for AI training, as when a renewable training license expires or the flow of new content ceases. If the platform trained AI on her content, exit reduces AI quality by $\tau_jq_j$ and eliminates the associated reduction in prompting effort. For the main analysis, we set $\tau_H=\tau_L=\tfrac12$, giving both creators' content equal weight in determining AI quality. We consider heterogeneous transfer rates ($\tau_H\neq\tau_L$) in Section~\ref{subsec:heterogeneous-data}.

Consumers can already access general-purpose AI outside the platform. To attract paying consumers, the platform's AI must offer an advantage over these alternatives. On creative content marketplaces, training on creators' content can provide this advantage by improving AI quality and reducing prompting effort. This motivates the following assumption:

\begin{assumption}[Training data dependence]
\label{assumption:training-data-dependence}
Without training on creators' content, the platform's AI cannot sustain positive sales revenue.
\end{assumption}

When the platform does not train AI on either creator's content, consumers incur prompting effort $\kappa_0+\kappa_{\textrm{data}}=1$ under our main specification. Assumption~\ref{assumption:training-data-dependence} then requires $q^{\textrm{base}}\leq1$.

\subsection{Creators}
\label{subsec:creators}

Each creator $j\in\{L,H\}$ decides whether to participate on the platform, $e_j\in\{0,1\}$. A creator who exits ($e_j=0$) earns zero and withdraws her content and training data from the platform. A participating creator ($e_j=1$) incurs a fixed cost $\phi q_j$, where $\phi>0$, reflecting the higher cost of maintaining higher-quality content. Marginal production costs are normalized to zero.

We consider two regimes that differ in whether the platform compensates creators for training data. In both regimes, creators receive a share $\alpha\in(0,1)$ of revenue from their own content sales. In the \textit{no compensation regime} (NC), the platform trains AI on participating creators' content without additional payment.\footnote{Without compensation, allowing creators to opt out eliminates their incentive to share training data: doing so improves AI without increasing their direct sales revenue. Hence, we model the no compensation regime as one in which platforms obtain training rights as a condition of participation, as in Reddit and Meta \citep{reddit2026useragreement,meta2023privacy}.} Training data use is a condition of participation, so $s_j=e_j$. In the \textit{compensation regime} (C), participating creators can also earn revenue by contributing their content for AI training. Each creator independently chooses whether to contribute, $s_j\in\{0,1\}$, with $s_j\leq e_j$. The platform shares a fraction $\alpha$ of AI revenue equally among contributors. Creators who do not contribute can continue selling their content but receive no AI revenue.

A participating creator's gross payoff is:
\begin{equation}
V_j =
\begin{cases}
\alpha D_j P,
& \text{under NC}, \\[6pt]
\alpha\left(D_j P + \dfrac{s_j}{s_H+s_L}D_{AI}P_{AI}\right),
& \text{under C},
\end{cases}
\label{eq:creator-payoff}
\end{equation}
where $D_j$ and $D_{AI}$ denote demand for creator $j$'s content and AI content, respectively.\footnote{We adopt the convention $0/0=0$. We extend the analysis to quality-weighted revenue sharing in Section~\ref{subsec:quality-weighted}.} A participating creator's net payoff is $V_j-\phi q_j$, while an exiting creator earns zero.

Our focus is on whether AI can fully displace a creator who remains on the platform. We say a participating creator is \emph{fully displaced} when $D_j=0$. Under compensation, such a creator can remain on the platform as a training \emph{data supplier} if her AI revenue covers her participation cost. Without compensation, a creator who loses all direct demand cannot cover her participation cost and exits. This distinction separates displacement from exit: a displaced creator may participate and continue supplying training data, whereas an exiting creator withdraws both her content and her training data from the platform.

\subsection{Platform}
\label{subsec:platform}

The platform sets a common price $P\geq0$ for creators' content (following prevailing norms on creative content marketplaces) and a price $P_{AI}\geq0$ for AI-generated content. It retains a fraction $1-\alpha$ of revenue from creators' content, where $\alpha$ is a fixed platform-specific parameter.

The platform's share of AI revenue depends on the compensation regime. Under NC, it retains all AI revenue. Under C, it pays a fraction $\alpha$ of AI revenue to creators who contribute training data.\footnote{We use the same revenue sharing rate for creators' content and AI-generated content to avoid introducing an additional parameter and to isolate the effect of compensating creators for their training data.} Letting $D=D_H+D_L$ denote total demand for creators' content, the platform's profit in each regime is given by:
\begin{equation}
\pi =
\begin{cases}
(1-\alpha)DP+D_{AI}P_{AI},
& \text{under NC}, \\[6pt]
(1-\alpha)(DP+D_{AI}P_{AI}),
& \text{under C}.
\end{cases}
\label{eq:platformprofit}
\end{equation}
When the platform does not train AI on either creator's content, AI generates no positive sales revenue under Assumption~\ref{assumption:training-data-dependence}, so the expression in \eqref{eq:platformprofit} under C still applies.

\subsection{Timing and Equilibrium}
\label{subsec:timing}

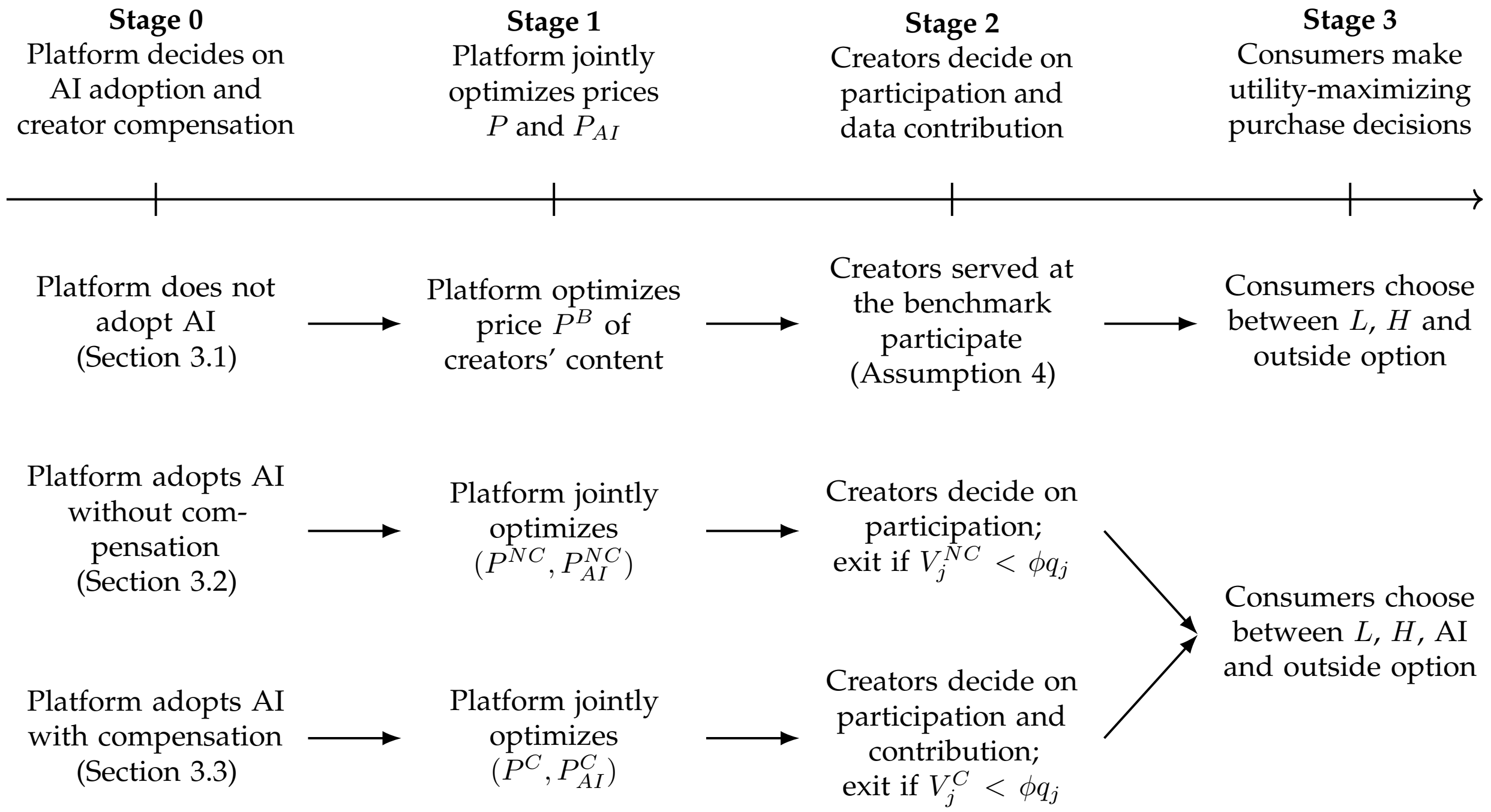
\begin{figure}[h]
\vspace{18pt}
\centering
\captionsetup{position=top,skip=10pt}
\caption{Stages of the Game}
\label{fig:timeline}
\resizebox{\textwidth}{!}{\begin{tikzpicture}[
font=\small,
arrow/.style={-Latex, thick},
stage/.style={align=center},
nodebox/.style={align=center, text width=3.4cm}
]

\node (stage0) [stage] at (0,0)
{\textbf{Stage 0} \\ Platform decides on \\ AI adoption and \\ creator compensation};

\node (stage1) [stage] at (4.8,0)
{\textbf{Stage 1} \\ Platform jointly \\ optimizes prices \\ $P$ and $P_{AI}$};

\node (stage2) [stage] at (9.6,0)
{\textbf{Stage 2} \\ Creators decide on \\ participation and \\ data contribution};

\node (stage3) [stage] at (14.4,0)
{\textbf{Stage 3} \\ Consumers make \\ utility-maximizing \\ purchase decisions};

\draw[->, thick] (-1.8,-1.5) -- (16.0,-1.5);
\foreach \x in {0,4.8,9.6,14.4} {
\draw[thick] (\x,-1.7) -- (\x,-1.3);
}

\node (noAI0) [nodebox] at (0,-3.0)
{Platform does not \\ adopt AI \\ (Section~\ref{subsec:no-ai})};

\node (NC0) [nodebox] at (0,-5.5)
{Platform adopts AI \\ without compensation \\ (Section~\ref{subsec:nocompensation})};

\node (C0) [nodebox] at (0,-8.0)
{Platform adopts AI \\ with compensation \\ (Section~\ref{subsec:withcompensation})};

\node (priceB) [nodebox] at (4.8,-3.0)
{Platform optimizes \\ price $P^{B}$ of \\ creators' content};

\node (priceNC) [nodebox] at (4.8,-5.5)
{Platform jointly \\ optimizes \\ $(P^{NC},P_{AI}^{NC})$};

\node (priceC) [nodebox] at (4.8,-8.0)
{Platform jointly \\ optimizes \\ $(P^{C},P_{AI}^{C})$};

\draw[arrow] (noAI0.east) -- (priceB.west);
\draw[arrow] (NC0.east) -- (priceNC.west);
\draw[arrow] (C0.east) -- (priceC.west);

\node (creatorB) [nodebox] at (9.6,-3.0)
{Creators served at \\ the benchmark participate \\ (Assumption~\ref{assumption:benchmark-participation})};

\node (creatorNC) [nodebox] at (9.6,-5.5)
{Creators decide on \\ participation; \\ exit if $V_j^{NC}<\phi q_j$};

\node (creatorC) [nodebox] at (9.6,-8.0)
{Creators decide on \\ participation and \\ contribution; \\ exit if $V_j^{C}<\phi q_j$};

\draw[arrow] (priceB.east) -- (creatorB.west);
\draw[arrow] (priceNC.east) -- (creatorNC.west);
\draw[arrow] (priceC.east) -- (creatorC.west);

\node (chooseB) [nodebox] at (14.4,-3.0)
{Consumers choose \\ between $L$, $H$ and \\ outside option};

\node (chooseAI) [nodebox] at (14.4,-6.75)
{Consumers choose \\ between $L$, $H$, AI \\ and outside option};

\draw[arrow] (creatorB.east) -- (chooseB.west);
\draw[arrow] (creatorNC.east) -- (chooseAI.west);
\draw[arrow] (creatorC.east) -- (chooseAI.west);

\end{tikzpicture}}\par
\vspace{1em}
\end{figure}

The game proceeds in four stages, as illustrated in Figure~\ref{fig:timeline}. In Stage~0, the platform decides whether to integrate AI and, if so, whether to compensate creators for their data.\footnote{Section~\ref{subsec:welfare} also considers mandated compensation when comparing private and social incentives.} In Stage~1, the platform sets the price $P$ for creators' content and, if AI is integrated, the price $P_{AI}$. In Stage~2, creators simultaneously decide whether to participate and, under C, whether to contribute their content for training. In Stage~3, consumers choose the option that maximizes their utility.

We solve for subgame-perfect equilibria by backward induction. Given the platform's prices, creators' decisions in Stage~2 form a simultaneous-move game, for which we consider pure-strategy Nash equilibria, and a mixed-strategy equilibrium where no pure-strategy equilibrium satisfies Assumption \ref{assumption:weak-participation}.\footnote{In the benchmark and NC regimes, creators have a pure-strategy equilibrium at every price (Appendix~\ref{apx:stage2}). In the C regime, some prices have no pure-strategy equilibrium. At these prices, we consider a mixed-strategy equilibrium in which creators randomize their participation and training data contribution decisions. However, in every equilibrium, the platform chooses prices that maximize its profit and at which creators' decisions form a pure-strategy equilibrium.} The platform anticipates these decisions and the resulting consumer demand when setting prices to maximize profit. Since creator exit withdraws training data, the platform's pricing decisions also affect AI quality and consumers' prompting effort.

The Stage~2 game requires two equilibrium selection rules. The first concerns participation when a creator's gross payoff $V_j$ equals her participation cost $\phi q_j$. Since the platform sets prices before creators decide whether to participate, its optimal prices may leave a creator indifferent between participating and exiting. If an indifferent creator exits, the platform loses access to her content and training data. We resolve this indifference in favor of participation, allowing the platform to retain creators whose revenue just covers their participation costs.

\begin{assumption}[Weak participation]
\label{assumption:weak-participation}
A creator participates when she is indifferent between participating and exiting.
\end{assumption}

The second rule concerns coordination between creators over their training data contribution decisions under compensation. At some prices, AI generates positive sales revenue only when both creators contribute. If neither contributes, a creator who contributes alone earns no additional AI revenue, so neither has an incentive to deviate from $(s_H,s_L)=(0,0)$. However, when both contribute, AI generates revenue that is shared between them, making both creators better off. Both $(0,0)$ and $(1,1)$ may then be Nash equilibria. We assume that creators coordinate on the equilibrium that Pareto dominates the others when one exists.

\begin{assumption}[Pareto coordination]
\label{assumption:coordination}
When the Stage~2 game has multiple pure-strategy Nash equilibria that satisfy Assumption \ref{assumption:weak-participation} and one Pareto dominates the others (no creator is worse off and some creator is better off), creators select that equilibrium.
\end{assumption}

We use the market before the platform integrates AI as the \textit{benchmark}. Our interest is in whether AI displaces creators who would otherwise sell content on the platform. We focus on markets in which both creators initially participate and attract positive demand. Two factors can prevent this outcome. If the high-quality creator has a sufficiently large quality advantage, the platform may prefer to set a price at which only her content attracts demand. Even when both creators attract demand, a creator may exit if her sales revenue does not cover her participation cost. We impose two restrictions to focus on markets in which both creators are active before AI integration.

\begin{assumption}[Benchmark participation]
\label{assumption:benchmark-participation}
Creator qualities satisfy $q_H/q_L<\rho^{\star}\equiv1+\sqrt{2}$, and the participation cost parameter satisfies:
\[
\phi\leq\phi^B\equiv\frac{\alpha(3q_L-q_H)(q_H+q_L)}{16q_L}.
\]
\end{assumption}

\noindent The restriction $q_H/q_L<\rho^{\star}$ ensures that the platform's optimal price generates positive demand for both creators in the benchmark. We derive both thresholds in Proposition~\ref{prop:benchmark} and its proof.

%% file: sections/analysis.tex
\section{Main Analysis}
\label{sec:analysis}

We characterize how compensation for training data changes the platform's pricing decisions, creator participation, and the allocation of consumer demand between creators' content and AI-generated content. We first establish the benchmark before AI integration and then derive the equilibria without and with compensation, before examining the platform's compensation decision and comparing welfare across the two regimes. We mark equilibrium objects with the superscripts $B$, $NC$, and $C$ for the benchmark, the regime without compensation, and the regime with compensation, as in $\pi^{B}$, $P^{NC}$, and $V_j^{C}$. For clarity, we also summarize the notation used throughout the paper in Appendix~\ref{apx:notation}.

\subsection{Benchmark: The Market Before AI Integration}
\label{subsec:no-ai}

Before the platform integrates a generative AI system, consumers choose between purchasing content from $H$ or $L$ and the outside option (e.g., prompting a general-purpose generative AI). The platform sets a common price $P\geq0$, creators decide whether to participate, and the platform earns $\pi^B(P)=(1-\alpha)(D_H+D_L)P$. When both creators participate, the consumer indifferent between them is located at $x_{HL}=1/2-(q_H-q_L)/2$ and obtains utility $(q_H+q_L-1)/2-P$. The market is fully covered when $P\leq(q_H+q_L-1)/2$ and not fully covered otherwise.

The platform faces a tradeoff between price and market coverage. Consumers near the middle of the Hotelling line incur greater mismatch costs, so attracting them requires lowering $P$. The platform may prefer to let some consumers choose the outside option rather than reduce the price. When the quality advantage of $H$ is sufficiently large, maintaining positive demand for $L$ requires a further price reduction, making it more profitable to set a price at which only $H$ attracts demand.

\begin{proposition}[Benchmark]
\label{prop:benchmark}
There is a quality ratio threshold $\rho^{\star}=1+\sqrt{2}$ such that the platform's optimal price generates positive demand for both creators under partial coverage when $q_H/q_L<\rho^{\star}$ and $\phi \leq\phi^B$, and positive demand only for the high-quality creator when $q_H/q_L>\rho^{\star}$ and $\phi \leq \alpha q_H/4$. When $q_H/q_L=\rho^{\star}$, the platform earns the same maximum profit whether both creators attract positive demand or only $H$ does, provided $\phi\leq\alpha q_L/8$.
\end{proposition}

\begin{proof}
See Appendix~\ref{apx:proof-benchmark}.
\end{proof}

Proposition~\ref{prop:benchmark} distinguishes two cases according to which creators participate. When $q_H/q_L<\rho^{\star}$, the platform sets $P^B=(q_H+q_L)/4$, generating demands $D_L^B=(3q_L-q_H)/4$ and $D_H^B=(3q_H-q_L)/4$. A mass $1-(q_H+q_L)/2$ of consumers chooses the outside option. Platform profit is $\pi^B=(1-\alpha)(q_H+q_L)^2/8$, and each creator earns $V_j^B=\alpha D_j^B P^B$. If $L$'s revenue covers her participation cost, $H$'s does as well, since $V_L^B/q_L<V_H^B/q_H$. When $q_H/q_L>\rho^{\star}$, the platform instead sets $P^B=q_H/2$, generating demand $D_H^B=q_H/2$ and profit $\pi^B=(1-\alpha)q_H^2/4$, conditional on $H$'s participation. The low-quality creator attracts no demand and exits.

By bounding $q_H/q_L$ and $\phi$ from above, Assumption~\ref{assumption:benchmark-participation} restricts our attention to the first case, in which both creators participate and attract positive demand before AI integration. This allows us to examine whether AI displaces creators who would otherwise sell content and whether they remain as training data suppliers.

\subsection{Without Compensation (NC)}
\label{subsec:nocompensation}

Under the NC regime, the platform trains AI on participating creators' content and retains all AI revenue, whereas creators earn revenue only from their own content. Hence, $s_j=e_j$ and $V_j^{NC}=\alpha D_jP$. When consumers purchase AI-generated content instead of purchasing from creators, creators lose revenue and may exit, withdrawing their training data. The platform sets prices to balance AI revenue against the need to retain creators' training data. We formalize this tradeoff by first deriving consumer demand assuming both creators participate, then showing that the platform earns lower profit if a creator exits, and finally deriving its optimal prices.

We begin by deriving consumer demand at given prices when the platform trains AI on both creators' content, $(s_H, s_L)=(1,1)$. Since $q_{AI}=q^{\textrm{base}}+(q_H+q_L)/2$ and $\kappa=0$, every consumer obtains $U_{AI}=q_{AI}-P_{AI}$ from the AI. The following lemma characterizes consumer demand conditional on both creators participating.

\begin{lemma}[Consumer demand under NC]
\label{lemma:demands-nc}
Suppose both creators participate, so $e_H=e_L=1=s_H=s_L$. Let $\widetilde P_{AI}\equiv P_{AI}-q^{\textrm{base}}$, $\delta\equiv\widetilde P_{AI}-P$, and $g\equiv(q_H-q_L)/2$. If $\widetilde P_{AI}>(q_H+q_L)/2$, consumers do not purchase AI-generated content, and demand for creators' content follows the benchmark before AI integration at the same price $P$ (Section~\ref{subsec:no-ai}). If instead $\widetilde P_{AI}\leq(q_H+q_L)/2$, the market is fully covered, with consumer demand given by:
\begin{center}
\normalfont
\renewcommand{\arraystretch}{1.2}
\begin{tabular}{lccc}
\toprule
Range of $\delta$ & $D_L$ & $D_{AI}$ & $D_H$ \\
\midrule
$\delta\geq1/2$ & $1/2-g$ & $0$ & $1/2+g$ \\
$g<\delta<1/2$ & $\delta-g$ & $1-2\delta$ & $\delta+g$ \\
$-g<\delta\leq g$ & $0$ & $1-\delta-g$ & $\delta+g$ \\
$\delta\leq-g$ & $0$ & $1$ & $0$ \\
\bottomrule
\end{tabular}
\end{center}
\end{lemma}
\begin{proof}
See Appendix~\ref{apx:proof-lemma-demands-nc}.
\end{proof}

We next show that the platform obtains higher profit by retaining both creators than by allowing either creator to exit. Suppose $L$ exits (an analogous argument applies if $H$ exits). The platform can no longer sell $L$'s content or train AI on $L$'s content. AI quality falls by $q_L/2$, by Equation~\amseqref{eq:q-ai}. A consumer at $x$ incurs mismatch cost $1-x$ when purchasing from $H$ and prompting effort $1-x$ when purchasing AI-generated content, by Equations~\amseqref{eq:consumer-utility} and~\amseqref{eq:kappa}. Consumers thus choose between purchasing from $H$ and purchasing AI-generated content based only on quality and price, which do not vary across consumers.

If the platform prices AI so that consumers obtain strictly higher utility from AI-generated content than from $H$'s content, every consumer who would purchase from $H$ purchases AI-generated content instead. With no demand for her content, $H$ cannot cover the cost of participation and would prefer to exit. If neither creator participates, the platform has no content to sell and cannot train AI on any platform content. Consumers choose the outside option instead of paying for AI-generated content under Assumption~\ref{assumption:training-data-dependence}, so the platform's profit is zero.

To retain $H$, the platform must set prices so that consumers obtain at least as much utility from $H$'s content as from AI-generated content. At these prices, consumers purchase from $H$ when indifferent between the two offerings,\textsuperscript{\ref{fn:consumer-tie-breaking}} so $D_{AI}=0$. With demand only for $H$'s content, the platform's profit is at most $(1-\alpha)q_H^2/4$. This bound is strictly below the benchmark profit $\pi^B=(1-\alpha)(q_H+q_L)^2/8$ under Assumption~\ref{assumption:benchmark-participation}.

Thus, the platform is more profitable when it retains both creators than when either creator exits. After $L$'s exit, the platform's profit is strictly below $\pi^B$ whether $H$ remains or exits. Analogously, the platform's profit is strictly below $\pi^B$ if $H$ exits. The platform can obtain a profit of $\pi^B$ by retaining both creators, pricing their content at $P^B$, and pricing AI out of the market.

Retaining both creators allows the platform to additionally earn AI revenue without reducing demand for creators' content. The platform can price creators' content and AI as follows:
\[
P=P^B=\frac{q_H+q_L}{4},
\qquad
P_{AI}=q^{\textrm{base}}+\frac{q_H+q_L}{2}.
\]
At these prices, consumers who previously purchased from creators continue to do so. Consumers who previously chose the outside option now purchase AI-generated content. Both creators receive the same revenue as in the benchmark before AI integration. This corresponds to row 2 of the table in Lemma~\ref{lemma:demands-nc}, which we call the \textit{three-product region}.

We now use consumer demand in the three-product region of Lemma~\ref{lemma:demands-nc} to derive how the platform prices creators' and AI-generated content to maximize profit while retaining both creators.

\begin{proposition}[Equilibrium prices and demand under NC]
\label{prop:nc-pricing}
Let $s\equiv\sqrt{q_L^2-4\phi q_L/\alpha}$. In every subgame-perfect equilibrium under NC (Section~\ref{subsec:timing}), both creators participate. The platform sets prices such that $L$'s participation constraint binds, with $V_L^{NC}=\phi q_L$. $H$'s constraint is slack, with $V_H^{NC}=\alpha(q_L+s)(2q_H-q_L-s)/4>\phi q_H$. The platform sets prices within the three-product region, with $\delta=(q_H-s)/2\in(g,1/2)$. Both creators and AI-generated content attract positive demand.
The equilibrium prices and demand are:
\begin{align*}
P^{NC}&=\frac{q_L+s}{2}, &
P_{AI}^{NC}&=q^{\textrm{base}}+\frac{q_H+q_L}{2},\\
D_L^{NC}&=\frac{q_L-s}{2}, &
D_H^{NC}&=\frac{2q_H-q_L-s}{2}, &
D_{AI}^{NC}&=1-q_H+s.
\end{align*}
\end{proposition}

\begin{proof}
See Appendix~\ref{apx:proof-nc}.
\end{proof}

Thus, when the platform does not compensate creators for training data, it sets the price of creators' content such that the low-quality creator just covers her participation cost, while the high-quality creator earns a positive net payoff. Although AI can perfectly match consumers' tastes, the platform prices creators' content below AI-generated content to preserve their demand and retain their training data. At the equilibrium AI price, the platform cannot raise the price of creators' content further without inducing the low-quality creator to exit and withdraw her training data.

Finally, we derive the platform's equilibrium profit in Corollary~\ref{cor:nc-profit} by substituting the equilibrium prices and demand from Proposition~\ref{prop:nc-pricing} into the NC profit expression in Equation~\amseqref{eq:platformprofit}.

\begin{corollary}[Platform profit under NC]
\label{cor:nc-profit}
The platform's profit $\pi^{NC}$ is given by:
\begin{equation}
\label{eq:nc-profit}
\pi^{NC}=(1-\alpha)\frac{(q_H-s)(q_L+s)}{2}
+(1-q_H+s)\left(q^{\textrm{base}}+\frac{q_H+q_L}{2}\right).
\end{equation}
$\pi^{NC}$ is strictly decreasing in $\phi$ and strictly increasing in $q^{\textrm{base}}$. Further, $\partial\pi^{NC}/\partial q^{\textrm{base}}=D_{AI}^{NC}$.
\end{corollary}
\begin{proof}
See Appendix~\ref{apx:proof-nc}.
\end{proof}
\noindent Corollary~\ref{cor:nc-profit} shows that higher participation costs reduce the platform's profit. To retain $L$, the platform lowers the price of creators' content, attracting enough additional consumers to raise her revenue and cover her higher participation cost. However, the platform receives less revenue per purchase from creators. In addition, consumers who switch from purchasing AI-generated content to purchasing from creators pay a lower price, and the platform now shares that revenue with creators. Retaining creators' training data thus becomes less profitable for the platform as their participation costs rise.

Proposition~\ref{prop:nc-pricing} and Corollary~\ref{cor:nc-profit} also show that better AI (higher $q^{\textrm{base}}$) increases the platform's profit, but does not further displace creators' demand when they are not compensated for their training data. As AI improves, consumers are willing to pay more for AI-generated content, but the platform must still preserve creators' demand to retain their training data. The platform maximizes profit by raising the AI price to capture the full increase in willingness to pay while charging the same price for creators' content.

\paragraph{Comparison with the benchmark.} Relative to the benchmark before AI integration, the platform earns higher profit under NC: $\pi^{NC}>\pi^B$ (Corollary~\ref{cor:profit-ranking}). It also prices creators' content weakly higher ($P^{NC}-P^B=(s-g)/2\geq0$). AI serves all consumers who chose the outside option before AI integration, a mass $1-(q_H+q_L)/2$, along with a mass $s-g\geq0$ of consumers who previously purchased from creators (where $s=g$ when $\phi=\phi^B$). The low-quality creator's payoff under NC is weakly lower than in the benchmark before AI integration, whereas the high-quality creator's payoff can rise or fall depending on whether the higher price offsets the reduction in demand.

\subsection{With Compensation (C)}
\label{subsec:withcompensation}

We now consider the compensation regime (C), in which the platform shares AI revenue with creators who contribute training data. A contributing creator can remain on the platform after losing all demand for her content if her share of AI revenue covers her participation cost. The platform can thus increase demand for AI-generated content while retaining creators and their training data. Participating creators choose whether to contribute training data, so the platform may receive data from both creators, one creator, or neither creator. We begin by characterizing consumer demand for each combination of creators' contribution decisions in Lemma~\ref{lemma:demands-c}.

\begin{lemma}[Consumer demand under C]
\label{lemma:demands-c}
Let $(s_H,s_L)$ indicate which creators contribute training data. For each combination of these decisions, the table shows AI quality, prompting effort, and consumer demand:
\par\nobreak\vspace{8pt}
\noindent\hfill\begin{minipage}{0.78\linewidth}
\normalfont
\setstretch{1}
\renewcommand{\arraystretch}{1.15}
\begin{tabular}{@{}
>{\raggedright\arraybackslash}m{0.13\textwidth}
>{\raggedright\arraybackslash}m{0.34\textwidth}
>{\raggedright\arraybackslash}m{0.12\textwidth}
>{\raggedright\arraybackslash}m{\dimexpr0.41\textwidth-6\tabcolsep\relax}
@{}}
\toprule
$(s_H,s_L)$ & $q_{AI}$ & $\kappa(x)$ & Consumer demand \\
\midrule
$(1,1)$ & $q^{\textrm{base}}+(q_H+q_L)/2$ & $0$ & As in Lemma~\ref{lemma:demands-nc} \\
\addlinespace[5pt]
$(1,0)$ & $q^{\textrm{base}}+q_H/2$ & $1-x$ & $D_HD_{AI}=0$ \\
\addlinespace[5pt]
$(0,1)$ & $q^{\textrm{base}}+q_L/2$ & $x$ & $D_LD_{AI}=0$ \\
\addlinespace[5pt]
$(0,0)$ & $q^{\textrm{base}}$ & $1$ & $D_{AI}P_{AI}=0$ \\
\bottomrule
\end{tabular}
\end{minipage}\hfill\null
\par\nobreak\vspace{8pt}
\noindent When only $H$ contributes training data ($(s_H,s_L)=(1,0)$), $H$'s content and AI-generated content cannot both attract positive demand. Analogously, when only $L$ contributes training data ($(s_H,s_L)=(0,1)$), $L$'s content and AI-generated content cannot both attract positive demand. When neither creator contributes training data, the platform earns no revenue from AI-generated content.
\end{lemma}
\begin{proof}
See Appendix~\ref{apx:proof-lemma-demands-c}.
\end{proof}

As in our discussion of Lemma~\ref{lemma:demands-nc}, Lemma~\ref{lemma:demands-c} helps explain why the platform obtains its highest profit when both creators contribute training data. Training on both creators' content improves AI quality and reduces the prompting effort required to generate content matching consumers' tastes. The platform can capture consumers' greater willingness to pay by charging a higher price for AI-generated content than when only one creator contributes training data. It retains the same share of AI revenue with two contributors as with one. The platform's payments to creators for training data can cover their participation costs even when their content attracts no demand.

We now use consumer demand in Lemma~\ref{lemma:demands-c} to derive how the platform prices creators' and AI-generated content to maximize profit while retaining both creators' training data. Without compensation, the platform preserves demand for both creators' content (Proposition~\ref{prop:nc-pricing}). Proposition~\ref{prop:c-pricing} shows that, with compensation, the platform instead sets prices at which AI fully displaces the low-quality creator. Both creators remain on the platform and contribute training data.

\begin{proposition}[Equilibrium prices and demand under C]
\label{prop:c-pricing}
In every equilibrium, both creators participate on the platform and contribute their content for AI training, so $e_H=e_L=s_H=s_L=1$. The following two cases give the equilibrium prices and demand depending on whether the AI model's quality before training on creators' content, $q^{\textrm{base}}$, is strictly below or greater than or equal to $(q_H-q_L)/2=\overline{q}^{\textrm{base}}$:

\begin{enumerate}
    \item \textbf{\textit{Partial displacement (CQ1):}}
    If $q^{\textrm{base}}<\overline{q}^{\textrm{base}}$, the platform sets prices at which AI fully displaces demand for the low-quality creator's content while consumers continue purchasing from the high-quality creator. The equilibrium prices and demand are:
    \begin{align*}
    P^{C}&=\frac{3q_H+q_L}{4}+\frac{q^{\textrm{base}}}{2}, &
    P_{AI}^{C}&=q^{\textrm{base}}+\frac{q_H+q_L}{2},\\
    D_L^{C}&=0, &
    D_H^{C}&=\frac{q_H-q_L}{4}-\frac{q^{\textrm{base}}}{2}, &
    D_{AI}^{C}&=1-D_H^{C}.
    \end{align*}

    \item \textbf{\textit{Full displacement (CQ2):}}
    If $q^{\textrm{base}}\geq\overline{q}^{\textrm{base}}$, the platform sets prices at which AI serves the entire market and neither creator's content attracts positive demand. The equilibrium prices and demand are:
    \begin{align*}
    P^{C}&\geq q_H, &
    P_{AI}^{C}&=q^{\textrm{base}}+\frac{q_H+q_L}{2},\\
    D_L^{C}&=0, &
    D_H^{C}&=0, &
    D_{AI}^{C}&=1.
    \end{align*}
    The platform can set any price $P^{C}\geq q_H$ for creators' content. Neither creator earns revenue from her own content at these prices, so raising the price further leaves consumer demand, the platform's profit, and creators' payoffs unchanged.
\end{enumerate}
\end{proposition}
\begin{proof}
See Appendix~\ref{apx:proof-c}.
\end{proof}
Proposition~\ref{prop:c-pricing} formalizes why compensation allows the platform to retain creators' training data without preserving demand for their content. The platform can price creators' content high enough to fully displace the low-quality creator's demand because the low-quality creator's share of AI revenue covers her participation cost. When AI-generated content has lower quality than the high-quality creator's content, consumers with tastes close to the high-quality creator's style pay more to purchase from the high-quality creator than to purchase AI-generated content.

As AI improves, the platform raises the AI price to capture consumers' greater willingness to pay for AI-generated content. The platform then retains more revenue from each consumer purchasing AI-generated content, making it more profitable to raise the price of creators' content and accept lower demand for the high-quality creator's content. Once AI-generated content matches the high-quality creator's content in quality, AI serves the entire market, and both creators remain as training data suppliers because their shares of AI revenue cover their participation costs. At the equilibrium prices, each creator prefers to contribute when the other contributes. A creator who opts out of AI training loses her share of AI revenue without increasing demand for her content.

\paragraph{Comparison with the benchmark.} Relative to the benchmark before AI integration, the platform earns higher profit under C: $\pi^{C}>\pi^B$ (Corollary~\ref{cor:profit-ranking}). It also prices creators' content strictly higher ($P^{C}>P^B$). AI serves all consumers who chose the outside option before AI integration, a mass $1-(q_H+q_L)/2$, along with a mass $(q_H+q_L)/2-D_H^{C}>0$ of consumers who previously purchased from creators (including at $\phi=\phi^B$). Both creators' payoffs under C are strictly higher than in the benchmark before AI integration because their shares of AI revenue more than offset the reduction in revenue from their own content.

\paragraph{Comparison with the regime without compensation (NC).} Whether the platform earns higher profit under C or NC depends on creators' participation costs and AI quality (Corollary~\ref{cor:profit-ranking}). It prices creators' content strictly higher under C ($P^C>P^{NC}$). The platform trains AI on both creators' content in both regimes ($s_H=s_L=1$). The quality and price of AI-generated content are also identical in both regimes ($P_{AI}^C=P_{AI}^{NC}=q^{\textrm{base}}+(q_H+q_L)/2$).

We now examine how compensation changes both demand for creators' content and the revenue that sustains their participation on the platform. We first compare demand for creators' and AI-generated content under C and NC (Corollary~\ref{cor:compensation-displacement}). Under C, a creator who loses all demand for her content must cover her participation cost through her share of AI revenue. We thus examine whether displaced creators earn enough to remain on the platform and how their payoffs change as AI improves (Corollary~\ref{cor:data-supplier}).

\begin{corollary}[Compensation increases AI displacement]
\label{cor:compensation-displacement}
Equilibrium demand under C and NC satisfies:
\[
D_L^{C}=0<D_L^{NC},
\qquad
D_H^{C}<D_H^{NC},
\qquad
D_{AI}^{C}>D_{AI}^{NC}.
\]
\end{corollary}
\begin{proof}
See Appendix~\ref{apx:proof-c}.
\end{proof}

\noindent Corollary~\ref{cor:compensation-displacement} shows that compensation increases AI displacement even though both creators contribute their content for AI training in both regimes, and the quality and price of AI-generated content are identical across regimes. Without compensation, the platform must preserve enough demand for the low-quality creator's content for her revenue to cover her participation cost; otherwise, she exits and withdraws her training data. With compensation, her share of AI revenue covers her participation cost even when consumers no longer purchase from her. The platform can thus raise the price of creators' content, inducing more consumers to purchase AI-generated content while retaining creators' training data.

Displaced creators no longer earn revenue from their own content and must cover their participation costs through their share of AI revenue. We next characterize displaced creators' net payoffs and how AI improvements affect the low-quality creator's payoff in Corollary~\ref{cor:data-supplier}.

\begin{corollary}[Displaced creators remain as training data suppliers]
\label{cor:data-supplier}
At the equilibrium prices of Proposition~\ref{prop:c-pricing}, each creator $j\in\{H,L\}$ earns strictly more than her participation cost, $V_j^C>\phi q_j$, under both partial displacement (CQ1) and full displacement (CQ2). In particular, the low-quality creator's content attracts no demand, but she earns a positive net payoff, with $D_L^C=0$, $V_L^C>\phi q_L$, and $V_L^C$ strictly increasing in $q^{\textrm{base}}$. Under full displacement (CQ2), neither creator's content attracts positive demand, and both creators earn only their shares of AI revenue, with $V_H^C=V_L^C=\alpha q_{AI}/2$.
\end{corollary}
\begin{proof}
See Appendix~\ref{apx:proof-c}.
\end{proof}

\begin{figure}[t]
\centering
\captionsetup{position=top,skip=10pt}
\caption{Consumer Demand With and Without Compensation}
\label{fig:market-split}
\includegraphics[width=\textwidth]{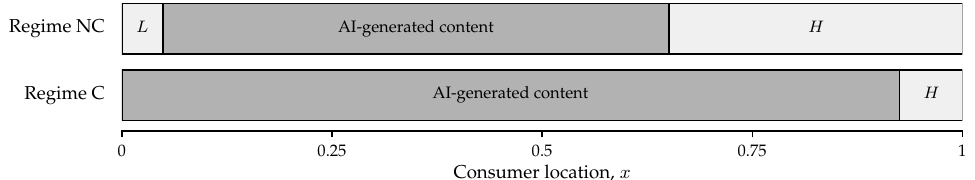}
\par\vspace{6pt}
\parbox{\textwidth}{\raggedright
\small\textit{Note.}
Equilibrium consumer demand at $q_H=0.7$, $q_L=0.4$, $\alpha=0.3$, $\phi=\phi^B/2$, and $q^{\textrm{base}}=0$. Segments labeled L and H show consumers purchasing from the corresponding creator. Both creators contribute training data in both regimes. Under C, consumers no longer purchase from $L$, who remains on the platform as a training data supplier.}
\end{figure}

\noindent Corollary~\ref{cor:data-supplier} shows that compensation allows creators to earn positive net payoffs from supplying training data after losing all demand for their content. Consumers no longer purchase from the low-quality creator under compensation, as illustrated in Figure~\ref{fig:market-split}, but the platform continues training AI on her content and sharing AI revenue with her. Her share of AI revenue exceeds her participation cost, making it profitable to remain on the platform solely as a training data supplier.

As AI improves, the low-quality creator benefits even though consumers no longer purchase from her. Consumers become willing to pay more for AI-generated content, and the platform raises the AI price. Under partial displacement (CQ1), the platform also raises the price of creators' content, inducing more consumers to purchase AI-generated content instead of purchasing from the high-quality creator. The higher AI price and greater demand for AI-generated content increase the AI revenue shared with the low-quality creator and thus her net payoff. Once AI serves the entire market under full displacement (CQ2), the high-quality creator also earns revenue solely from supplying training data. Further AI improvements allow the platform to charge more for AI-generated content, so both creators earn more from supplying training data.

\subsection{The Platform's Compensation Decision}
\label{subsec:platform-choice}
\label{subsec:comparison-welfare}

We now examine when the platform chooses to compensate creators for training data. Without compensation, the platform must preserve demand for creators' content to retain their training data. With compensation, the platform shares AI revenue with creators and can retain their training data even when consumers no longer purchase from them. Corollary~\ref{cor:profit-ranking} compares the platform's profit across the two regimes and the benchmark before AI integration.

\begin{corollary}[The platform's compensation decision]
\label{cor:profit-ranking}
The platform earns higher profit under both C and NC than in the benchmark before AI integration, with $\pi^C>\pi^B$ and $\pi^{NC}>\pi^B$. Moreover, $\pi^{NC}$ strictly decreases in $\phi$, whereas $\pi^C$ is independent of $\phi$. For each $q^{\textrm{base}}$, there is a threshold $\phi^\star(q^{\textrm{base}})\in[0,\phi^B]$ such that the platform prefers NC when $\phi<\phi^\star$ and C when $\phi>\phi^\star$. If $0<\phi^\star<\phi^B$, the platform is indifferent between the two regimes at $\phi=\phi^\star$, with $\pi^{NC}(\phi^\star)=\pi^C$. If $\phi^\star=0$, the platform prefers C for every $0<\phi\leq\phi^B$; if $\phi^\star=\phi^B$, it weakly prefers NC throughout this range.
\end{corollary}
\begin{proof}
See Appendix~\ref{apx:proof-welfare}.
\end{proof}

\noindent Corollary~\ref{cor:profit-ranking} shows that the platform may prefer not to compensate creators (NC) even though compensating creators (C) allows it to raise the price of creators' content and increase demand for AI-generated content. Under C, the platform shares AI revenue with creators, reducing the revenue it retains from each consumer purchasing AI-generated content. Under NC, it retains all AI revenue but must preserve enough demand for creators' content to cover their participation costs and retain their training data. When participation costs are low, the low-quality creator can cover her cost even when few consumers purchase from her. The platform can then prefer NC because preserving the demand needed to retain creators reduces its profit by less than sharing AI revenue under C.

\begin{figure}[p]
\centering
\caption{Platform Profit With and Without Compensation}
\label{fig:profit-ranking}
\includegraphics[width=0.60\textwidth]{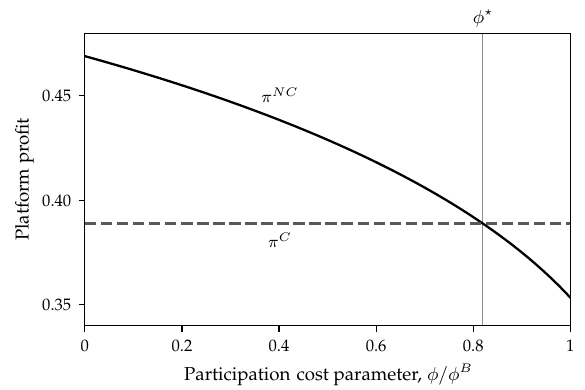}
\vspace{3pt}

\parbox{\textwidth}{\raggedright\small
\textit{Note.} Equilibrium platform profit as $\phi/\phi^B$ varies, with $q_H=0.7$, $q_L=0.4$, $\alpha=0.3$, and $q^{\textrm{base}}=0$. B denotes the benchmark before AI integration. The vertical line marks $\phi=\phi^\star$, where the platform earns the same profit under NC and C. The platform prefers NC when $\phi<\phi^\star$ and C when $\phi>\phi^\star$.
}
\end{figure}

Higher creator participation costs can make the platform prefer C. Under NC, the platform lowers the price of creators' content so the low-quality creator can cover her higher participation cost. More consumers then purchase from creators instead of purchasing AI-generated content. They pay a lower price, and the platform retains only the share $1-\alpha$ of that revenue, reducing its profit under NC. Under C, the platform can retain creators without changing its optimal prices or the AI revenue shared with them throughout $0<\phi\leq\phi^B$, so its profit remains unchanged. The platform thus prefers C when $\phi>\phi^\star$. Figure~\ref{fig:profit-ranking} illustrates an example of this comparison.

The platform's compensation decision also depends on AI quality. As AI improves, the platform raises the price of AI-generated content by the same amount in both regimes. Under C, more consumers purchase AI-generated content than under NC, but the platform retains a smaller share of AI revenue. When the platform is initially indifferent between NC and C, the greater demand under C allows it to benefit more from the higher price despite sharing AI revenue with creators. Better AI can thus make the platform prefer C at lower creator participation costs.\footnote{Appendix~\ref{apx:proof-welfare} establishes this result formally.}

\begin{figure}[p]
\centering
\caption{The Platform's Compensation Decision and Creator Displacement}
\label{fig:regime-map}
\includegraphics[width=0.60\textwidth]{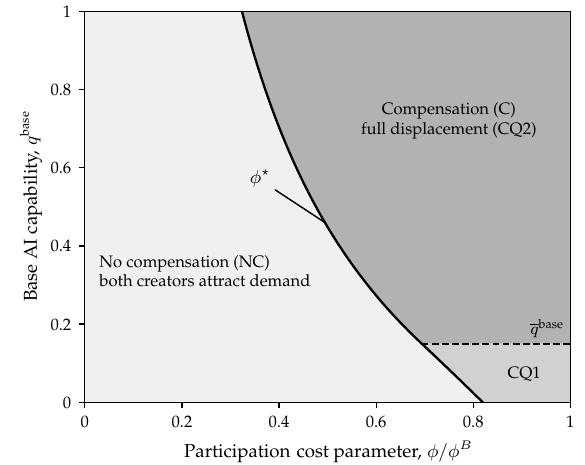}
\vspace{3.25pt}

\parbox{\textwidth}{\raggedright\small
\textit{Note.} The platform's compensation decision and creator displacement as $\phi/\phi^B$ and $q^{\textrm{base}}$ vary, with $q_H=0.7$, $q_L=0.4$, and $\alpha=0.3$. The solid curve marks $\phi=\phi^\star$, where the platform earns the same profit under NC and C. The platform prefers NC when $\phi<\phi^\star$ and C when $\phi>\phi^\star$. Under NC, both creators attract demand. Under C, the dashed line marks $q^{\textrm{base}}=\overline{q}^{\textrm{base}}$, separating partial displacement (CQ1), where only $H$ attracts positive demand, from full displacement (CQ2), where AI serves the entire market. Both creators contribute training data in both regimes.
}
\end{figure}

Figure~\ref{fig:regime-map} illustrates when the platform chooses to compensate creators and which creators continue to attract demand. The platform prefers NC at lower creator participation costs and C at higher costs, with the solid curve marking where it is indifferent. Where the platform prefers C, the dashed line separates partial displacement (CQ1) from full displacement (CQ2).

\subsection{Welfare Analysis}
\label{subsec:welfare}

We now examine the welfare consequences of the greater AI displacement induced by compensating creators for training data. We first compare creators' net payoffs and consumer surplus under C and NC, then assess whether the resulting allocation increases total surplus. We determine when the platform's private compensation decision selects the regime with lower total surplus.

For each regime $r\in\{NC,C\}$, let $\mathrm{CS}^r$ denote consumer surplus, let $R_j^r\equiv V_j^r-\phi q_j$ denote creator $j$'s net payoff, and define total surplus as $\mathrm{TS}^r\equiv\pi^r+R_H^r+R_L^r+\mathrm{CS}^r$.

\begin{proposition}[Welfare effects of compensation]
\label{prop:welfare}
Compensation has the following effects on welfare:
\begin{enumerate}
    \item \textbf{\textit{Creators' net payoffs.}} Both creators earn strictly higher net payoffs under C than NC:
    \[
    R_L^{NC}=0<R_L^C,\qquad R_H^{NC}<R_H^C.
    \]
    Both differences in net payoffs increase strictly with $q^{\textrm{base}}$.

    \item \textbf{\textit{Consumer surplus.}} Consumer surplus is strictly lower under C than NC:
    \[
    \mathrm{CS}^C<\mathrm{CS}^{NC}.
    \]
    The reduction in consumer surplus increases strictly with $q^{\textrm{base}}$ for $q^{\textrm{base}}<\overline{q}^{\textrm{base}}$, where $\overline{q}^{\textrm{base}}=(q_H-q_L)/2$. For $q^{\textrm{base}}\geq\overline{q}^{\textrm{base}}$, AI serves the entire market under C, and the platform captures consumers' full willingness to pay, so $\mathrm{CS}^C=0$ and the reduction remains constant.

    \item \textbf{\textit{Total surplus.}} Total surplus is strictly higher under C than NC:
    \[
    \mathrm{TS}^C>\mathrm{TS}^{NC}.
    \]
    The increase in total surplus grows strictly with $q^{\textrm{base}}$.
\end{enumerate}
\end{proposition}
\begin{proof}
See Appendix~\ref{apx:proof-welfare}.
\end{proof}

\noindent Compensation increases both creators' net payoffs despite reducing demand for their content. Under NC, the platform sets the price of creators' content so that the low-quality creator's revenue exactly covers her participation cost, leaving her with zero net payoff. Under C, she loses all demand for her content, but her share of AI revenue covers her participation cost and leaves her with a positive net payoff. The high-quality creator also earns a higher net payoff under C, as her share of AI revenue more than offsets the reduction in revenue from her own content.

Compensation reduces consumer surplus. The platform charges the same price for AI-generated content of identical quality in both regimes, but a higher price for creators' content under C. Consumers who continue purchasing from creators pay more, while some consumers who previously obtained positive surplus from creators' content instead purchase AI-generated content. The platform prices AI-generated content to capture consumers' full willingness to pay, so those who switch to AI receive no surplus. Compensation thus reduces consumer surplus through the higher price of creators' content and the resulting consumer demand (not AI quality or price).

Total surplus nevertheless increases. Both regimes serve the entire market, retain both creators and their training data, and offer AI-generated content of identical quality. Since prices and revenue sharing payments are transfers, the difference in total surplus arises from which content consumers purchase. Under NC, the platform preserves demand for the low-quality creator's content to cover her participation cost. AI trained on both creators' content offers higher quality and perfectly matches those consumers' tastes. Under C, the platform can retain her training data without preserving her sales, allowing more consumers to purchase AI-generated content. Across consumers who switch from creators' content to AI-generated content, the gains in total surplus exceed the losses from reduced purchases of the high-quality creator's content.

As AI capability improves, the differences in both creators' net payoffs and total surplus between C and NC increase strictly. Under NC, the platform raises the price of AI-generated content without changing the price or demand for creators' content, leaving creators' net payoffs and consumer surplus unchanged. Under C, the platform raises the price of AI-generated content and progressively reduces demand for the high-quality creator's content, increasing the AI revenue shared with creators. More consumers purchase AI-generated content under C than NC, so improvements in AI quality contribute more to total surplus under C. Consumer surplus under C declines with demand for the high-quality creator's content and reaches zero once AI serves the entire market. The consumer surplus loss relative to NC remains constant thereafter.

The platform's privately preferred regime need not generate higher total surplus. Under C, the platform treats the AI revenue it shares with creators as a cost, whereas revenue sharing is a transfer when calculating total surplus. The platform also does not capture the full increase in total surplus from greater demand for AI-generated content. When participation costs are low, the platform can prefer NC because retaining creators through content sales is less costly than sharing AI revenue. Corollary~\ref{cor:profit-ranking} further shows that improving AI can make the platform prefer C at lower participation costs, reducing the set of participation costs for which it chooses NC.

\begin{figure}[t]
\centering
\caption{Welfare Differences Between Compensation and No Compensation}
\label{fig:welfare}
\includegraphics[width=\textwidth]{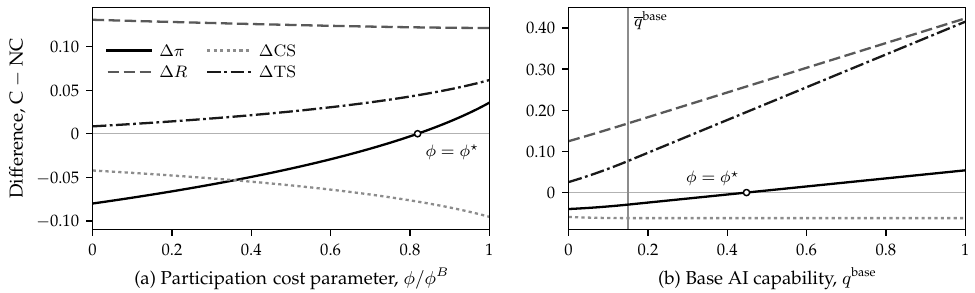}
\vspace{3pt}

\parbox{\textwidth}{\raggedright\small
\textit{Note.} Both panels show differences in platform profit ($\Delta\pi$), creators' net payoffs ($\Delta R$), consumer surplus ($\Delta\mathrm{CS}$), and total surplus ($\Delta\mathrm{TS}$) between C and NC. Parameters are $q_H=0.7$, $q_L=0.4$, and $\alpha=0.3$. Panel~(a) varies $\phi/\phi^B$ at $q^{\textrm{base}}=0$. Panel~(b) varies $q^{\textrm{base}}$ at $\phi=\phi^B/2$, with the full displacement threshold $\overline{q}^{\textrm{base}}$ marked.
}
\end{figure}

When $\phi<\phi^\star(q^{\textrm{base}})$, the platform prefers NC, but a requirement to compensate creators would induce the platform to adopt C. Requiring compensation would increase both creators' net payoffs and total surplus while reducing consumer surplus and platform profit. A policymaker seeking to maximize total surplus would favor requiring compensation when the platform would otherwise choose NC, whereas a policymaker placing greater weight on consumer surplus or platform profit might prefer NC. Figure~\ref{fig:welfare} illustrates these comparisons: compensation increases creators' net payoffs and total surplus and reduces consumer surplus throughout, while the difference in platform profit changes sign at the compensation threshold.

%% file: sections/extension.tex
\section{Extensions}
\label{sec:extensions}

The main analysis shows compensating creators for training data allows the platform to displace their direct sales while retaining them as data suppliers. This result relies on the AI's ability to attract consumers away from creators and on compensation being sufficient to sustain creator participation. We extend the analysis by relaxing three assumptions of the baseline model: perfect AI steerability, equal transfer rates of creators' data, and equal splitting of AI revenue among contributors. These extensions clarify when compensation continues to facilitate AI displacement and when displaced creators can remain on the platform as data suppliers.

\subsection{Imperfect Steerability}
\label{subsec:imperfect-steerability}

The main analysis assumes perfect AI steerability ($\kappa_0=0$), so consumers incur no prompting effort when AI is trained on both creators' content. In this section, we relax this assumption by allowing $\kappa_0>0$, which captures residual prompting effort due to the model's inherent limitations, even when both creators contribute data. A higher $\kappa_0$ means a less steerable model, rendering AI less attractive than creators' content and possibly limiting the platform's ability to displace their sales.

When both creators contribute, consumer utility from AI is $U_{AI}=q^{\text{base}}+(q_H+q_L)/2-P_{AI}-\kappa_0$. Since base AI capability and prompting effort enter utility with opposite signs, we define effective AI capability as $\theta\equiv q^{\text{base}}-\kappa_0$. For $\theta\geq 0$, the results of Section~\ref{sec:analysis} still hold with $\theta$ replacing $q^{\text{base}}$. We therefore focus on $\theta<0$, where prompting effort more than offsets base AI capability.

We first consider the NC regime, where creators’ participation depends only on their own sales. We ask how prompting effort
changes the platform’s prices and the allocation of demand.

\begin{proposition}[Equilibrium under NC with imperfect steerability]
\label{prop:imperfect-steerability-nc}
There exists a threshold $\widehat{\kappa}_0(\phi)>q^{\text{base}}$ such that, for $\kappa_0<\widehat{\kappa}_0(\phi)$, the equilibrium is that of Proposition~\ref{prop:nc-pricing} with $\theta$ replacing $q^{\text{base}}$. In particular, the human price, consumer demands, and creators' net payoffs remain unchanged, while the AI price decreases one for one with $\kappa_0$ and the platform's profit decreases at a rate equal to AI demand $D_{AI}^{NC}$.
\end{proposition}
\begin{proof}
See Appendix~\ref{apx:proof-imperfect-steerability-nc}.
\end{proof}

Proposition \ref{prop:imperfect-steerability-nc} shows that under NC, the platform alone bears the cost of imperfect steerability because the low-quality creator's participation constraint $\alpha D_LP\geq\phi q_L$ does not depend on AI revenue or prompting effort. The platform therefore maintains the same human price and consumer allocation as in Proposition~\ref{prop:nc-pricing}, while lowering the AI price to offset the additional prompting effort. This holds when $\kappa_0<\widehat{\kappa}_0(\phi)$, i.e., as long as AI sales remain profitable enough for the low-quality creator's participation constraint to determine the platform's optimal prices.

Under C, prompting effort also affects creators: it reduces both the AI's appeal to consumers and the AI revenue that compensates them. Our next proposition reveals whether the platform can continue to displace the low-quality creator while retaining her as a training data supplier.

\begin{proposition}[Equilibrium under C with imperfect steerability]
\label{prop:imperfect-steerability}
Define $\underline{\theta}_{C}\equiv -(1+\sqrt{2})(q_H-q_L)/2$. Let $\underline{\theta}_{L}(\phi)\in(-(q_H+q_L)/2,0)$ be the value of $\theta$ at which the low-quality creator's share of AI revenue equals her participation cost $\phi q_L$, with compensation evaluated using the equilibrium prices and demand in Proposition~\ref{prop:c-pricing}, replacing $q^{\textrm{base}}$ with $\theta$. The equilibrium depends on effective AI capability $\theta$ as follows:

\begin{enumerate}
    \item \textbf{\textit{Displacement as in Proposition~\ref{prop:c-pricing} (CQ1 if $\theta<\overline{q}^{\textrm{base}}$, CQ2 otherwise):}}
    If $\theta\geq\max\{\underline{\theta}_{C},\underline{\theta}_{L}(\phi)\}$, the equilibrium is that of Proposition~\ref{prop:c-pricing} with $\theta$ replacing $q^{\textrm{base}}$: the low-quality creator is fully displaced but remains a training data supplier, earning a strictly positive net payoff from AI compensation for every such $\theta$ except $\theta=\underline{\theta}_{L}(\phi)$. At $\theta=\underline{\theta}_{C}$, the three-product configuration (item~2) is also an equilibrium.

    \item \textbf{\textit{Three-product configuration:}}
    If $-(q_H+q_L)/2<\theta<\underline{\theta}_{C}$ or $\theta=\theta_C<\theta_L(\phi)$, the platform continues to offer AI, but the low-quality creator regains positive demand for her own content.

    \item \textbf{\textit{No AI revenue:}}
    If $\kappa_0\geq\kappa_0^*\equiv q^{\textrm{base}}+(q_H+q_L)/2$, i.e., $\theta\leq-(q_H+q_L)/2$, AI earns no revenue and the outcome is the benchmark of Proposition~\ref{prop:benchmark}; for $\kappa_0>\kappa_0^*$, AI makes no sales.
\end{enumerate}

\end{proposition}
\begin{proof}
See Appendix~\ref{apx:proof-imperfect-steerability}.
\end{proof}

Proposition \ref{prop:imperfect-steerability} shows that compensation still increases AI displacement under moderate prompting effort: for $\kappa_0<\widehat{\kappa}_0(\phi)$,
$\theta>\theta_C$, and $\theta>\theta_L(\phi)$, the inequalities in Corollary~\ref{cor:compensation-displacement} still hold, and the low-quality creator remains a data supplier after losing all direct sales. When $\kappa_0$ increases, the platform lowers the AI price to offset the extra effort. Because AI sales become less profitable, it also lowers the human price, allowing the high-quality creator to retain more consumers.

Less steerable AI limits this effect in two ways. First, when effective AI capability falls below $\underline{\theta}_{C}$, the platform instead prefers a configuration in which all three products attract demand. The low-quality creator then regains positive direct sales while still contributing data. Second, as prompting effort increases, lower AI revenue also reduces the compensation of contributing creators. If the displaced creator's participation constraint binds before the platform reaches the three-product configuration, the prices of Proposition~\ref{prop:c-pricing} can no longer sustain her participation, and we do not characterize the equilibrium for $\theta_C<\theta<\theta_L(\phi)$. Which occurs first depends on creator qualities and participation costs: when the quality gap is large enough, the participation constraint binds before the platform switches to the three-product configuration.

\subsection{Heterogeneous Transfer Rates}
\label{subsec:heterogeneous-data}

The main analysis assumes equal transfer rates ($\tau_H=\tau_L=1/2$), so each creator's contribution to AI quality is proportional to her content quality. We relax this assumption to allow creators' data to differ in how effectively they improve AI quality. Specifically, we fix $\tau_H=1/2$ and allow $0<\tau_L\leq 1/2$, while maintaining equal splitting of AI revenue among contributors. When both creators contribute, AI quality becomes $q_{AI}=q^{\text{base}}+\tau_Hq_H+\tau_Lq_L$.

A lower $\tau_L$ reduces AI quality, which can change the platform's prices, the split of demand between AI and human content, and creator compensation. We therefore ask whether the platform can still retain both creators' data while fully displacing the low-quality creator. We focus on $q_H/q_L\geq\sqrt{2}$, a sufficient condition under which the platform still prefers full displacement and equal revenue sharing still covers the low-quality creator’s participation cost, even as her contribution to AI quality approaches zero.

\begin{proposition}[Heterogeneous transfer rates]
\label{prop:heterogeneous-data}
Suppose $q_H/q_L\geq\sqrt{2}$, $\tau_H=1/2$, and $0<\tau_L\leq 1/2$, and maintain equal splitting of AI revenue. Let $q_{AI}=q^{\text{base}}+\tau_Hq_H+\tau_Lq_L$. In equilibrium, both creators contribute, $P_{AI}=q_{AI}$, and $D_L^C=0$. If $q_{AI}<q_H$ (CQ1), $P=(q_H+q_{AI})/2$ and $D_H^C=(q_H-q_{AI})/2$; otherwise (CQ2), $D_{AI}^C=1$ at every human price $P\geq q_H$. In particular:

\begin{enumerate}

    \item A lower $\tau_L$ reduces the equilibrium AI price and the platform's profit.

    \item In CQ1, a lower $\tau_L$ reduces the human price and AI demand while increasing demand for the high-quality creator's content. In CQ2, AI continues to serve the entire market.

    \item Both creators still contribute their data, and the low-quality creator remains fully displaced from direct sales ($D_L^C=0$), earning a strictly positive net payoff from AI compensation for every such $\tau_L$.

\end{enumerate}
\end{proposition}
\begin{proof}
See Appendix~\ref{apx:proof-heterogeneous-data}.
\end{proof}

Proposition~\ref{prop:heterogeneous-data} shows that heterogeneous transfer rates change how demand and payoffs are distributed but not creators' roles: the platform still displaces the low-quality creator's sales while retaining her data, even when that data contributes little to AI quality. A lower transfer rate reduces AI quality, making AI less competitive with human content. The platform responds by lowering the AI price. In CQ1, it also lowers the human price, shifting some consumers from AI to the high-quality creator's content. The resulting decline in AI revenue also lowers platform profit.

Both creators' payoffs decrease as $\tau_L$ falls, even though their shares of AI revenue are unchanged. Under equal splitting, each creator receives half of the creators' AI revenue, so a lower $\tau_L$ reduces the compensation received by both creators. In CQ1, the high-quality creator also gains direct demand, but this does not offset the decline in her AI compensation and the lower human price.

Despite these changes, the low-quality creator continues to participate. Although her contribution to AI quality becomes smaller, her data remains essential for AI steerability: without it, the AI attracts no consumers at the equilibrium prices. Moreover, equal splitting ties her compensation to total AI revenue rather than to her individual contribution to AI quality. This payment remains sufficient to cover her participation cost even as $\tau_L$ approaches zero (Appendix~\ref{apx:proof-heterogeneous-data}).

\subsection{Quality-Weighted Compensation}
\label{subsec:quality-weighted}

The main analysis assumes the creators' share of AI revenue is split equally among contributors, regardless of how much each creator's data improves AI quality. Under compensation, this rule also determines the marginal participant: since creator $j$ participates if and only if $V_j/q_j\geq\phi$, this is the creator with the smaller ratio. Without compensation, $L$ is the marginal participant: the platform holds her exactly at her participation constraint (Proposition~\ref{prop:nc-pricing}), while $H$, with her larger market, has slack. The next corollary asks whether this remains true under C with equal splitting.

\begin{corollary}[Reversal under equal splitting]
\label{cor:reversal}
Without compensation, $V_L^{NC}/q_L=\phi<V_H^{NC}/q_H$, whereas with compensation and equal splitting of AI revenue, $V_H^{C}/q_H<V_L^{C}/q_L$.
\end{corollary}
\begin{proof}
See Appendix~\ref{apx:proof-reversal}.
\end{proof}

Corollary~\ref{cor:reversal} shows that compensation reverses the order under equal splitting: both contributors receive the same share of AI revenue, but participation costs rise with quality, and $H$'s remaining revenue from direct sales is too small to make up for the difference.

We now consider an alternative compensation rule that allocates the creators' share of AI revenue in proportion to each creator's contribution to AI quality. Specifically, we replace $s_j/(s_H+s_L)$ in \eqref{eq:creator-payoff} with $\tau_jq_js_j/(\tau_Hq_Hs_H+\tau_Lq_Ls_L)$, and we mark equilibrium objects under this rule with the superscript $\mathrm{QW}$. Under the baseline assumption of equal transfer rates ($\tau_H=\tau_L=1/2$), creator $j$ receives a share $q_j/(q_H+q_L)$ of the creators' AI revenue when both contribute. The next proposition examines whether this redistribution undoes the reversal in Corollary~\ref{cor:reversal} and whether it also changes the equilibrium allocation.

\begin{proposition}[Quality-weighted compensation]
\label{prop:quality-weighted}
Suppose $\tau_H=\tau_L=1/2$. Under quality-weighted compensation, the equilibrium prices, demands, platform profit, consumer surplus, and creators' total net payoff remain unchanged from Proposition~\ref{prop:c-pricing}. Both creators still contribute their data and the low-quality creator remains fully displaced from direct sales. Quality weighting redistributes AI revenue from the low-quality creator to the high-quality creator, while both continue to earn strictly positive net payoffs.

When transfer rates are heterogeneous, suppose $q_H/q_L\geq\sqrt{2}$, $\tau_H=1/2$, and $0<\tau_L\leq 1/2$. The outcome of Proposition~\ref{prop:heterogeneous-data} remains an equilibrium under quality weighting if and only if $\tau_L\geq\tau_L^{\star}$, where $\tau_L^{\star}\in(0,1/2)$ depends on $\phi$ and $\alpha$ only through $\phi/\alpha$. The low-quality creator gains a strictly positive net payoff when $\tau_L>\tau_L^{\star}$, and her participation constraint binds at $\tau_L=\tau_L^{\star}$.
\end{proposition}

\begin{proof}
See Appendices~\ref{apx:proof-heterogeneous-data} and~\ref{apx:proof-quality-weighted}.
\end{proof}

Proposition~\ref{prop:quality-weighted} shows that quality weighting changes how creators share AI revenue but not the equilibrium allocation. Quality weighting does not alter the platform's profit under any contribution profile, since its retained share of AI revenue remains $1-\alpha$. At the equilibrium prices of Proposition~\ref{prop:c-pricing}, neither creator can generate AI revenue alone, and both still earn positive net payoffs when contributing together. Relative to equal splitting, quality weighting transfers $\alpha(q_H-q_L)D_{AI}^C P_{AI}^C/[2(q_H+q_L)]$ from $L$ to $H$, leaving creators' total net payoff unchanged.

More importantly, this redistribution undoes the reversal in Corollary~\ref{cor:reversal}. Under quality weighting, both creators receive the same AI compensation per unit of quality, but $H$ also earns revenue from direct sales in CQ1. Hence, $V_H^{C,\mathrm{QW}}/q_H-V_L^{C,\mathrm{QW}}/q_L=\alpha D_H^C P^C/q_H\geq 0$, with strict inequality in CQ1. In CQ1, $L$ is again the marginal participant; in CQ2, the two participation constraints are equally tight. Thus, the reversal is driven by equal splitting rather than compensation itself.

With heterogeneous transfer rates, quality weighting sustains the outcome of Proposition~\ref{prop:heterogeneous-data} only if $\tau_L\geq\tau_L^{\star}$, whereas equal splitting sustains it for every $\tau_L$. As $\tau_L$ decreases, $L$ receives a smaller share of declining AI revenue. Below $\tau_L^{\star}$, her compensation at the prices of Proposition~\ref{prop:heterogeneous-data} no longer covers her participation cost and we do not discuss the resulting equilibrium.

%% file: sections/conclusion.tex
\section{Conclusion}
\label{sec:conclusion}

We study compensation for training data in a model of a creative platform where consumers choose between purchasing from creators and prompting an integrated generative AI system. We compare a regime that makes training data use a condition of platform participation with one that compensates creators and allows them to opt out of AI training. In the absence of compensation, creators cover their participation costs through their content sales, which constrains the platform to preserve their demand. With compensation, the low-quality creator chooses to remain solely as a training data supplier while AI fully displaces her demand. The platform sets a higher price for creators' content with compensation, although both creators contribute training data and the quality and price of AI-generated content are identical in the two regimes.

As AI improves, the platform raises the price of AI-generated content identically with and without compensation. Without compensation, it leaves the price of creators' content unchanged to preserve their demand and retain their training data. With compensation, it also raises the price of creators' content, progressively displacing the remaining demand for creators until AI serves the entire market. Both creators nevertheless remain on the platform, earning positive net payoffs from supplying training data. Further AI improvements increase their revenue from training data even after AI fully displaces their demand.

Requiring compensation increases creators' net payoffs and total surplus but reduces consumer surplus. The increase in total surplus comes from reallocating demand toward AI-generated content that better matches consumers' tastes, without expanding market coverage or improving AI quality. Consumers lose surplus as the platform raises the price of creators' content and captures the willingness to pay of consumers purchasing AI-generated content. The platform's compensation decision depends on creator participation costs and AI capability: retaining creators through content sales can be more profitable than sharing revenue from AI-generated content sales when participation costs are low. Revenue sharing is a cost to the platform but a transfer when calculating total surplus. Requiring compensation in this case increases total surplus, although consumers lose surplus and the platform loses profit. As AI improves, the range of creator participation costs for which the platform prefers not to compensate can narrow.

A platform can depend on creators' training data while competing with their content if it can compensate them without preserving their sales. For policymakers, the relevant comparison must account for how the platform changes prices in response to compensation. Requiring compensation changes what consumers purchase and how revenue is divided between creators and the platform. Preserving demand for creators' content and sustaining creators' net payoffs are different objectives, and neither alone captures the consequences for consumers or total surplus. Policymakers may prefer different compensation requirements depending on how they value total surplus and the payoffs of creators, consumers, and the platform.

We extend the model to consider imperfect AI steerability, heterogeneous transfer rates, and alternative compensation rules. Compensation-induced displacement persists with moderate prompting effort, although sufficiently high prompting effort can restore demand for creators' content. Greater prompting effort also reduces AI revenue and can leave compensation below a displaced creator's participation cost at the prices derived for imperfect steerability. When compensation is weighted by contributions to AI quality, displacement persists under equal transfer rates, but the distribution of revenue between creators changes. With quality-weighted compensation and heterogeneous transfer rates, the low-quality creator may contribute too little to AI quality for her share of revenue to cover her participation cost at the prices characterized in our analysis. Equal revenue sharing can sustain her participation because the platform pays her a fixed share of total AI revenue. These extensions show that compensation-induced displacement depends on consumers' willingness to purchase AI-generated content and whether compensation covers displaced creators' participation costs.

Our work has several limitations. We assume that creators withdraw their training data when they exit the platform. This assumption is most relevant when training licenses must be renewed or creators continue producing new content. If the platform retains perpetual rights to previously supplied content, creator exit may have a weaker effect on AI quality and prompting effort, reducing the platform's incentive to preserve demand for creators' content without compensation. We also consider two creators whose contributions affect AI quality and prompting effort, and a common price for creators' content. Future research could allow creators to choose content quality and the platform to choose separate revenue shares for creators' content and AI-generated content. A dynamic model could then examine whether compensation sustains the production of new content as AI improves. Greater AI displacement reduces revenue from direct content sales, while compensation may provide revenue from supplying new content for training. How these incentives affect future content production remains an open question.

%% file: sections/appendix.tex
\appendix
\numberwithin{equation}{section}

\section{Notation}
\label{apx:notation}

We collect notation, case labels and thresholds of Sections~\ref{sec:model}--\ref{sec:extensions} and the shorthand of Appendix~\ref{apx:proofs} in Table~\ref{tab:notation-full}. All these symbols are also defined where they first appear in the paper.

{\small
\renewcommand{\arraystretch}{1.12}
\begin{longtable}{@{}p{0.27\textwidth}p{0.68\textwidth}@{}}
\caption{Notation and Case Labels}\label{tab:notation-full}\\
\toprule
\endfirsthead
\multicolumn{2}{@{}l}{\textit{Table~\ref{tab:notation-full} (continued)}}\\
\toprule
\endhead
\bottomrule
\endfoot
\multicolumn{2}{@{}l}{\textit{Primitives}}\\
\midrule
$q_{H},\,q_{L}$ & Content quality of the high- and low-quality creator, $0<q_{L}<q_{H}<1$.\\
$\alpha\in(0,1)$ & Creators' share of own-content revenue and, under C, contributors' share of AI revenue; the platform keeps the rest.\\
$\phi>0$ & Participation cost parameter: creator $j$'s participation cost is $\phi q_{j}$.\\
$q^{\textrm{base}}\geq0$ & Base AI capability: quality of AI-generated content before training on creators' content; $q^{\textrm{base}}\leq1$ by Assumption~\ref{assumption:training-data-dependence} when $\kappa_{0}=0$.\\
$\kappa_{0},\ \kappa_{\textrm{data}}$ & Base (steerability) and data components of prompting effort $\kappa(x;s_{H},s_{L})$, \eqref{eq:kappa}; $\kappa_{\textrm{data}}=1$, and $\kappa_{0}=0$ except in Section~\ref{subsec:imperfect-steerability}.\\
$\tau_{H},\tau_{L}\in(0,1)$ & Transfer rates of creators' content into AI quality, \eqref{eq:q-ai}; $\tau_{H}=\tau_{L}=\tfrac12$ except in Sections~\ref{subsec:heterogeneous-data}--\ref{subsec:quality-weighted}, which allow $0<\tau_{L}\leq\tau_{H}=\tfrac12$.\\
\addlinespace
\multicolumn{2}{@{}l}{\textit{Decisions, demands, payoffs}}\\
\midrule
$P,\ P_{AI}$ & Common price of creators' content; price of AI-generated content (Stage~1).\\
$e_{j},\ s_{j}\in\{0,1\}$ & Participation of creator $j$ and training of the AI on her content (Stage~2); $s_{j}=e_{j}$ under NC, while $j$ chooses $s_{j}\leq e_{j}$ under C.\\
$q_{AI}$ & Quality of AI-generated content, $q_{AI}=q^{\textrm{base}}+\tau_{H}q_{H}s_{H}+\tau_{L}q_{L}s_{L}$, \eqref{eq:q-ai}.\\
$D_{H},D_{L},D_{AI}$ & Demand for $H$'s, $L$'s, and AI-generated content; $D\equiv D_{H}+D_{L}$.\\
$V_{j},\ R_{j}$ & Gross payoff of participating creator $j$, \eqref{eq:creator-payoff}; net payoff $R_{j}\equiv V_{j}-\phi q_{j}$.\\
$\pi,\ \mathrm{CS},\ \mathrm{TS}$ & Platform profit, \eqref{eq:platformprofit}; consumer surplus; total surplus $\mathrm{TS}\equiv\pi+R_{H}+R_{L}+\mathrm{CS}$.\\
\addlinespace
\multicolumn{2}{@{}l}{\textit{Shorthand}}\\
\midrule
$(x)^{+}$ & Positive part of $x$: $(x)^{+}\equiv\max\{x,0\}$.\\
$g,\ t$ & $g\equiv(q_{H}-q_{L})/2\in(0,\tfrac12)$ (Lemma~\ref{lemma:demands-nc}) and $t\equiv q_{H}+q_{L}\in(0,2)$.\\
$\widetilde{P}_{AI},\ \delta$ & Net AI price $\widetilde{P}_{AI}\equiv P_{AI}-q^{\textrm{base}}$ and price gap $\delta\equiv\widetilde{P}_{AI}-P$ (Lemma~\ref{lemma:demands-nc}); in a covered market, demands depend on prices only through $\delta$. Appendix~\ref{apx:proofs} uses $\widetilde{P}_{AI}\equiv P_{AI}-\theta$, which is the same when $\theta=q^{\textrm{base}}$.\\
$s$ & Cost index $s\equiv\sqrt{q_{L}^{2}-4\phi q_{L}/\alpha}$ (Proposition~\ref{prop:nc-pricing}), distinct from $s_{j}$; it strictly decreases in $\phi$ and equals $g$ at $\phi=\phi^{B}$, so $g\leq s<q_{L}$.\\
$\theta$ & Effective AI capability $\theta\equiv q^{\textrm{base}}-\kappa_{0}$ (Section~\ref{subsec:imperfect-steerability}); $\theta=q^{\textrm{base}}$ when $\kappa_{0}=0$. Appendix~\ref{apx:proofs} uses the general form $\theta\equiv z-(q_{H}+q_{L})/2$, which equals $q^{\textrm{base}}-\kappa_{0}$ when $\tau_{H}=\tau_{L}=\tfrac12$.\\
$z$ & Quality of AI-generated content, net of $\kappa_{0}$, when both creators contribute, \eqref{eq:z}; $z=q^{\textrm{base}}+(q_{H}+q_{L})/2$ in the main analysis.\\
\addlinespace
\multicolumn{2}{@{}l}{\textit{Symbols used only in Appendix~\ref{apx:proofs}}}\\
\midrule
$(1,1)$; $\mathcal{P}^{*},\ \pi^{*}$ & Profile in which both creators participate and contribute; a set of price vectors and the profit that every allowed Stage-2 play gives on it (Lemma~\ref{lemma:stage2}).\\
$d_{k}(x),\ y_{k}$ & Distance of a consumer at $x$ from creator $k$'s location; quality of AI-generated content net of $\kappa_{0}$ when only creator $k$ contributes, $y_{k}\equiv z-\tau_{j}q_{j}$ for $j\neq k$ (Lemma~\ref{lemma:one-contributor}).\\
$\Pi_{HL},\ \Pi_{H}$ & Bounds on profit without the AI when both creators sell, $(1-\alpha)t^{2}/8$, and when at most one does, $(1-\alpha)q_{H}^{2}/4$ (Lemma~\ref{lemma:no-ai}).\\
$V_{j}^{B}$ & Creator $j$'s revenue at the benchmark price, $\alpha(q_{j}-\tfrac{t}{4})\tfrac{t}{4}$.\\
$\delta^{NC},\ \widetilde{\pi}(\delta)$ & Equilibrium price gap under NC, $(q_{H}-s)/2$; the platform's profit under NC at price gap $\delta$ when the AI is priced at $z$, \eqref{eq:pi-tilde}.\\
$\pi^{NC}(\phi,\theta)$ & Equilibrium profit under NC as a function of $\phi$ and $\theta$, \eqref{eq:pi-I}.\\
$D_{H}^{C}(z),\ F(z),\ G(z)$ & At the displacement prices: $H$'s demand $\tfrac12(q_{H}-z)^{+}$, the platform's profit divided by $1-\alpha$, and AI revenue (Appendix~\ref{apx:pricing-c}).\\
$\mathcal{P}(z)$ & Displacement prices: $(\tfrac{q_{H}+z}{2},z)$ if $z<q_{H}$, and $(P,z)$ with $P\geq q_{H}$ otherwise.\\
$\Psi(\delta),\ M$ & Profit divided by $1-\alpha$ under $(1,1)$ at price gap $\delta$ when the AI is priced at $z$, and its maximum (Lemma~\ref{lemma:c-both}); $\delta_{c}$ and $\delta_{3}$ are the price gaps at its two peaks.\\
$V_{H}(z),\ A^{C}$ & $H$'s gross payoff at the displacement prices under equal splitting (Lemma~\ref{lemma:c-payoffs}); each creator's AI payment in the equilibrium of Proposition~\ref{prop:c-pricing}.\\
$\Lambda(s,q^{\textrm{base}}),\ s^{\star}$ & Profit difference $\pi^{NC}-\pi^{C}$ and its zero in $s$ (proof of Corollary~\ref{cor:profit-ranking}).\\
$W^{r}$ & Consumers' gross utility in regime $r$ (proof of Proposition~\ref{prop:welfare}).\\
$\widehat{\theta},\ \overline{\pi}$ & $\widehat{\theta}\equiv q^{\textrm{base}}-\widehat{\kappa}_{0}(\phi)$ (proof of Proposition~\ref{prop:imperfect-steerability-nc}); the platform's maximal profit under $(1,1)$ (proof of Proposition~\ref{prop:imperfect-steerability}).\\
$h,\ \ell,\ \eta(\tau_{L})$ & $h\equiv q_{H}/2$, $\ell\equiv\tau_{L}q_{L}$, and $L$'s payoff under quality weighting divided by $\alpha$ (proof of Proposition~\ref{prop:quality-weighted}).\\
\addlinespace
\multicolumn{2}{@{}l}{\textit{Benchmark and regimes (superscripts on equilibrium objects)}}\\
\midrule
$B$ & Benchmark before AI integration (Section~\ref{subsec:no-ai}).\\
$NC$ & No compensation (Section~\ref{subsec:nocompensation}): training on a participant's content is a condition of participation, and creators earn only $\alpha D_{j}P$.\\
$C$ & Compensation (Section~\ref{subsec:withcompensation}): each participant chooses $s_{j}$, and contributors split the share $\alpha$ of AI revenue equally. The superscript $\mathrm{QW}$ marks quality-weighted compensation, which splits it in proportion to $\tau_{j}q_{j}s_{j}$ (Section~\ref{subsec:quality-weighted}).\\
\addlinespace
\multicolumn{2}{@{}l}{\textit{Market structure cases}}\\
\midrule
$\overline{q}^{\textrm{base}}\equiv(q_{H}-q_{L})/2$ & Full displacement threshold separating CQ1 from CQ2 (Proposition~\ref{prop:c-pricing}); when both creators contribute, $q^{\textrm{base}}<\overline{q}^{\textrm{base}}$ is equivalent to $q_{AI}<q_{H}$, the form Proposition~\ref{prop:heterogeneous-data} uses for general $\tau_{L}$; Section~\ref{subsec:imperfect-steerability} uses $\theta$ in place of $q^{\textrm{base}}$.\\
Partial displacement (CQ1) & $q^{\textrm{base}}<\overline{q}^{\textrm{base}}$: AI fully displaces $L$ ($D_{L}^{C}=0$), while $D_{H}^{C}>0$.\\
Full displacement (CQ2) & $q^{\textrm{base}}\geq\overline{q}^{\textrm{base}}$: AI serves the entire market ($D_{AI}^{C}=1$).\\
Three-product configuration & All three products attract demand (row~2 of the table in Lemma~\ref{lemma:demands-nc}): the equilibrium configuration under NC (Proposition~\ref{prop:nc-pricing}; Proposition~\ref{prop:imperfect-steerability-nc} for $\kappa_{0}<\widehat{\kappa}_{0}(\phi)$) and under C for $-(q_{H}+q_{L})/2<\theta<\underline{\theta}_{C}$ or $\theta=\theta_C<\theta_L(\phi)$ (Proposition~\ref{prop:imperfect-steerability}).\\
\addlinespace
\multicolumn{2}{@{}l}{\textit{Thresholds}}\\
\midrule
$\rho^{\star}\equiv1+\sqrt{2}$ & Quality ratio at which serving both creators and serving only $H$ give the same maximal benchmark profit (Proposition~\ref{prop:benchmark}); Assumption~\ref{assumption:benchmark-participation} requires $q_{H}/q_{L}<\rho^{\star}$.\\
$\phi^{B}$ & $\phi^{B}\equiv\alpha(3q_{L}-q_{H})(q_{H}+q_{L})/(16q_{L})$, the largest $\phi$ at which both creators participate at $P^{B}=(q_{H}+q_{L})/4$; Assumption~\ref{assumption:benchmark-participation} requires $\phi\leq\phi^{B}$.\\
$\phi^{\star}(q^{\textrm{base}})\in[0,\phi^{B}]$ & The platform prefers NC for $\phi<\phi^{\star}$ and C for $\phi>\phi^{\star}$ (Corollary~\ref{cor:profit-ranking}).\\
$\widehat{\kappa}_{0}(\phi)$ & Value of $\kappa_{0}$, \eqref{eq:kappa-hat}, below which the NC equilibrium is that of Proposition~\ref{prop:nc-pricing} with $\theta$ in place of $q^{\textrm{base}}$ (Proposition~\ref{prop:imperfect-steerability-nc}); $q^{\textrm{base}}<\widehat{\kappa}_{0}(\phi)\leq\kappa_{0}^{*}$.\\
$\underline{\theta}_{C}\equiv-(1+\sqrt{2})g$ & Under C, the equilibrium is the three-product configuration for $-(q_{H}+q_{L})/2<\theta<\underline{\theta}_{C}$ or $\theta=\theta_C<\theta_L(\phi)$, and that of Proposition~\ref{prop:c-pricing} with $\theta$ in place of $q^{\textrm{base}}$ for $\theta\geq\max\{\underline{\theta}_{C},\underline{\theta}_{L}(\phi)\}$, with both at $\theta=\underline{\theta}_{C}>\theta_L(\phi)$ (Proposition~\ref{prop:imperfect-steerability}).\\
$\underline{\theta}_{L}(\phi)$ & Value of $\theta\in(-(q_{H}+q_{L})/2,0)$ at which $L$'s AI payment at the prices of Proposition~\ref{prop:c-pricing} with $\theta$ in place of $q^{\textrm{base}}$ equals her participation cost $\phi q_{L}$ (Proposition~\ref{prop:imperfect-steerability}).\\
$\kappa_{0}^{*}\equiv q^{\textrm{base}}+(q_{H}+q_{L})/2$ & AI makes no sales when $\kappa_{0}>\kappa_{0}^{*}$, i.e., $\theta<-(q_{H}+q_{L})/2$ (Proposition~\ref{prop:imperfect-steerability}).\\
$\tau_{L}^{\star}\in(0,\tfrac12)$ & For $q_{H}/q_{L}\geq\sqrt{2}$, the outcome of Proposition~\ref{prop:heterogeneous-data} is an equilibrium under quality-weighted compensation if and only if $\tau_{L}\geq\tau_{L}^{\star}$; $L$'s participation constraint binds at $\tau_{L}^{\star}$ (Proposition~\ref{prop:quality-weighted}).\\
\end{longtable}
}
{\footnotesize\noindent\textit{Note.} Unless stated otherwise, we use $q_{H}=0.7$, $q_{L}=0.4$, $\alpha=0.3$, $\phi=\phi^{B}/2$, and $q^{\textrm{base}}=0$ in numerical illustrations.}

\section{Proofs}
\label{apx:proofs}

In this section, we prove all results stated in the main paper. The equilibrium results (Propositions~\ref{prop:benchmark}--\ref{prop:c-pricing} and \ref{prop:imperfect-steerability-nc}--\ref{prop:quality-weighted}) share one logic. The platform sets prices anticipating creators' participation and data-contribution decisions. We must therefore compare its profit across everything creators may do at every price and then verify what they do at the stated prices. To formally state this logic, we first introduce additional notation and the proof strategy commonly used in these results, and then collect the properties of creators' Stage-2 play (Appendix~\ref{apx:stage2}) and of consumer demand (Appendix~\ref{apx:demand}) that the proofs share. Then, we prove these results in the order shown in the main paper, with auxiliary lemmas placed before proofs that use them.

\noindent\textbf{Notation.} We use two shorthands for the creators' qualities throughout:

\[
t\equiv q_{H}+q_{L}\in(0,2),\qquad g\equiv\frac{q_{H}-q_{L}}{2}\in\bigl(0,\tfrac12\bigr).
\]
We also write $(x)^{+}\equiv\max\{x,0\}$ for the positive part of a number $x$. We cover the main analysis and the extensions of Section~\ref{sec:extensions} with one argument. We thus let $z$ denote the quality of AI-generated content, net of the base prompting effort $\kappa_{0}$, when the AI is trained on both creators' content:
\begin{equation}
z\equiv q^{\textrm{base}}-\kappa_{0}+\tau_{H}q_{H}+\tau_{L}q_{L}.
\label{eq:z}
\end{equation}
By \eqref{eq:kappa} and \eqref{eq:q-ai}, the utility that a consumer at $x$ obtains from AI-generated content depends on whose content the AI is trained on:
\[
U_{AI}=\begin{cases}
z-P_{AI}, & \text{both creators' content},\\[2pt]
z-\tau_{j}q_{j}-P_{AI}-d_{k}(x), & \text{only creator $k$'s content ($j\neq k$)},\\[2pt]
q^{\textrm{base}}-\kappa_{0}-1-P_{AI}, & \text{neither creator's content}.
\end{cases}
\]
In the second case, the AI loses creator $j$'s contribution $\tau_{j}q_{j}$ to its quality, and the consumer incurs extra prompting effort equal to her distance from $k$'s location, $d_{L}(x)=x$ or $d_{H}(x)=1-x$, as $\kappa_{\textrm{data}}=1$. In the third case, $U_{AI}\leq-P_{AI}$, as Assumption~\ref{assumption:training-data-dependence} requires $q^{\textrm{base}}-\kappa_{0}\leq1$.

We further use three symbols for prices:
\[
\theta\equiv z-\frac{t}{2},\qquad
\widetilde{P}_{AI}\equiv P_{AI}-\theta,\qquad
\delta\equiv\widetilde{P}_{AI}-P.
\]
When the AI is trained on both creators' content, $U_{AI}=\tfrac{t}{2}-\widetilde{P}_{AI}$, and $\delta$ is the gap between the net AI price $\widetilde{P}_{AI}$ and the price of creators' content. In the main analysis ($\kappa_{0}=0$ and $\tau_{H}=\tau_{L}=\tfrac12$), $z=q^{\textrm{base}}+\tfrac{t}{2}$ and $\theta=q^{\textrm{base}}$, so $\widetilde{P}_{AI}$ and $\delta$ are those of Lemma~\ref{lemma:demands-nc}. In Section~\ref{subsec:imperfect-steerability}, $\theta=q^{\textrm{base}}-\kappa_{0}$, as defined there; in Sections~\ref{subsec:heterogeneous-data} and~\ref{subsec:quality-weighted}, where $\kappa_{0}=0$ and $\tau_{H}=\tfrac12$, $\theta=q^{\textrm{base}}-(\tfrac12-\tau_{L})q_{L}$.

\smallskip
\noindent\textbf{Proof strategy.} We prove each equilibrium result (Propositions~\ref{prop:benchmark}--\ref{prop:c-pricing} and \ref{prop:imperfect-steerability-nc}--\ref{prop:quality-weighted}) in three steps. First, we bound the platform's profit at every price vector under every Stage-2 profile (under NC, in every pure-strategy Stage-2 equilibrium). Second, we show that the bound is attained only at the stated prices and under the stated profile. Whenever the AI sells, this is the profile in which both creators participate and contribute data. Third, we show that at these prices this profile is the only Stage-2 play consistent with Assumptions~\ref{assumption:weak-participation} and~\ref{assumption:coordination}. Lemma~\ref{lemma:stage2} turns these facts into the equilibrium characterization, and we collect in Appendix~\ref{apx:demand} the demand properties that the three steps share.

\subsection{Equilibrium Properties}
\label{apx:stage2}

We call a pure-strategy choice of $e_{j}$ (it's $s_{j}$ under C) by each creator a \emph{profile}. We say that creator $j$ \emph{contributes} if $s_{j}=1$; under NC, this means that she participates. We write $(1,1)$ for the profile in which both creators participate and contribute if the AI is offered. As in Section~\ref{subsec:timing}, we call a Stage-2 play at a price vector \emph{allowed} if it is a pure-strategy Nash equilibrium that satisfies Assumption~\ref{assumption:weak-participation} and that Assumption~\ref{assumption:coordination} selects when it applies. If no such equilibrium exists, an allowed play is a mixed-strategy Nash equilibrium, which the platform values at its expected profit. 

By an \emph{equilibrium} we mean a subgame-perfect equilibrium in which the platform chooses a price vector and creators use an allowed play at every price vector. Under C, a pure-strategy Nash equilibrium of the Stage-2 game can fail to exist at some price vectors.\footnote{For example, at $(P,P_{AI})=(0.07,0.05)$ with $(q_{H},q_{L},\alpha,\phi,q^{\textrm{base}},\kappa_{0})=(0.9,0.8,0.5,0.02,0.8,0)$, which satisfies Assumption~\ref{assumption:benchmark-participation}, the creators' best replies are unique and cycle.}

We use Lemma~\ref{lemma:stage2} to prove every equilibrium result (Propositions~\ref{prop:benchmark}--\ref{prop:c-pricing} and \ref{prop:imperfect-steerability-nc}--\ref{prop:quality-weighted}) to pass from Stage-2 play to the platform's equilibrium prices. The lemma has three parts. Part~(a) rules out mixed play in the benchmark and under NC. Part~(b) reduces the search for equilibria to comparing profits across price vectors, and part~(c) extends profit bounds from profiles to mixed plays.

\begin{applemma}[Stage-2 play]
\label{lemma:stage2}
(a) In the benchmark and under NC (for every $\kappa_{0}\geq0$), at every price vector some pure-strategy Nash equilibrium satisfies Assumption~\ref{assumption:weak-participation}, so every allowed play is pure.

(b) Fix the benchmark or a regime (under C, with equal splitting or quality weighting). Suppose that, for a nonempty set $\mathcal{P}^{*}$ of price vectors, every allowed play gives profit $\pi^{*}$ on $\mathcal{P}^{*}$ and strictly less elsewhere. Then the equilibria are exactly the strategy profiles in which creators use an allowed play at every price vector and the platform chooses a vector in $\mathcal{P}^{*}$. In particular, an equilibrium exists and yields profit $\pi^{*}$.

(c) At any given price vector, if every profile gives profit at most $\pi^{*}$, then so does every allowed play, with equality only if the play puts probability one on the profiles that give $\pi^{*}$.
\end{applemma}

\begin{proof}
(a) A creator sells weakly more when the other creator exits, so a creator who participates against a participating rival also participates against an absent one; this monotonicity yields a pure-strategy equilibrium. We fix prices and creators $j\neq j'$, and let $D_{j}(e_{j'})$ denote $j$'s demand when she participates. If $j'$ exits, $j$'s content loses a competitor. Under NC, the AI also loses $j'$'s training data, which lowers every consumer's AI utility. By \eqref{eq:kappa} and \eqref{eq:q-ai}, the fall is $\tau_{j'}q_{j'}+\kappa_{\textrm{data}}|x-x_{j}|\geq0$ for every $\kappa_{0}$, where $x_{L}=0$ and $x_{H}=1$ are the creators' locations. As ties favor creators' content over the AI and platform products over the outside option (footnote~\ref{fn:consumer-tie-breaking}), every consumer who bought from $j$ still does, so her demand weakly rises: $D_{j}(0)\geq D_{j}(1)$.

By Assumption~\ref{assumption:weak-participation}, $j$ participates if and only if her sales revenue covers her cost: her best reply is $\beta_{j}(e_{j'})=1$ if $\alpha D_{j}(e_{j'})P\geq\phi q_{j}$ and $\beta_{j}(e_{j'})=0$ otherwise. Hence $\beta_{j}(1)\leq\beta_{j}(0)$, and the pure-strategy equilibria satisfying Assumption~\ref{assumption:weak-participation} are the profiles with $e_{j}=\beta_{j}(e_{j'})$ for both creators. We now show that such a profile exists. If $(1,1)$ is not one, then $\beta_{j}(1)=0$ for some $j$. If $\beta_{j'}(0)=1$, the profile in which $j$ exits and $j'$ participates is one. Otherwise $\beta_{j'}(1)\leq\beta_{j'}(0)=0$, so $j'$ exits whatever $j$ does, and the profile in which $j'$ exits and $j$ plays $\beta_{j}(0)$ is one. As a pure-strategy equilibrium satisfying Assumption~\ref{assumption:weak-participation} thus exists, no mixed play is allowed there.

(b) An allowed play exists at every price vector. It is a pure-strategy equilibrium that satisfies Assumption~\ref{assumption:weak-participation} and that Assumption~\ref{assumption:coordination} selects whenever it applies. If there is none, it is a mixed-strategy equilibrium of the finite Stage-2 game, which exists by Nash's theorem. By definition, allowed plays are Nash equilibria of the Stage-2 subgames. A strategy profile in which creators use them is therefore an equilibrium if and only if the platform's price vector maximizes profit given those plays. By hypothesis, this holds if and only if the price vector lies in $\mathcal{P}^{*}$.

(c) An allowed play induces a distribution $\mu$ over the finitely many profiles of the Stage-2 game. We let $\pi(\omega)$ denote the platform's profit under profile $\omega$. The play's expected profit $\sum_{\omega}\mu(\omega)\pi(\omega)$ is at most $\pi^{*}$, with equality only if $\pi(\omega)=\pi^{*}$ for every profile $\omega$ with $\mu(\omega)>0$.
\end{proof}

\subsection{Consumer Demand}
\label{apx:demand}

In this section, we describe consumer demand and the profit it allows according to which creators contribute. The lemmas hold for every $\kappa_{0}\geq0$ and all transfer rates.

We start with the market in which the AI is not offered. We use Lemma~\ref{lemma:no-ai} to prove Proposition~\ref{prop:benchmark}. Its profit bounds also apply whenever the AI earns no revenue (Lemma~\ref{lemma:no-ai-revenue}).

\begin{applemma}[Demand and profit without the AI]
\label{lemma:no-ai}
Suppose that the platform does not offer the AI, and fix a price $P\geq0$ for creators' content. Let $\Pi_{HL}\equiv(1-\alpha)t^{2}/8$ and $\Pi_{H}\equiv(1-\alpha)q_{H}^{2}/4$.

(a) A participating creator $j$ sells only to consumers with nonnegative utility from her content, a set of measure $(q_{j}-P)^{+}$. The two sets are disjoint if $P>\tfrac{t-1}{2}$ and cover $[0,1]$ otherwise. She sells to her whole set if she participates alone or $P>\tfrac{t-1}{2}$; if both participate and $P\leq\tfrac{t-1}{2}$, every consumer buys.

(b) Under every participation profile, profit is at most $\max\{\Pi_{HL},\Pi_{H}\}$. Equality requires either that both creators participate and $P=\tfrac{t}{4}<q_{L}$, or that $P=\tfrac{q_{H}}{2}$ and only $H$ sells.

(c) $\Pi_{HL}-\Pi_{H}$ has the sign of $\rho^{\star}-q_{H}/q_{L}$. Hence, under every participation profile, profit is at most $\Pi_{HL}$ if $q_{H}/q_{L}<\rho^{\star}$, with equality only if both creators participate and $P=\tfrac{t}{4}$.
\end{applemma}

\begin{proof}
(a) By \eqref{eq:consumer-utility}, $U_{L}\geq0$ if and only if $x\leq q_{L}-P$, and $U_{H}\geq0$ if and only if $x\geq1-q_{H}+P$. Each set has measure $(q_{j}-P)^{+}$ as $P\geq0$ and $q_{j}<1$. The two sets are disjoint if $q_{L}-P<1-q_{H}+P$, that is, if $P>\tfrac{t-1}{2}$. Otherwise they cover $[0,1]$. A creator who participates alone or whose set is disjoint from the other's faces no competition there and sells to all of it. If both participate and the sets cover $[0,1]$, every consumer has nonnegative utility from some content and buys.

(b) Profit is $(1-\alpha)P$ times total demand. Suppose first that both creators sell, which requires that both participate and $P<q_{L}$. If the market is covered ($P\leq\tfrac{t-1}{2}$), profit is $(1-\alpha)P\leq(1-\alpha)\tfrac{t-1}{2}=\Pi_{HL}-(1-\alpha)\tfrac{(2-t)^{2}}{8}<\Pi_{HL}$. If it is not, total demand is $t-2P$ by part~(a), and
\[
\Pi_{HL}-(1-\alpha)P(t-2P)=2(1-\alpha)\Bigl(P-\frac{t}{4}\Bigr)^{2},
\]
so profit equals $\Pi_{HL}$ only at $P=\tfrac{t}{4}$. This price always exceeds $\tfrac{t-1}{2}$, as $t<2$, and it is below $q_{L}$ if and only if $q_{H}<3q_{L}$. Suppose next that at most one creator sells. Profit is zero if none does and, by part~(a), at most $(1-\alpha)P(q_{j}-P)^{+}$ if only creator $j$ does. For $j=H$, $\Pi_{H}-(1-\alpha)P(q_{H}-P)=(1-\alpha)(P-\tfrac{q_{H}}{2})^{2}$, so profit equals $\Pi_{H}$ only at $P=\tfrac{q_{H}}{2}$; for $j=L$, profit is at most $(1-\alpha)q_{L}^{2}/4<\Pi_{H}$.

(c) The sign follows from
\[
\Pi_{HL}-\Pi_{H}=(1-\alpha)\frac{q_{L}^{2}+2q_{H}q_{L}-q_{H}^{2}}{8}=\frac{(1-\alpha)q_{L}^{2}}{8}\Big(\rho^{\star}-\frac{q_{H}}{q_{L}}\Big)\Big(\frac{q_{H}}{q_{L}}+\sqrt{2}-1\Big).
\]
If $q_{H}/q_{L}<\rho^{\star}$, then $\Pi_{H}<\Pi_{HL}$ and $q_{H}<3q_{L}$, so part~(b) gives the second claim.
\end{proof}

Lemma~\ref{lemma:both-contribute} gives demand when the AI is trained on both creators' content. It is Lemma~\ref{lemma:demands-nc} of the main paper stated for a general $\theta$. We use it for pricing under both regimes (Lemmas~\ref{lemma:nc-pricing} and~\ref{lemma:c-both}).

\begin{applemma}[Demand when both creators contribute]
\label{lemma:both-contribute}
Suppose both creators participate and contribute. If $\widetilde{P}_{AI}>\tfrac{t}{2}$, no consumer buys AI-generated content, and consumers choose among creators' content and the outside option as they would without the AI at the same price $P$. If $\widetilde{P}_{AI}\leq\tfrac{t}{2}$, the market is covered and demands are those in the table of Lemma~\ref{lemma:demands-nc}; in particular, they depend on prices only through $\delta$.
\end{applemma}

\begin{proof}
Every consumer obtains $U_{AI}=z-P_{AI}=\tfrac{t}{2}-\widetilde{P}_{AI}$ from the AI. If $\widetilde{P}_{AI}>\tfrac{t}{2}$, this utility is negative, so no consumer buys AI-generated content. Let $\widetilde{P}_{AI}\leq\tfrac{t}{2}$. Then $U_{AI}\geq0$ and, as ties with the outside option go to platform products (footnote~\ref{fn:consumer-tie-breaking}), every consumer buys. By \eqref{eq:consumer-utility},
\[
U_{L}-U_{AI}=(\delta-g)-x,\qquad
U_{H}-U_{AI}=x-(1-g-\delta),\qquad
U_{L}-U_{H}=2(x_{HL}-x),
\]
where $x_{HL}\equiv\tfrac12-g$ is the consumer who is indifferent between the two creators. Ties between a creator's content and the AI go to the creator. Hence a consumer at $x\neq x_{HL}$ buys from $L$ if and only if $x\leq\delta-g$ and $x<x_{HL}$, buys from $H$ if and only if $x\geq1-g-\delta$ and $x>x_{HL}$, and buys AI-generated content otherwise. With $m\equiv\min\{\delta,\tfrac12\}$ and up to endpoints, $L$'s buyers are the consumers with $x\leq m-g$ and $H$'s are those with $x\geq1-g-m$. As $m-g<1$ and $1-g-m>0$,
\[
D_{L}=(m-g)^{+},\qquad
D_{H}=(m+g)^{+},\qquad
D_{AI}=1-D_{L}-D_{H},
\]
which gives the four rows of the table: $m=\tfrac12$ if $\delta\geq\tfrac12$, and $m=\delta$ otherwise.
\end{proof}

Lemma~\ref{lemma:one-contributor} covers the remaining profiles, in which at most one creator contributes. It proves the last three rows of Lemma~\ref{lemma:demands-c}. We also use it to bound the platform's profit when a creator exits under NC (Lemma~\ref{lemma:nc-pricing}) or withholds her data under C (Lemma~\ref{lemma:c-one}).

\begin{applemma}[Demand with at most one contributor]
\label{lemma:one-contributor}
(a) Suppose only creator $k$ contributes, and let $j\neq k$ and $y_{k}\equiv z-\tau_{j}q_{j}$. Then $U_{k}-U_{AI}=q_{k}-y_{k}+P_{AI}-P$ at every location, so $k$'s content and AI-generated content cannot both attract demand: $D_{k}D_{AI}=0$. If the AI sells, then $D_{k}=0$, $P_{AI}<y_{k}$, and
\[
D_{AI}P_{AI}\leq\tfrac14y_{k}^{2},
\qquad
D_{j}P\leq\tfrac14q_{j}^{2}.
\]

(b) If neither creator contributes, the platform earns no AI revenue at any price.
\end{applemma}

\begin{proof}
(a) The AI is trained on $k$'s content, so $k$ participates. Both $U_{k}=q_{k}-P-d_{k}(x)$ and $U_{AI}=y_{k}-P_{AI}-d_{k}(x)$ fall one for one with distance from $k$, because prompting effort mirrors the mismatch cost. Their difference is thus the same for all consumers. If it is nonnegative, every consumer weakly prefers $k$'s content, and $D_{AI}=0$ because ties go to creators' content. If it is negative, every consumer strictly prefers the AI, and $D_{k}=0$. A buyer of AI content has $U_{AI}\geq0$, that is, $d_{k}(x)\leq y_{k}-P_{AI}$, so $D_{AI}\leq(y_{k}-P_{AI})^{+}$. If the AI sells, then $P_{AI}<y_{k}$ and $D_{AI}P_{AI}\leq(y_{k}-P_{AI})P_{AI}\leq\tfrac14y_{k}^{2}$. A buyer of $j$'s content has $U_{j}\geq0$, so that $D_{j}P\leq(q_{j}-P)^{+}P\leq\tfrac14q_{j}^{2}$.

(b) Then $U_{AI}\leq-P_{AI}$, so no consumer buys AI-generated content unless $P_{AI}=0$.
\end{proof}

Lemma~\ref{lemma:no-ai-revenue} bounds the platform's profit whenever the AI earns nothing. We use it to rule out such outcomes in Lemmas~\ref{lemma:nc-pricing}, \ref{lemma:c-both} and~\ref{lemma:c-one}, and to prove part~3 of Proposition~\ref{prop:imperfect-steerability}.

\begin{applemma}[Profit without AI revenue]
\label{lemma:no-ai-revenue}
Suppose that $q_{H}/q_{L}<\rho^{\star}$. Consider either regime, a price vector and a profile such that AI revenue $D_{AI}P_{AI}$ is zero. The platform's profit is then at most $\Pi_{HL}=(1-\alpha)t^{2}/8$, which is the benchmark profit $\pi^{B}$ under Assumption~\ref{assumption:benchmark-participation} (Proposition~\ref{prop:benchmark}). Equality requires that $P=\tfrac{t}{4}$ and that both creators participate and sell their benchmark demands $q_{j}-\tfrac{t}{4}$.
\end{applemma}

\begin{proof}
Without AI revenue, the platform's profit is $(1-\alpha)DP$ in both regimes by \eqref{eq:platformprofit}. A buyer of creator $j$'s content obtains nonnegative utility from it, so $D$ is at most the measure of consumers who obtain nonnegative utility from some creator's content. By Lemma~\ref{lemma:no-ai}(a), this measure is total demand at price $P$ when both creators participate and the AI is not offered. Lemma~\ref{lemma:no-ai}(c) then gives $(1-\alpha)DP\leq\Pi_{HL}$, with equality only if $P=\tfrac{t}{4}$ and $D=\tfrac{t}{2}$. At $P=\tfrac{t}{4}>\tfrac{t-1}{2}$, the two creators' sets of potential buyers are disjoint, with measures $q_{j}-\tfrac{t}{4}$ that sum to $\tfrac{t}{2}$. Hence $D=\tfrac{t}{2}$ requires each creator to participate and sell to her entire set of potential buyers.
\end{proof}

We close this section with prices at which the AI is priced at its quality $z$, so that its buyers obtain zero utility. Every equilibrium price pair characterized below at which the AI earns revenue has this form. At such prices the AI is viable only with both creators' data, which is what ties each creator's AI payment to her own contribution. We use Lemma~\ref{lemma:ai-at-quality} in the last step of the proofs of Propositions~\ref{prop:nc-pricing}, \ref{prop:c-pricing} and \ref{prop:imperfect-steerability-nc}--\ref{prop:quality-weighted}, where it shows which Stage-2 play the equilibrium prices induce.

\begin{applemma}[The AI priced at its quality]
\label{lemma:ai-at-quality}
Let $z>0$ and consider prices with $P_{AI}=z$ and $P>\tfrac{t-1}{2}$.

(a) In either regime, AI-generated content sells only under $(1,1)$, and under every profile a participating creator $j$ sells $(q_{j}-P)^{+}$. Under $(1,1)$, $D_{AI}=1-(q_{H}-P)^{+}-(q_{L}-P)^{+}>0$.

(b) Under C, with equal splitting or quality weighting, suppose that under $(1,1)$ both creators' net payoffs are nonnegative and at least one is positive. Then $(1,1)$ is the only allowed play.
\end{applemma}

\begin{proof}
(a) Under $(1,1)$, every consumer obtains $U_{AI}=z-P_{AI}=0$ from the AI. If only creator $k$ contributes, $U_{AI}\leq-\tau_{j}q_{j}<0$ for $j\neq k$, and if neither contributes, $U_{AI}\leq-z<0$. So the AI sells nothing under every profile other than $(1,1)$. As $P>\tfrac{t-1}{2}$, no consumer obtains nonnegative utility from both creators' content (Lemma~\ref{lemma:no-ai}(a)). Under every profile, the consumers who obtain nonnegative utility from a participating creator $j$'s content therefore buy it, as the AI offers them at most zero and ties go to creators' content; these consumers have measure $(q_{j}-P)^{+}$. Under $(1,1)$, all other consumers buy AI-generated content, which gives them zero utility, so $D_{AI}=1-(q_{H}-P)^{+}-(q_{L}-P)^{+}$. This is positive because $P>\tfrac{t-1}{2}$ and $q_{H}<1$.

(b) We let $R_{j}$ denote creator $j$'s net payoff under $(1,1)$ and $A_{j}$ her AI payment, which is positive by part~(a) as each contributor's share of AI revenue is positive under both rules. By part~(a), under any other profile a participating creator earns the same sales revenue but no AI payment. Her net payoff is then $R_{j}-A_{j}$, and a creator who exits earns zero. So no creator gains by deviating from $(1,1)$: withholding data forfeits $A_{j}$, and exit yields $0\leq R_{j}$. So $(1,1)$ is a Nash equilibrium, and it satisfies Assumption~\ref{assumption:weak-participation} because both creators participate. In every other pure-strategy equilibrium, creator $j$ earns at most $\max\{R_{j}-A_{j},0\}\leq R_{j}$, with strict inequality if $R_{j}>0$. Hence $(1,1)$ Pareto dominates every other pure-strategy equilibrium, and Assumption~\ref{assumption:coordination} selects it. As a pure-strategy equilibrium satisfying Assumption~\ref{assumption:weak-participation} exists, no mixed play is allowed.
\end{proof}

\subsection{Proof of Proposition~\ref{prop:benchmark}}
\label{apx:proof-benchmark}

\begin{proof}
Lemma~\ref{lemma:no-ai} bounds the platform's profit by $\Pi_{HL}$ when both creators sell and by $\Pi_{H}$ when at most one does. The quality ratio decides which bound is larger. It then remains to show that the creators indeed participate at the price that attains the larger bound.

\noindent\textit{Step 1 (Stage-2 play at $P>\tfrac{t-1}{2}$).} By Lemma~\ref{lemma:no-ai}(a), each creator $j$ sells $(q_{j}-P)^{+}$ whatever the other does, so her decision does not depend on the other's. She participates if and only if $\alpha(q_{j}-P)^{+}P\geq\phi q_{j}$, where Assumption~\ref{assumption:weak-participation} settles equality. This is the only allowed play. Allowed plays are pure (Lemma~\ref{lemma:stage2}(a)), so a bound on profit under every profile also holds under every allowed play.

\noindent\textit{Step 2 (Case $q_{H}/q_{L}<\rho^{\star}$ and $\phi\leq\phi^{B}$).} By Lemma~\ref{lemma:no-ai}(c), every profile yields at most $\Pi_{HL}$, with equality only at $P^{B}=\tfrac{t}{4}$ with both creators participating. As $P^{B}>\tfrac{t-1}{2}$, Step~1 applies. Creator $j$ earns $V_{j}^{B}=\alpha(q_{j}-P^{B})P^{B}$, and
\[
\frac{V_{L}^{B}}{q_{L}}=\frac{\alpha(3q_{L}-q_{H})(q_{H}+q_{L})}{16q_{L}}=\phi^{B},\qquad
\frac{V_{H}^{B}}{q_{H}}-\frac{V_{L}^{B}}{q_{L}}=\frac{\alpha(q_{H}-q_{L})(q_{H}+q_{L})^{2}}{16q_{H}q_{L}}>0.
\]
The low-quality creator thus has the tighter participation constraint, and $\phi^{B}$ is the largest cost at which it holds. As $\phi\leq\phi^{B}$, both creators participate. Every allowed play therefore yields $\Pi_{HL}$ at $P^{B}$ and strictly less at every other price. By Lemma~\ref{lemma:stage2}(b), in every equilibrium both creators participate and
\[
P^{B}=\frac{q_{H}+q_{L}}{4},\quad D_{L}^{B}=\frac{3q_{L}-q_{H}}{4}>0,\quad D_{H}^{B}=\frac{3q_{H}-q_{L}}{4},\quad \pi^{B}=(1-\alpha)\frac{(q_{H}+q_{L})^{2}}{8},
\]
while the remaining mass $1-\tfrac{t}{2}>0$ of consumers takes the outside option.

\noindent\textit{Step 3 (Case $q_{H}/q_{L}>\rho^{\star}$ and $\phi\leq\alpha q_{H}/4$).} By Lemma~\ref{lemma:no-ai}(b) and~(c), every profile yields at most $\Pi_{H}$, with equality only at $P=\tfrac{q_{H}}{2}$ with only $H$ selling. As $\rho^{\star}>2$, $\tfrac{q_{H}}{2}>q_{L}>\tfrac{t-1}{2}$, so Step~1 applies. $L$ would sell nothing and exits because $\phi>0$. $H$ sells $\tfrac{q_{H}}{2}$ and participates because $\alpha q_{H}^{2}/4\geq\phi q_{H}$. By Lemma~\ref{lemma:stage2}(b), every equilibrium has $P^{B}=\tfrac{q_{H}}{2}$, $D_{H}^{B}=\tfrac{q_{H}}{2}$, $D_{L}^{B}=0$ and $\pi^{B}=\Pi_{H}$.

\noindent\textit{Step 4 (Case $q_{H}/q_{L}=\rho^{\star}$ and $\phi\leq\alpha q_{L}/8$).} Here $\Pi_{HL}=\Pi_{H}$ by Lemma~\ref{lemma:no-ai}(c), and $2<\rho^{\star}<3$ gives $\tfrac{t}{4}<q_{L}<\tfrac{q_{H}}{2}$. By Lemma~\ref{lemma:no-ai}(b), profit thus equals $\Pi_{HL}$ exactly at $\tfrac{t}{4}$ with both creators participating and at $\tfrac{q_{H}}{2}$ with only $H$ selling. As $(2-\sqrt{2})(2+\sqrt{2})=2$, $\phi^{B}=\alpha q_{L}(3-\rho^{\star})(1+\rho^{\star})/16=\alpha q_{L}/8<\alpha q_{H}/4$, so $\phi\leq\phi^{B}<\alpha q_{H}/4$ and the Stage-2 arguments of Steps~2 and~3 apply at $\tfrac{t}{4}$ and $\tfrac{q_{H}}{2}$. By Lemma~\ref{lemma:stage2}(b), the equilibrium prices are $\tfrac{t}{4}$, where both creators sell, and $\tfrac{q_{H}}{2}$, where only $H$ sells.
\end{proof}

\subsection{Proof of Lemma~\ref{lemma:demands-nc}}
\label{apx:proof-lemma-demands-nc}

\begin{proof}
In the main analysis $\theta=q^{\textrm{base}}$, so the lemma is Lemma~\ref{lemma:both-contribute}.
\end{proof}

\subsection{Proof of Proposition~\ref{prop:nc-pricing} and Corollary~\ref{cor:nc-profit}}
\label{apx:proof-nc}

Throughout this section, we consider regime NC with equal transfer rates ($\tau_{H}=\tau_{L}=\tfrac12$) and maintain Assumption~\ref{assumption:benchmark-participation}, so $\pi^{B}=(1-\alpha)t^{2}/8$ (Proposition~\ref{prop:benchmark}). Recall that $s\equiv\sqrt{q_{L}^{2}-4\phi q_{L}/\alpha}$. As $s^{2}$ strictly decreases in $\phi$ and equals $g^{2}$ at $\phi=\phi^{B}$, $0<\phi\leq\phi^{B}$ gives $g\leq s<q_{L}$, with $s=g$ only at $\phi=\phi^{B}$. We let $\delta^{NC}\equiv\tfrac{q_{H}-s}{2}$, which lies in $(g,\tfrac12)$ because $s<q_{L}$ and $q_{H}<1$, take $P^{NC}$ and $D_{j}^{NC}$ from Proposition~\ref{prop:nc-pricing}, and define
\begin{equation}
\pi^{NC}(\phi,\theta)\equiv(1-\alpha)(q_{H}-s)\frac{q_{L}+s}{2}+(1-q_{H}+s)\Bigl(\frac{q_{H}+q_{L}}{2}+\theta\Bigr).
\label{eq:pi-I}
\end{equation}
In the main analysis, $\pi^{NC}(\phi,q^{\textrm{base}})$ is the profit \eqref{eq:nc-profit} of Corollary~\ref{cor:nc-profit}.

We use Lemma~\ref{lemma:nc-pricing} on pricing under NC to prove Proposition~\ref{prop:nc-pricing} and Corollary~\ref{cor:nc-profit}. We state it for a general $\theta$ so that it also serves Proposition~\ref{prop:imperfect-steerability-nc}. Its message is that the platform keeps the full price of AI-generated content but only the share $1-\alpha$ of creators' revenue. Yet it must leave $L$ enough sales to cover her participation cost, because her exit would remove her training data.

\begin{applemma}[Pricing under NC]
\label{lemma:nc-pricing}
Let $\kappa_{0}\geq0$, so that $\theta=q^{\textrm{base}}-\kappa_{0}$ and $z=\theta+\tfrac{t}{2}$.

(a) In a pure-strategy Stage-2 equilibrium in which a creator exits or the AI sells nothing, profit is at most $\pi^{B}$. If a creator exits, it is at most $(1-\alpha)q_{H}^{2}/4$, which is strictly below $\pi^{B}$.

(b) Consider a pure-strategy Stage-2 equilibrium in which both creators participate and the AI sells. Then all three products sell with $g<\delta<\tfrac12$ and $\delta\geq\delta^{NC}$, and profit is at most
\begin{equation}
\widetilde{\pi}(\delta)\equiv\frac{t}{2}-\alpha t\delta-2(1-\alpha)\delta^{2}+(1-2\delta)\theta,
\label{eq:pi-tilde}
\end{equation}
with equality if and only if the platform prices the AI at its quality, $P_{AI}=z$.

(c) Suppose $z>0$. At $(P^{NC},z)$, the only allowed play has both creators participate. Demands are $D_{j}^{NC}=q_{j}-P^{NC}$ and $D_{AI}^{NC}=1-q_{H}+s$, $L$'s participation constraint binds ($V_{L}=\phi q_{L}$), $H$'s is slack ($V_{H}=\alpha(q_{L}+s)(2q_{H}-q_{L}-s)/4>\phi q_{H}$), and profit is $\widetilde{\pi}(\delta^{NC})=\pi^{NC}(\phi,\theta)$.

(d) If $\theta\geq0$, then $\partial\pi^{NC}(\phi,\theta)/\partial s=(q_{H}-s)+\alpha(s-g)+\theta>0$ and $\pi^{NC}(\phi,\theta)>\pi^{B}$.
\end{applemma}

\begin{proof}
(a) If neither creator participates, AI revenue is zero (Lemma~\ref{lemma:one-contributor}(b)). If only creator $j$ participates, the AI is trained on her content alone. In equilibrium her sales cover her cost, $\alpha D_{j}P\geq\phi q_{j}>0$, so $D_{j}>0$ and hence $D_{AI}=0$ (Lemma~\ref{lemma:one-contributor}(a)). Her buyers obtain nonnegative utility from her content, so $D_{j}\leq(q_{j}-P)^{+}$. In these two cases profit is therefore at most $(1-\alpha)P(q_{j}-P)^{+}\leq(1-\alpha)q_{H}^{2}/4=\Pi_{H}$, and $\Pi_{H}<\Pi_{HL}=\pi^{B}$ by Lemma~\ref{lemma:no-ai}(c). When both creators participate but the AI sells nothing, AI revenue is zero, and Lemma~\ref{lemma:no-ai-revenue} gives the bound.

(b) Both creators participate, so each covers her cost and $D_{L}>0$. As $D_{AI}>0$ as well, Lemma~\ref{lemma:both-contribute} gives $\widetilde{P}_{AI}\leq\tfrac{t}{2}$ and $g<\delta<\tfrac12$, with $D_{H}=\delta+g$, $D_{L}=\delta-g$ and $D_{AI}=1-2\delta$. Substituting $P_{AI}=P+\delta+\theta$ into \eqref{eq:platformprofit}, profit is
\[
\pi(P,\delta)\equiv(1-\alpha)\,2\delta P+(1-2\delta)\bigl(P+\delta+\theta\bigr).
\]
Now raise both prices by the same amount, which leaves $\delta$ and hence all demands unchanged. Profit rises, as $\partial\pi/\partial P=1-2\alpha\delta>0$, and so does each creator's revenue. The only limit is that consumers must remain willing to buy AI-generated content, $P+\delta=\widetilde{P}_{AI}\leq\tfrac{t}{2}$. Hence $\pi(P,\delta)\leq\pi(\tfrac{t}{2}-\delta,\delta)=\widetilde{\pi}(\delta)$, with equality if and only if $P_{AI}=z$, the price at which AI buyers obtain zero utility. For the same reason, $L$ covers her cost at $(P,\delta)$ only if she does at $(\tfrac{t}{2}-\delta,\delta)$, that is, only if $\alpha(\delta-g)(\tfrac{t}{2}-\delta)\geq\phi q_{L}$. Using $\phi q_{L}/\alpha=(q_{L}^{2}-s^{2})/4$,
\[
\alpha(\delta-g)\Bigl(\frac{t}{2}-\delta\Bigr)-\phi q_{L}=-\alpha\Bigl(\delta-\frac{q_{H}-s}{2}\Bigr)\Bigl(\delta-\frac{q_{H}+s}{2}\Bigr),
\]
so $\delta\geq\delta^{NC}$: the price gap must leave $L$ enough buyers to cover her cost.

(c) At $(P^{NC},z)$, $\delta=\tfrac{t}{2}-P^{NC}=\delta^{NC}$, so under $(1,1)$ Lemma~\ref{lemma:both-contribute} gives
\[
D_{L}=\delta^{NC}-g=\frac{q_{L}-s}{2},\qquad
D_{H}=\delta^{NC}+g=\frac{2q_{H}-q_{L}-s}{2},\qquad
D_{AI}=1-2\delta^{NC}=1-q_{H}+s,
\]
all positive, with $D_{j}=q_{j}-P^{NC}$. Hence $V_{L}=\alpha(q_{L}^{2}-s^{2})/4=\phi q_{L}$ and $V_{H}-\phi q_{H}=\alpha(q_{H}-q_{L})(q_{L}+s)^{2}/(4q_{L})>0$, and profit is $\pi(P^{NC},\delta^{NC})=\widetilde{\pi}(\delta^{NC})$, which equals \eqref{eq:pi-I}. As $P^{NC}\geq\tfrac{q_{L}+g}{2}=\tfrac{t}{4}>\tfrac{t-1}{2}$, Lemma~\ref{lemma:ai-at-quality}(a) applies: each participating creator sells $q_{j}-P^{NC}$ whether or not the other participates. Participating thus gives $L$ a net payoff of zero and $H$ a positive one whatever the other creator does. The pure-strategy equilibria are therefore $(1,1)$ and the profile in which only $H$ participates. The latter violates Assumption~\ref{assumption:weak-participation} because the indifferent $L$ exits. As allowed plays are pure under NC (Lemma~\ref{lemma:stage2}(a)), the profile $(1,1)$ is the only allowed play.

(d) Differentiating \eqref{eq:pi-I} gives the stated derivative, which is positive because $g\leq s<q_{H}$ and $\theta\geq0$. So $\pi^{NC}(\phi,\theta)$ is at least its value at $s=g$, where $q_{H}-s=q_{L}+s=\tfrac{t}{2}$:
\[
\pi^{NC}(\phi,\theta)\geq\pi^{B}+\Bigl(1-\frac{t}{2}\Bigr)\Bigl(\theta+\frac{t}{2}\Bigr)>\pi^{B}.
\]
At $s=g$, the human price and creators' demands are those of the benchmark, and the AI adds the revenue from the consumers whom the benchmark leaves unserved.
\end{proof}

\begin{proof}[Proof of Proposition~\ref{prop:nc-pricing}]
In the main analysis, $\theta=q^{\textrm{base}}\in[0,1]$ and $z=q^{\textrm{base}}+\tfrac{t}{2}=P_{AI}^{NC}>0$. We show that the platform sells as much AI-generated content as $L$'s participation allows.

\noindent\textit{Step 1 (The platform leaves $L$ the minimum).} Consider a pure-strategy Stage-2 equilibrium in which both creators participate and the AI sells. By Lemma~\ref{lemma:nc-pricing}(b), profit is at most $\widetilde{\pi}(\delta)$ and $\delta\geq\delta^{NC}$. As $\theta\geq0$, $\widetilde{\pi}$ strictly decreases in $\delta$,
\[
\widetilde{\pi}'(\delta)=-\alpha t-4(1-\alpha)\delta-2\theta<0,
\]
because a wider price gap moves consumers from AI-generated content to creators' content, on which the platform earns less. The platform's profit is therefore at most $\widetilde{\pi}(\delta^{NC})=\pi^{NC}$, with equality only if $\delta=\delta^{NC}$ and $P_{AI}=z$, that is, only at the price pair $(P^{NC},P_{AI}^{NC})$.

\noindent\textit{Step 2 (Other outcomes yield less).} By Lemma~\ref{lemma:nc-pricing}(a), a pure-strategy Stage-2 equilibrium in which a creator exits or the AI sells nothing yields at most $\pi^{B}$, and $\pi^{B}<\pi^{NC}$ by Lemma~\ref{lemma:nc-pricing}(d). So every pure-strategy Stage-2 equilibrium at any other price pair yields strictly less than $\pi^{NC}$.

\noindent\textit{Step 3 (Equilibrium).} By Lemma~\ref{lemma:nc-pricing}(c), at $(P^{NC},P_{AI}^{NC})$ the only allowed play has both creators participate, and it yields $\pi^{NC}$. As allowed plays are pure under NC (Lemma~\ref{lemma:stage2}(a)), Step~2 covers all allowed plays at every other price pair. Lemma~\ref{lemma:stage2}(b) with $\mathcal{P}^{*}=\{(P^{NC},P_{AI}^{NC})\}$ and $\pi^{*}=\pi^{NC}$ then shows that an equilibrium exists and that every equilibrium has these prices, with the participation decisions, demands and payoffs stated in Lemma~\ref{lemma:nc-pricing}(c).

\smallskip
\noindent\textit{Proof of Corollary~\ref{cor:nc-profit}.} The platform's equilibrium profit is $\pi^{NC}(\phi,q^{\textrm{base}})$, which is \eqref{eq:nc-profit}. It depends on $\phi$ only through $s$, with $ds/d\phi=-2q_{L}/(\alpha s)<0$, so it strictly decreases in $\phi$ by Lemma~\ref{lemma:nc-pricing}(d). As $s$ does not depend on $q^{\textrm{base}}$, $\partial\pi^{NC}/\partial q^{\textrm{base}}=1-q_{H}+s=D_{AI}^{NC}>0$.
\end{proof}

\subsection{Proof of Lemma~\ref{lemma:demands-c}}
\label{apx:proof-lemma-demands-c}

\begin{proof}
The columns $q_{AI}$ and $\kappa(x)$ follow directly from \eqref{eq:kappa} and \eqref{eq:q-ai}. Row $(1,1)$ is Lemma~\ref{lemma:both-contribute}, rows $(1,0)$ and $(0,1)$ are Lemma~\ref{lemma:one-contributor}(a), and row $(0,0)$ is Lemma~\ref{lemma:one-contributor}(b). These lemmas hold whether a creator who does not contribute her data participates or exits the platform.
\end{proof}

\subsection{Pricing under Compensation}
\label{apx:pricing-c}

In this section, we characterize the platform's profit under C profile by profile, for the main analysis and the extensions at once. We maintain Assumption~\ref{assumption:benchmark-participation} throughout, so $\pi^{B}=(1-\alpha)t^{2}/8$ (Proposition~\ref{prop:benchmark}). For $z\geq0$, we define
\[
D_{H}^{C}(z)\equiv\tfrac12(q_{H}-z)^{+},\qquad
F(z)\equiv z+D_{H}^{C}(z)^{2},\qquad
G(z)\equiv z\,\bigl[1-D_{H}^{C}(z)\bigr],
\]
and we let $\mathcal{P}(z)$ denote the set of \emph{displacement prices}: the pair $(\tfrac{q_{H}+z}{2},z)$ if $z<q_{H}$, and the pairs $(P,z)$ with $P\geq q_{H}$ if $z\geq q_{H}$. Both $F$ and $G$ are continuous and strictly increasing: $F'(z)=1-D_{H}^{C}(z)>0$ and $G'(z)=1-\tfrac{q_{H}}{2}+z>0$ for $z<q_{H}$, and $F'=G'=1$ for $z>q_{H}$. We let $\underline{\theta}_{C}\equiv-(1+\sqrt{2})g$, so that $-3g<\underline{\theta}_{C}<-2g$; Assumption~\ref{assumption:benchmark-participation} gives $\underline{\theta}_{C}+\tfrac{t}{2}=\tfrac{(2+\sqrt{2})q_{L}-\sqrt{2}q_{H}}{2}>0$. Lemmas~\ref{lemma:c-both} and~\ref{lemma:c-payoffs} describe the displacement prices when $\theta\geq\underline{\theta}_{C}$ and both creators contribute. There the AI displaces $L$ and $H$ keeps demand $D_{H}^{C}(z)$. The platform earns $(1-\alpha)F(z)$ and AI revenue is $G(z)$.

Lemma~\ref{lemma:c-both} gives the platform's most profitable prices when both creators contribute their data. This lemma underlies Proposition~\ref{prop:c-pricing} and its extensions in Propositions~\ref{prop:imperfect-steerability}--\ref{prop:quality-weighted}.

\begin{applemma}[Profit when both creators contribute]
\label{lemma:c-both}
Let $z>0$, that is, $\theta>-\tfrac{t}{2}$. Under C and $(1,1)$, profit at any price pair is at most $(1-\alpha)M$, with $M$ and the price pairs attaining the bound as follows.

(i) \emph{Displacement.} If $\theta>\underline{\theta}_{C}$, then $M=F(z)$, and the bound is attained exactly at the displacement prices $\mathcal{P}(z)$, where demands are $D_{L}=0$, $D_{H}=D_{H}^{C}(z)$ and $D_{AI}=1-D_{H}^{C}(z)$.

(ii) \emph{Three products.} If $\theta<\underline{\theta}_{C}$, then $M=z+\tfrac12\theta^{2}$, and the bound is attained only at the three-product prices $(\tfrac{t+\theta}{2},z)$, where $D_{H}=g-\tfrac{\theta}{2}$, $D_{L}=-g-\tfrac{\theta}{2}>0$ and $D_{AI}=1+\theta>0$.

(iii) \emph{Tie.} If $\theta=\underline{\theta}_{C}$, then $M=F(z)=z+\tfrac12\theta^{2}$, and the bound is attained only at the pair in $\mathcal{P}(z)$ and at the three-product prices, with the demands stated in (i) and (ii), respectively.

\noindent In all three cases, $M\geq F(z)\geq z$ and the bound exceeds the benchmark profit, $(1-\alpha)M>\pi^{B}$.
\end{applemma}

\begin{proof}
Under $(1,1)$ the platform's choice reduces to the price gap $\delta$, which determines who buys what (Lemma~\ref{lemma:both-contribute}). We first show that the platform prices the AI at $z$, and then maximize over $\delta$.

If the AI sells nothing, AI revenue is zero and profit is at most $\pi^{B}$ (Lemma~\ref{lemma:no-ai-revenue}), which is below the bound by the last claim of the lemma. Suppose $D_{AI}>0$. Then $U_{AI}=z-P_{AI}\geq0$, so $P_{AI}\leq z$. By Lemma~\ref{lemma:both-contribute}, the market is covered and demands depend on prices only through $\delta$. Here $\delta<\tfrac12$ because $D_{AI}>0$, and $\delta=P_{AI}-\theta-P\leq z-\theta=\tfrac{t}{2}$ because $P\geq0$. Substituting $P=P_{AI}-\theta-\delta$ and $D_{H}+D_{L}+D_{AI}=1$ into \eqref{eq:platformprofit},
\[
\pi=(1-\alpha)\bigl[\Psi(\delta)-(z-P_{AI})\bigr],\qquad
\Psi(\delta)\equiv(D_{H}+D_{L})\Bigl(\frac{t}{2}-\delta\Bigr)+D_{AI}\,z.
\]
So $\pi\leq(1-\alpha)\Psi(\delta)$, with equality if and only if $P_{AI}=z$. In a covered market, raising both prices by the same amount loses no buyer, so the platform raises the AI price until AI buyers obtain zero utility. The human price is then $P=\tfrac{t}{2}-\delta$, and by Lemma~\ref{lemma:both-contribute},
\[
\Psi(\delta)=\begin{cases}
z, & \delta\leq-g\quad\text{(only the AI sells)},\\[2pt]
(\delta+g)(\tfrac{t}{2}-\delta)+(1-\delta-g)z, & -g\leq\delta\leq g\quad\text{(the AI and $H$ sell)},\\[2pt]
2\delta(\tfrac{t}{2}-\delta)+(1-2\delta)z, & g\leq\delta\leq\bar\delta\equiv\min\{\tfrac12,\tfrac{t}{2}\}\quad\text{(all three sell)}.
\end{cases}
\]
$\Psi$ is continuous, and on the second and third ranges it is strictly concave, with derivatives $-(2\delta+g+\theta)$ and $-2(2\delta+\theta)$. These vanish at $\delta_{c}\equiv-\tfrac{g+\theta}{2}$ and $\delta_{3}\equiv-\tfrac{\theta}{2}$, where the two expressions equal $z+\tfrac14(g-\theta)^{2}$ and $z+\tfrac12\theta^{2}$, and
\begin{equation}
\Bigl[z+\frac{(g-\theta)^{2}}{4}\Bigr]-\Bigl[z+\frac{\theta^{2}}{2}\Bigr]=\frac{g^{2}-2g\theta-\theta^{2}}{4}.
\label{eq:row3-row2}
\end{equation}
The right-hand side is a downward parabola in $\theta$ with roots $\underline{\theta}_{C}$ and $(\sqrt{2}-1)g$, so it is positive for $\underline{\theta}_{C}<\theta<(\sqrt{2}-1)g$ and negative for $\theta<\underline{\theta}_{C}$. Four cases cover $\theta>-\tfrac{t}{2}$.

\emph{Case $\theta\geq g$.} Both derivatives are negative for $\delta>-g$, so $\Psi$ is maximized exactly on $\delta\leq-g$, where it equals $z=F(z)$. With $P_{AI}=z$, the condition $\delta\leq-g$ means $P\geq q_{H}$.

\emph{Case $-2g\leq\theta<g$.} Then $\delta_{c}\in(-g,g)$ and $\delta_{3}\leq g$, so $\Psi$ decreases on $[g,\bar\delta]$ and is maximized only at $\delta_{c}$, where it equals $z+\tfrac14(g-\theta)^{2}=F(z)$, as $g-\theta=q_{H}-z>0$.

\emph{Case $-3g\leq\theta<-2g$.} Then $\delta_{c}\in(-g,g]$ and $\delta_{3}\in(g,\tfrac{t}{4})\subset(g,\bar\delta)$, as $\theta>-\tfrac{t}{2}$ and $t<2$. $\Psi$ then has one peak on each of the last two ranges, and \eqref{eq:row3-row2} ranks the two peaks. The maximum of $\Psi$ is at $\delta_{c}$ alone if $\theta>\underline{\theta}_{C}$, at $\delta_{3}$ alone if $\theta<\underline{\theta}_{C}$, and at both peaks if $\theta=\underline{\theta}_{C}$.

\emph{Case $\theta<-3g$.} Then $\delta_{c}>g$, so $\Psi$ increases on $[-g,g]$ and is maximized only at $\delta_{3}\in(g,\tfrac{t}{4})$.

As $-3g<\underline{\theta}_{C}<-2g$, these four cases give $M$ as stated in (i)--(iii). Recall that $P_{AI}=z$ and $P=\tfrac{t}{2}-\delta$. The gap $\delta_{c}$ lies in $(-g,g]$ whenever it is a maximizer. It gives $P=\tfrac{q_{H}+z}{2}$ and, by Lemma~\ref{lemma:both-contribute}, $D_{L}=0$ and $D_{H}=\delta_{c}+g=\tfrac{q_{H}-z}{2}=D_{H}^{C}(z)$. The gap $\delta_{3}$ gives $P=\tfrac{t+\theta}{2}$, $D_{H}=g-\tfrac{\theta}{2}$, $D_{L}=-g-\tfrac{\theta}{2}$ and $D_{AI}=1+\theta$, where $D_{L}>0$ because $\theta<-2g$ in this case.

Finally, $M\geq F(z)\geq z$ in every case, by \eqref{eq:row3-row2} when $\theta<\underline{\theta}_{C}$. Moreover, $\tfrac{t}{4}\in(g,\bar\delta)$ because $q_{H}<3q_{L}$, so $M\geq\Psi(\tfrac{t}{4})=\tfrac{t^{2}}{8}+(1-\tfrac{t}{2})z>\tfrac{t^{2}}{8}$ and $(1-\alpha)M>\pi^{B}$. Intuitively, the platform already earns more than in the benchmark by keeping the benchmark price $\tfrac{t}{4}$ for creators' content and selling AI-generated content to the consumers whom the benchmark leaves unserved.
\end{proof}

With a single contributor, the AI competes head-on with that creator's content. Lemma~\ref{lemma:c-one} shows that the platform then earns at most the benchmark profit if the AI earns nothing, and less than $(1-\alpha)z$ otherwise. Under $(1,1)$, it can secure $(1-\alpha)z$ by letting the AI serve the whole market at price $z$. We use this lemma in Lemma~\ref{lemma:c-bound} and in the proof of Proposition~\ref{prop:imperfect-steerability} to show that the platform prefers a profile in which both creators contribute their data.

\begin{applemma}[Profit with at most one contributor]
\label{lemma:c-one}
Consider regime C in either of two settings: equal transfer rates ($\tau_{H}=\tau_{L}=\tfrac12$) with any $\kappa_{0}\geq0$, or $\kappa_{0}=0$ with $\tau_{H}=\tfrac12$ and $0<\tau_{L}\leq\tfrac12$. Let $z>0$. At every price pair and under every profile with at most one contributor, the platform's profit is at most the benchmark profit $\pi^{B}$ if AI revenue is zero, and strictly below $(1-\alpha)z$ otherwise.
\end{applemma}

\begin{proof}
If AI revenue is zero, Lemma~\ref{lemma:no-ai-revenue} applies. Otherwise exactly one creator $k$ contributes (Lemma~\ref{lemma:one-contributor}(b)) and the AI sells. With $j\neq k$, Lemma~\ref{lemma:one-contributor}(a) gives $D_{k}=0$ and bounds the two remaining revenues, so profit is at most $(1-\alpha)(q_{j}^{2}+y_{k}^{2})/4$, where $0<y_{k}=z-\tau_{j}q_{j}$. By Assumption~\ref{assumption:training-data-dependence}, $y_{k}=q^{\textrm{base}}-\kappa_{0}+\tau_{k}q_{k}<2$, so $y_{k}^{2}/4<y_{k}/2$ and
\[
z-\frac{q_{j}^{2}+y_{k}^{2}}{4}>z-\frac{q_{j}^{2}}{4}-\frac{z-\tau_{j}q_{j}}{2}=\frac{z}{2}+\frac{q_{j}(2\tau_{j}-q_{j})}{4}.
\]
The last expression is strictly positive if $\tau_{j}=\tfrac12$, because $q_{j}<1$. In the remaining case, $\kappa_{0}=0$ and $j=L$ with $\tau_{L}<\tfrac12$, the expression is at least $\tfrac{z}{2}-\tfrac{q_{L}^{2}}{4}>0$, because $z>\tfrac{q_{H}}{2}>\tfrac{q_{L}^{2}}{2}$.
\end{proof}

Lemma~\ref{lemma:c-payoffs} gives payoffs at the displacement prices. We use it to show that both creators participate (Propositions~\ref{prop:imperfect-steerability}--\ref{prop:quality-weighted}) and to compare payoffs across regimes (Corollary~\ref{cor:data-supplier} and Proposition~\ref{prop:welfare}).

\begin{applemma}[Creators' payoffs at the displacement prices]
\label{lemma:c-payoffs}
Let $\theta\geq\underline{\theta}_{C}$ and $z>0$, and consider $(1,1)$ at a price pair in $\mathcal{P}(z)$. AI revenue is $G(z)$. Under equal splitting, $L$'s gross payoff is $\alpha G(z)/2$, and $H$'s is
\[
V_{H}(z)\equiv\begin{cases}
\alpha\bigl[q_{H}^{2}+(2-q_{H})z\bigr]/4, & z<q_{H},\\[2pt]
\alpha z/2, & z\geq q_{H},
\end{cases}
\]
which is continuous and strictly increasing in $z$ and satisfies $V_{H}(z)>\alpha q_{H}^{2}/4>\phi^{B}q_{H}$. Hence $H$'s net payoff $V_{H}(z)-\phi q_{H}$ is strictly positive for every $z>0$ and every $\phi\leq\phi^{B}$.
\end{applemma}

\begin{proof}
By Lemma~\ref{lemma:c-both}, $D_{L}=0$, $D_{H}=D_{H}^{C}(z)$ and $D_{AI}=1-D_{H}^{C}(z)$, so AI revenue is $D_{AI}z=G(z)$. As $L$ sells nothing, her payoff is half of the creators' share of this revenue. If $z\geq q_{H}$, $H$ also sells nothing and earns $\alpha z/2\geq\alpha q_{H}/2$. If $z<q_{H}$, she sells $\tfrac{q_{H}-z}{2}$ at the price $\tfrac{q_{H}+z}{2}$, so by \eqref{eq:creator-payoff},
\[
\frac{V_{H}(z)}{\alpha}=\frac{(q_{H}-z)(q_{H}+z)}{4}+\frac12\Bigl(1-\frac{q_{H}-z}{2}\Bigr)z=\frac{q_{H}^{2}+(2-q_{H})z}{4}.
\]
Both expressions increase in $z$ and equal $\alpha q_{H}/2$ at $z=q_{H}$. They exceed $\alpha q_{H}^{2}/4$ because $z>0$ and $q_{H}<1$, and $\tfrac{\alpha q_{H}}{4}-\phi^{B}=\tfrac{\alpha(q_{H}-q_{L})(q_{H}+3q_{L})}{16q_{L}}>0$. As $\phi\leq\phi^{B}$, $V_{H}(z)>\phi q_{H}$.
\end{proof}

In Lemma~\ref{lemma:c-bound}, we combine Lemmas~\ref{lemma:c-both} and~\ref{lemma:c-one} when there is no base prompting effort, as in the main analysis and in Sections~\ref{subsec:heterogeneous-data} and~\ref{subsec:quality-weighted}. It is the first step in proving Propositions~\ref{prop:c-pricing}, \ref{prop:heterogeneous-data} and~\ref{prop:quality-weighted}.

\begin{applemma}[Profit bound under C without base prompting effort]
\label{lemma:c-bound}
Let $\kappa_{0}=0$, $\tau_{H}=\tfrac12$ and $0<\tau_{L}\leq\tfrac12$, and suppose $q_{H}/q_{L}\geq\sqrt{2}$ if $\tau_{L}<\tfrac12$. Then $\theta>\underline{\theta}_{C}$. Under either sharing rule, the platform's profit is at most $(1-\alpha)F(z)$ at every price pair and under every profile. Equality holds if and only if the profile is $(1,1)$ and the price pair lies in $\mathcal{P}(z)$. At these pairs, $D_{L}=0$, $D_{H}=D_{H}^{C}(z)$ and $D_{AI}=1-D_{H}^{C}(z)$.
\end{applemma}

\begin{proof}
Neither consumer demand nor the platform's profit depends on how creators split their share of AI revenue. Here $z=q^{\textrm{base}}+\tfrac{q_{H}}{2}+\tau_{L}q_{L}>0$ and $\theta=q^{\textrm{base}}-(\tfrac12-\tau_{L})q_{L}$. If $\tau_{L}=\tfrac12$, then $\theta=q^{\textrm{base}}\geq0>\underline{\theta}_{C}$. If $\tau_{L}<\tfrac12$, then $\theta>-\tfrac{q_{L}}{2}$ and
\[
-\frac{q_{L}}{2}-\underline{\theta}_{C}=\frac{1+\sqrt{2}}{2}\bigl(q_{H}-\sqrt{2}\,q_{L}\bigr)\geq0,
\]
so $\theta>\underline{\theta}_{C}$: when $q_{H}/q_{L}\geq\sqrt{2}$, a low transfer rate never makes the platform prefer the three-product prices. Lemma~\ref{lemma:c-both}(i) thus gives the claim for $(1,1)$, and $(1-\alpha)F(z)>\pi^{B}$. Under every other profile, profit is at most $\pi^{B}$ or strictly below $(1-\alpha)z\leq(1-\alpha)F(z)$ by Lemma~\ref{lemma:c-one}.
\end{proof}

\subsection{Proof of Proposition~\ref{prop:c-pricing} and Corollaries~\ref{cor:compensation-displacement} and~\ref{cor:data-supplier}}
\label{apx:proof-c}

\begin{proof}
In the main analysis, $z=q^{\textrm{base}}+\tfrac{t}{2}=P_{AI}^{C}$. As $\overline{q}^{\textrm{base}}=g$ and $q_{H}=g+\tfrac{t}{2}$, CQ1 is the case $z<q_{H}$ and CQ2 the case $z\geq q_{H}$, and the price pairs in the statement are the displacement prices $\mathcal{P}^{C}\equiv\mathcal{P}(z)$. We write $D_{H}^{C}\equiv D_{H}^{C}(z)$ and $\pi^{C}\equiv(1-\alpha)F(z)=(1-\alpha)[P_{AI}^{C}+(D_{H}^{C})^{2}]$.

Under compensation, a creator no longer needs direct sales to cover her participation cost, because her share of AI revenue can cover it instead. The platform can therefore choose the prices that maximize its profit when both creators contribute, which fully displace $L$ (Step~1), provided that the AI payment keeps both creators on the platform (Step~2).

\noindent\textit{Step 1 (Profit bound).} By Lemma~\ref{lemma:c-bound} with $\tau_{L}=\tfrac12$, every profile yields at most $\pi^{C}$ at every price pair. Equality holds only under $(1,1)$ at the pairs in $\mathcal{P}^{C}$, where $D_{L}=0$, $D_{H}=D_{H}^{C}$ and $D_{AI}=1-D_{H}^{C}$. In CQ1, the pair is $(P^{C},P_{AI}^{C})$ with $P^{C}=\tfrac{q_{H}+z}{2}=\tfrac{3q_{H}+q_{L}}{4}+\tfrac{q^{\textrm{base}}}{2}$, and $D_{H}^{C}=\tfrac{q_{H}-z}{2}=\tfrac{g-q^{\textrm{base}}}{2}>0$. In CQ2, $D_{H}^{C}=0$ and the AI serves the whole market at every $P\geq q_{H}$.

\noindent\textit{Step 2 (The AI payment covers both creators' costs).} Under $(1,1)$ at a pair in $\mathcal{P}^{C}$, each creator receives the AI payment $A^{C}\equiv\tfrac{\alpha}{2}D_{AI}^{C}P_{AI}^{C}$. As $D_{AI}^{C}=1-D_{H}^{C}\geq1-\tfrac{g}{2}>\tfrac34$ and $P_{AI}^{C}\geq\tfrac{t}{2}$, $A^{C}>\tfrac{3\alpha t}{16}$, and
\[
\frac{3\alpha t}{16}-\phi^{B}q_{H}=\frac{\alpha t\,\bigl[3q_{L}(1-q_{H})+q_{H}^{2}\bigr]}{16\,q_{L}}>0.
\]
So $V_{j}^{C}\geq A^{C}>\phi^{B}q_{H}\geq\phi q_{j}$ by Assumption~\ref{assumption:benchmark-participation}: even $L$, with no sales, prefers to participate.

\noindent\textit{Step 3 (Equilibrium).} At every pair in $\mathcal{P}^{C}$, $P>q_{L}>\tfrac{t-1}{2}$, since $P^{C}-q_{L}=\tfrac{3g+q^{\textrm{base}}}{2}>0$ in CQ1 and $P\geq q_{H}$ in CQ2. As both net payoffs are positive (Step~2), Lemma~\ref{lemma:ai-at-quality}(b) shows that $(1,1)$ is the only allowed play there, and it yields $\pi^{C}$. By Step~1 and Lemma~\ref{lemma:stage2}(c), every allowed play yields strictly less at every pair outside $\mathcal{P}^{C}$. Lemma~\ref{lemma:stage2}(b) with $\mathcal{P}^{*}=\mathcal{P}^{C}$ and $\pi^{*}=\pi^{C}$ proves the proposition: in every equilibrium the prices lie in $\mathcal{P}^{C}$, creators play $(1,1)$, and demands are those of Step~1. In CQ2, every $P\geq q_{H}$ is an equilibrium price, with the same demands, profit and payoffs.

\smallskip
\noindent\textit{Proof of Corollary~\ref{cor:compensation-displacement}.} By Propositions~\ref{prop:nc-pricing} and~\ref{prop:c-pricing}, and as $s<q_{L}$ for $\phi>0$, $D_{L}^{C}=0<\tfrac{q_{L}-s}{2}=D_{L}^{NC}$ and $D_{H}^{C}\leq\tfrac{g}{2}<2g<2g+\tfrac{q_{L}-s}{2}=D_{H}^{NC}$. As the market is covered in both regimes, the AI's demand is the remainder in each, so $D_{AI}^{C}-D_{AI}^{NC}=(D_{H}^{NC}-D_{H}^{C})+D_{L}^{NC}>0$.

\smallskip
\noindent\textit{Proof of Corollary~\ref{cor:data-supplier}.} Step~2 above gives $V_{j}^{C}>\phi q_{j}$ for both creators in CQ1 and CQ2. As $D_{L}^{C}=0$, $L$'s payoff is her AI payment alone, $V_{L}^{C}=\tfrac{\alpha}{2}G(z)$ (Lemma~\ref{lemma:c-payoffs}). $G$ is continuous and strictly increasing and $z=q^{\textrm{base}}+\tfrac{t}{2}$, so $V_{L}^{C}$ strictly increases in $q^{\textrm{base}}$ on $[0,1]$. In CQ2, $D_{H}^{C}=0$ as well, so $V_{H}^{C}=V_{L}^{C}=\tfrac{\alpha}{2}z=\tfrac{\alpha}{2}q_{AI}$, because $q_{AI}=z$ when both creators contribute.
\end{proof}

\subsection{Proof of Corollary~\ref{cor:profit-ranking} and Proposition~\ref{prop:welfare}}
\label{apx:proof-welfare}

\begin{proof}
Both results compare the equilibria of Propositions~\ref{prop:nc-pricing} and~\ref{prop:c-pricing}, in which both creators participate and contribute their data. We use the following facts from these two propositions. The quality and the price of AI-generated content are the same in the two regimes, $q_{AI}=P_{AI}^{NC}=P_{AI}^{C}=z=q^{\textrm{base}}+\tfrac{t}{2}>0$. For $r\in\{NC,C\}$, we let $D^{r}\equiv D_{H}^{r}+D_{L}^{r}$, so that $D^{r}P^{r}$ is human-content revenue and $D_{AI}^{r}=1-D^{r}$. Under NC, $D^{NC}=q_{H}-s$, $D_{H}^{NC}-D_{L}^{NC}=2g$ and $P^{NC}=\tfrac{q_{L}+s}{2}$, none of which depends on $q^{\textrm{base}}$. Under C, $D^{C}=D_{H}^{C}=\tfrac12\max\{g-q^{\textrm{base}},0\}\leq\tfrac{g}{2}$, so $(D_{H}^{C})^{2}$ is differentiable in $q^{\textrm{base}}$ with derivative $-D_{H}^{C}$; in CQ1, $P^{C}=q_{H}-D_{H}^{C}$. Finally, $\phi=\alpha(q_{L}^{2}-s^{2})/(4q_{L})$ strictly decreases in $s$ and equals $\phi^{B}$ at $s=g$, so the range $\phi\in(0,\phi^{B}]$ corresponds to $s\in[g,q_{L})$.

\smallskip
\noindent\textit{Proof of Corollary~\ref{cor:profit-ranking}.} The two regimes respond differently to $\phi$ and to $q^{\textrm{base}}$. A higher participation cost lowers profit only under NC, where the platform must then leave $L$ more sales; this single crossing in $\phi$ defines the threshold $\phi^{\star}$ (Steps~1 and~2). A more capable AI raises the AI price one for one in both regimes, and wherever the platform weakly prefers C this raises its profit by more under C than under NC; this is why the threshold falls as $q^{\textrm{base}}$ rises (Step~3).

\noindent\textit{Step 1 (Both regimes beat the benchmark).} By Lemma~\ref{lemma:nc-pricing}(d) and Corollary~\ref{cor:nc-profit}, $\pi^{NC}>\pi^{B}$ and $\pi^{NC}$ strictly decreases in $\phi$. By Lemma~\ref{lemma:c-both} with $\theta=q^{\textrm{base}}$, $\pi^{C}=(1-\alpha)F(z)=(1-\alpha)[P_{AI}^{C}+(D_{H}^{C})^{2}]>\pi^{B}$, and the profit $\pi^{C}$ does not depend on the participation cost parameter $\phi$.

\noindent\textit{Step 2 (The threshold).} For $s\in[g,q_{L}]$, we let $\Lambda(s,q^{\textrm{base}})\equiv\pi^{NC}-\pi^{C}$, with $\pi^{NC}$ given by \eqref{eq:pi-I} (at $s=q_{L}$, its limit as $\phi\downarrow0$). By Lemma~\ref{lemma:nc-pricing}(d), $\partial\Lambda/\partial s=(q_{H}-s)+\alpha(s-g)+q^{\textrm{base}}>0$. Fix $q^{\textrm{base}}$. If $\Lambda(g,q^{\textrm{base}})\geq0$, we set $\phi^{\star}=\phi^{B}$; then $\pi^{NC}>\pi^{C}$ for $\phi<\phi^{B}$ and $\pi^{NC}\geq\pi^{C}$ at $\phi^{B}$. If $\Lambda(q_{L},q^{\textrm{base}})\leq0$, which excludes the previous case, we set $\phi^{\star}=0$; then $\pi^{NC}<\pi^{C}$ on $(0,\phi^{B}]$. Otherwise $\Lambda(\cdot,q^{\textrm{base}})$ is negative at $g$ and positive at $q_{L}$, so it has a unique zero $s^{\star}\in(g,q_{L})$; we set $\phi^{\star}=\alpha(q_{L}^{2}-(s^{\star})^{2})/(4q_{L})\in(0,\phi^{B})$, so that $\pi^{NC}\gtrless\pi^{C}$ as $\phi\lessgtr\phi^{\star}$, with equality at $\phi^{\star}$.

\noindent\textit{Step 3 (A better AI lowers the threshold).} We show that $\phi^{\star}$ is nonincreasing in $q^{\textrm{base}}$, and strictly decreasing wherever $0<\phi^{\star}<\phi^{B}$, as claimed in Section~\ref{subsec:platform-choice}. The argument has three parts.

First, human-content revenue is higher under NC. On $[g,q_{L}]$, $D^{NC}P^{NC}=(q_{H}-s)(q_{L}+s)/2$ has $s$-derivative $g-s\leq0$, so it is at least $2gq_{L}$, its value at $s=q_{L}$. In CQ1, $D^{C}P^{C}=(g-q^{\textrm{base}})(3q_{H}+q_{L}+2q^{\textrm{base}})/8$ has $q^{\textrm{base}}$-derivative $-(t+2q^{\textrm{base}})/4<0$, so it is at most $g(3q_{H}+q_{L})/8$, its value at $q^{\textrm{base}}=0$; in CQ2 it is zero. Hence, for all $s\in[g,q_{L}]$ and $q^{\textrm{base}}\in[0,1]$,
\begin{equation}
D^{NC}P^{NC}-D^{C}P^{C}\geq2gq_{L}-\frac{g(3q_{H}+q_{L})}{8}=\frac{3g(5q_{L}-q_{H})}{8}>0,
\label{eq:welf-revenue-gap}
\end{equation}
where the last inequality holds because $q_{H}/q_{L}<\rho^{\star}<5$ by Assumption~\ref{assumption:benchmark-participation}.

Second, wherever the platform weakly prefers C, a better AI strengthens this preference: $\partial\Lambda/\partial q^{\textrm{base}}<0$ wherever $\Lambda\leq0$. In each regime, profit rises with $q^{\textrm{base}}$ at the rate of the AI demand whose revenue the platform keeps: $\partial\pi^{NC}/\partial q^{\textrm{base}}=D_{AI}^{NC}$ by Corollary~\ref{cor:nc-profit}, and $\partial\pi^{C}/\partial q^{\textrm{base}}=(1-\alpha)(1-D_{H}^{C})=(1-\alpha)D_{AI}^{C}$. Writing $\pi^{NC}=(1-\alpha)D^{NC}P^{NC}+D_{AI}^{NC}z$ and $\pi^{C}=(1-\alpha)(D^{C}P^{C}+D_{AI}^{C}z)$ by \eqref{eq:platformprofit} and subtracting,
\[
z\,\frac{\partial\Lambda}{\partial q^{\textrm{base}}}=z\bigl[D_{AI}^{NC}-(1-\alpha)D_{AI}^{C}\bigr]=\Lambda-(1-\alpha)\bigl(D^{NC}P^{NC}-D^{C}P^{C}\bigr),
\]
which is negative when $\Lambda\leq0$ by \eqref{eq:welf-revenue-gap}. In words, the platform earns more from human content under NC. It can therefore prefer C only if it keeps more AI revenue under C, and a higher AI price then benefits it more under C. In particular, $(1-\alpha)D_{AI}^{C}>D_{AI}^{NC}$ at indifference.

Third, let $q<q'$ be any two values of $q^{\textrm{base}}$. For each $s$, $\Lambda(s,q)\leq0$ implies $\Lambda(s,q')<\Lambda(s,q)$. Otherwise $\Lambda(s,\cdot)$, which falls just to the right of $q$, would first return to the level $\Lambda(s,q)\leq0$ at a point where its derivative is nonnegative. This would contradict the second part of the argument. It follows that $\Lambda(\cdot,q')<0$ wherever $\Lambda(\cdot,q)\leq0$. By Step~2, $\phi^{\star}(q)=0$ then implies $\phi^{\star}(q')=0$; $0<\phi^{\star}(q)<\phi^{B}$ implies $\Lambda(\cdot,q')<0$ on $[g,s^{\star}(q)]$, so that $\phi^{\star}(q')=0$ or $s^{\star}(q')>s^{\star}(q)$, and in both cases $\phi^{\star}(q')<\phi^{\star}(q)$; and $\phi^{\star}(q)=\phi^{B}$ implies $\phi^{\star}(q')\leq\phi^{B}=\phi^{\star}(q)$.

\smallskip
\noindent\textit{Proof of Proposition~\ref{prop:welfare}.} We now compare in turn creators' net payoffs (Step~4), consumer surplus (Step~5) and total surplus (Step~6) across the equilibria of the two regimes.

\noindent\textit{Step 4 (Creators' net payoffs).} By Proposition~\ref{prop:nc-pricing}, $R_{L}^{NC}=0$, so by Corollary~\ref{cor:data-supplier}, $R_{L}^{C}-R_{L}^{NC}=V_{L}^{C}-\phi q_{L}$ is positive and strictly increases in $q^{\textrm{base}}$. For $H$, participation costs cancel: $R_{H}^{C}-R_{H}^{NC}=V_{H}^{C}-V_{H}^{NC}$. Here $V_{H}^{NC}=\alpha(q_{L}+s)(2q_{H}-q_{L}-s)/4$ does not depend on $q^{\textrm{base}}$ (Proposition~\ref{prop:nc-pricing}), while $V_{H}^{C}=V_{H}(z)$ strictly increases in $q^{\textrm{base}}$ (Lemma~\ref{lemma:c-payoffs}). So $R_{H}^{C}-R_{H}^{NC}$ strictly increases on $[0,1]$. It is therefore positive, because at $q^{\textrm{base}}=0$, where $z=\tfrac{t}{2}<q_{H}$, the identity $(q_{L}+s)(2q_{H}-q_{L}-s)=q_{H}^{2}-(q_{H}-q_{L}-s)^{2}$ gives
\[
R_{H}^{C}-R_{H}^{NC}=\frac{\alpha}{8}\bigl[t(2-q_{H})+2(q_{H}-q_{L}-s)^{2}\bigr]>0 .
\]

\noindent\textit{Step 5 (Consumer surplus).} AI buyers obtain $q_{AI}-P_{AI}^{r}=0$, so consumer surplus comes only from buyers of creators' content. A buyer of $j$'s content at distance $d$ from $j$ obtains $q_{j}-P^{r}-d=D_{j}^{r}-d$, since $D_{j}^{r}=q_{j}-P^{r}$ whenever $D_{j}^{r}>0$ (Propositions~\ref{prop:nc-pricing} and~\ref{prop:c-pricing}). Integrating over $d\in[0,D_{j}^{r}]$ gives $\mathrm{CS}^{r}=[(D_{H}^{r})^{2}+(D_{L}^{r})^{2}]/2$, so $\mathrm{CS}^{C}<\mathrm{CS}^{NC}$ by Corollary~\ref{cor:compensation-displacement}. Moreover, $\mathrm{CS}^{NC}=[(q_{L}-s)^{2}+(2q_{H}-q_{L}-s)^{2}]/8$ does not depend on $q^{\textrm{base}}$, while $\mathrm{CS}^{C}=(g-q^{\textrm{base}})^{2}/8$ strictly decreases in CQ1 and $\mathrm{CS}^{C}=0$ in CQ2. The reduction therefore strictly increases below $\overline{q}^{\textrm{base}}$ and is constant above it.

\noindent\textit{Step 6 (Total surplus).} Prices are transfers: by \eqref{eq:platformprofit} and \eqref{eq:creator-payoff}, $\pi^{r}+V_{H}^{r}+V_{L}^{r}$ equals consumers' spending. As both creators participate, $\mathrm{TS}^{r}=W^{r}-\phi t$, where $W^{r}$ is consumers' gross utility: $q_{j}-d$ for a buyer of $j$'s content at distance $d$ from $j$, and $q_{AI}$ for an AI buyer. Every consumer buys, so $D_{AI}^{r}=1-D_{H}^{r}-D_{L}^{r}$, and with $q_{H}-q_{AI}=g-q^{\textrm{base}}$ and $q_{L}-q_{AI}=-g-q^{\textrm{base}}$,
\[
W^{r}-q_{AI}=(g-q^{\textrm{base}})D_{H}^{r}-(g+q^{\textrm{base}})D_{L}^{r}-\tfrac12\bigl[(D_{H}^{r})^{2}+(D_{L}^{r})^{2}\bigr].
\]
Under C this equals $\tfrac32(D_{H}^{C})^{2}$, since $D_{L}^{C}=0$ and $g-q^{\textrm{base}}=2D_{H}^{C}$ in CQ1, while $D_{H}^{C}=0$ in CQ2. Under NC, $D_{H}^{NC}-D_{L}^{NC}=2g$ and $D_{H}^{NC}+D_{L}^{NC}=q_{H}-s$ turn the first two terms into $2g^{2}-(q_{H}-s)q^{\textrm{base}}$ and the bracket into $4g^{2}+2D_{H}^{NC}D_{L}^{NC}$. Since $q_{AI}$ is the same in both regimes,
\[
\mathrm{TS}^{C}-\mathrm{TS}^{NC}=D_{H}^{NC}D_{L}^{NC}+(q_{H}-s)\,q^{\textrm{base}}+\tfrac32(D_{H}^{C})^{2}>0,
\]
because $D_{H}^{NC}D_{L}^{NC}>0$. As $(D_{H}^{C})^{2}$ has derivative $-D_{H}^{C}$, the difference has $q^{\textrm{base}}$-derivative $(q_{H}-s)-\tfrac32D_{H}^{C}>2g-\tfrac34g>0$, where we use $s<q_{L}$ and $D_{H}^{C}\leq\tfrac{g}{2}$. Hence it strictly increases on $[0,1]$.
\end{proof}

\subsection{Proof of Proposition~\ref{prop:imperfect-steerability-nc}}
\label{apx:proof-imperfect-steerability-nc}

\begin{proof}
With base prompting effort $\kappa_{0}$, consumers value AI-generated content as if base capability were $\theta=q^{\textrm{base}}-\kappa_{0}$, which may be negative. Lemma~\ref{lemma:nc-pricing} describes pricing under NC for every such $\theta$. The platform lowers the AI price one for one with $\kappa_{0}$ and changes nothing else as long as widening the price gap would lower its profit, which is the case for every $\kappa_{0}$ below $\widehat{\kappa}_{0}(\phi)$.

\noindent\textit{Step 1 (The threshold).} Consider a pure-strategy Stage-2 equilibrium in which both creators participate and the AI sells. By Lemma~\ref{lemma:nc-pricing}(b), profit is at most $\widetilde{\pi}(\delta)$, where $\delta^{NC}\leq\delta<\tfrac12$. The function $\widetilde{\pi}$ in \eqref{eq:pi-tilde} is strictly concave, and
\[
\widetilde{\pi}'(\delta^{NC})=-\alpha t-2(1-\alpha)(q_{H}-s)-2\theta=2(\widehat{\theta}-\theta),\qquad\widehat{\theta}\equiv q^{\textrm{base}}-\widehat{\kappa}_{0}(\phi),
\]
where
\begin{equation}
\widehat{\kappa}_{0}(\phi)\;\equiv\;q^{\textrm{base}}+\frac{\alpha(q_{H}+q_{L})}{2}+(1-\alpha)(q_{H}-s)\;>\;q^{\textrm{base}},
\label{eq:kappa-hat}
\end{equation}
the inequality holding as $q_{H}>s$. Let $\kappa_{0}<\widehat{\kappa}_{0}(\phi)$, that is, $\theta>\widehat{\theta}$. Then $\widetilde{\pi}$ strictly decreases on $[\delta^{NC},\tfrac12)$, so profit is at most $\widetilde{\pi}(\delta^{NC})=\pi^{NC}(\phi,\theta)$, with equality only if $\delta=\delta^{NC}$ and $P_{AI}=z$, that is, only at $(P^{NC},z)$. The AI price is strictly positive there: $z=\tfrac{t}{2}+\theta>\tfrac{t}{2}+\widehat{\theta}=(1-\alpha)(s-g)\geq0$. In particular, $\tfrac{t}{2}+\widehat{\theta}\geq0$, that is, $\widehat{\kappa}_{0}(\phi)\leq\kappa_{0}^{*}$, where $\kappa_{0}^{*}\equiv q^{\textrm{base}}+\tfrac{t}{2}$ is the level of base prompting effort in Proposition~\ref{prop:imperfect-steerability} at which the AI stops earning any revenue.

\noindent\textit{Step 2 (Other outcomes yield less).} Using $(q_{H}-s)(q_{L}+s)=\tfrac{t^{2}}{4}-(s-g)^{2}$ in \eqref{eq:pi-I}, the platform's gain over the benchmark is its AI revenue minus its loss on creators' content:
\[
\pi^{NC}(\phi,\theta)-\pi^{B}=D_{AI}^{NC}\,z-(1-\alpha)\frac{(s-g)^{2}}{2}
>(1-\alpha)(s-g)\Bigl(1-q_{H}+\frac{s+g}{2}\Bigr)\geq0,
\]
where the strict inequality uses $D_{AI}^{NC}>0$ and $z>(1-\alpha)(s-g)$ (Step~1), and the last uses $s\geq g$ and $q_{H}<1$. By Lemma~\ref{lemma:nc-pricing}(a), a pure-strategy Stage-2 equilibrium in which a creator exits or the AI sells nothing yields at most $\pi^{B}$. Together with Step~1, every pure-strategy Stage-2 equilibrium at any price pair other than $(P^{NC},z)$ yields strictly less than $\pi^{NC}(\phi,\theta)$.

\noindent\textit{Step 3 (Equilibrium).} As $z>0$, Lemma~\ref{lemma:nc-pricing}(c) shows that at $(P^{NC},z)$ the only allowed play has both creators participate, and it yields $\pi^{NC}(\phi,\theta)$. Allowed plays are pure under NC (Lemma~\ref{lemma:stage2}(a)), so Step~2 covers every other price pair. Lemma~\ref{lemma:stage2}(b) with $\mathcal{P}^{*}=\{(P^{NC},z)\}$ and $\pi^{*}=\pi^{NC}(\phi,\theta)$ then shows that every equilibrium is that of Proposition~\ref{prop:nc-pricing} with $\theta$ in place of $q^{\textrm{base}}$. Its human price, demands and net payoffs (zero for $L$, positive for $H$) do not involve $\kappa_{0}$. The AI price $z=q^{\textrm{base}}+\tfrac{t}{2}-\kappa_{0}$ falls one for one with $\kappa_{0}$, and $\partial\pi^{NC}(\phi,\theta)/\partial\kappa_{0}=-(1-q_{H}+s)=-D_{AI}^{NC}$.
\end{proof}

\subsection{Proof of Proposition~\ref{prop:imperfect-steerability}}
\label{apx:proof-imperfect-steerability}

\begin{proof}
Here transfer rates are equal, so $z=\theta+\tfrac{t}{2}$ with $\theta=q^{\textrm{base}}-\kappa_{0}$. The proposition follows what the platform does as $\theta$ falls. Lemma~\ref{lemma:c-both} gives the prices it prefers when both creators contribute: the displacement prices of Proposition~\ref{prop:c-pricing} while $\theta\geq\underline{\theta}_{C}$, and the three-product prices below it. We show that no other profile does better (Step~1), find when creators' payoffs support $(1,1)$ at these prices (Step~2), and conclude (Step~3). Step~4 treats the case in which the AI earns no revenue.

If $\theta\geq0$, consumers' utilities under every profile are those of the main analysis with $\theta$ in place of $q^{\textrm{base}}$, and $\theta\leq1$ by Assumption~\ref{assumption:training-data-dependence}. The game is thus that of Section~\ref{sec:analysis} with $\theta\in[0,1]$ in place of $q^{\textrm{base}}$. As $\underline{\theta}_{C}<0$ and $\underline{\theta}_{L}(\phi)<0$ (Step~2), every such $\theta$ falls under part~1. It then follows from Proposition~\ref{prop:c-pricing} and Corollary~\ref{cor:data-supplier}, which gives $V_{L}^{C}>\phi q_{L}$. In Steps~1--3, let $-\tfrac{t}{2}<\theta<0$, so $z\in(0,\tfrac{t}{2})$ and $z<q_{H}$. Steps~2 and~4 use creator $j$'s revenue at the benchmark price, $V_{j}^{B}\equiv\alpha(q_{j}-\tfrac{t}{4})\tfrac{t}{4}$. This revenue covers her cost because $V_{j}^{B}/q_{j}=\alpha(1-\tfrac{t}{4q_{j}})\tfrac{t}{4}$ is larger for $H$ and $V_{L}^{B}/q_{L}=\phi^{B}\geq\phi$.

\noindent\textit{Step 1 (Profit bound).} We let $\overline{\pi}\equiv(1-\alpha)F(z)$ if $\theta\geq\underline{\theta}_{C}$ and $\overline{\pi}\equiv(1-\alpha)(z+\tfrac12\theta^{2})$ if $\theta<\underline{\theta}_{C}$. By Lemma~\ref{lemma:c-both}, profit under $(1,1)$ is at most $\overline{\pi}$. If $\theta>\underline{\theta}_{C}$, the bound is attained only at $(P^{C},P_{AI}^{C})\equiv(\tfrac{q_{H}+z}{2},z)$, the prices of Proposition~\ref{prop:c-pricing} (CQ1) with $\theta$ in place of $q^{\textrm{base}}$. If $\theta<\underline{\theta}_{C}$, it is attained only at the three-product prices $(P^{\mathrm{3prod}},z)\equiv(\tfrac{t+\theta}{2},z)$. If $\theta=\underline{\theta}_{C}$, it is attained at both pairs. Moreover, $\overline{\pi}>\pi^{B}$ and $\overline{\pi}\geq(1-\alpha)z$. By Lemma~\ref{lemma:c-one}, every profile with at most one contributor yields at most $\pi^{B}$ or strictly less than $(1-\alpha)z$, hence less than $\overline{\pi}$. By Lemma~\ref{lemma:stage2}(c), an allowed play thus yields $\overline{\pi}$ only if it puts probability one on $(1,1)$ at a pair attaining the bound, and less otherwise.

\noindent\textit{Step 2 (Creators' payoffs under $(1,1)$).} We let $R_{j}$ denote creator $j$'s net payoff under $(1,1)$.

(a) \emph{Displacement prices.} Let $\theta\geq\underline{\theta}_{C}$. At $(P^{C},P_{AI}^{C})$, $R_{H}>0$ and $L$ sells nothing, so she earns $V_{L}^{C}=\tfrac{\alpha}{2}G(z)$ (Lemma~\ref{lemma:c-payoffs}). Her AI payment covers her cost if and only if the AI is capable enough. To see this, note that $G$ strictly increases, $\tfrac{\alpha}{2}G(0)=0<\phi q_{L}$, and
\[
\frac{\alpha}{2}G\Bigl(\frac{t}{2}\Bigr)-\phi q_{L}\geq\frac{\alpha}{2}G\Bigl(\frac{t}{2}\Bigr)-\phi^{B}q_{L}=\frac{\alpha t(2-q_{L})}{8}>0,
\]
so a unique $\underline{\theta}_{L}(\phi)\in(-\tfrac{t}{2},0)$ solves $\tfrac{\alpha}{2}G\bigl(\tfrac{t}{2}+\underline{\theta}_{L}(\phi)\bigr)=\phi q_{L}$, namely
\[
\underline{\theta}_{L}(\phi)=\tfrac12\Bigl[\sqrt{(2-q_{H})^{2}+16\phi q_{L}/\alpha}-(2+q_{L})\Bigr],
\]
and, for $\theta\geq\underline{\theta}_{C}$, $R_{L}$ has the sign of $\theta-\underline{\theta}_{L}(\phi)$. This threshold exceeds $\underline{\theta}_{C}$, so that $L$'s participation rather than the platform's pricing limits part~1, if and only if $\tfrac{\alpha}{2}G(\tfrac{t}{2}+\underline{\theta}_{C})<\phi q_{L}$. Since
\[
G\bigl(\tfrac{t}{2}+\underline{\theta}_{C}\bigr)=\Bigl(1-\tfrac{(2+\sqrt{2})g}{2}\Bigr)\,\tfrac{(2+\sqrt{2})q_{L}-\sqrt{2}q_{H}}{2}
\]
has positive factors (the first as $g<\tfrac12<2-\sqrt{2}$) that decrease in $q_{H}$, this holds if and only if $\phi$ or $q_{H}$ exceeds a cutoff given the other parameters. Either cutoff may lie outside the admissible range.

(b) \emph{Three-product prices.} Let $\theta\leq\underline{\theta}_{C}$. At $(P^{\mathrm{3prod}},z)$, each creator earns more than in the benchmark, so $R_{H}>0$ and $R_{L}>0$. Each creator sells $q_{j}-P^{\mathrm{3prod}}$ and receives half of the creators' share of AI revenue $(1+\theta)z$ (Lemma~\ref{lemma:c-both}). Substitution gives
\[
V_{H}-V_{H}^{B}=\tfrac{\alpha}{4}z\,(2-2q_{L}+z)>0,\qquad V_{L}-V_{L}^{B}=\tfrac{\alpha}{4}z\,(2-2q_{H}+z)>0.
\]
Hence $V_{j}>V_{j}^{B}\geq\phi q_{j}$ for both creators, so both net payoffs are positive.

\noindent\textit{Step 3 (Parts~1 and~2).} Both candidate prices of creators' content exceed $\tfrac{t-1}{2}$, as $P^{C}-\tfrac{t-1}{2}=\tfrac{1+g+\theta}{2}>0$ and $P^{\mathrm{3prod}}-\tfrac{t-1}{2}=\tfrac{1+\theta}{2}>0$. By Lemma~\ref{lemma:ai-at-quality}(b), $(1,1)$ is therefore the only allowed play at any candidate pair at which $R_{H}>0$ and $R_{L}\geq0$. If this holds at every pair attaining the bound of Step~1, Lemma~\ref{lemma:stage2}(b) with $\pi^{*}=\overline{\pi}$ then shows that an equilibrium exists and that every equilibrium has Stage-2 play $(1,1)$ and prices that attain the profit bound of Step~1.

Let $\max\{\underline{\theta}_{C},\underline{\theta}_{L}(\phi)\}\leq\theta<0$. The bound of Step~1 is attained at $(P^{C},P_{AI}^{C})$, where Step~2(a) gives $R_{H}>0$ and $R_{L}\geq0$. If $\theta=\underline{\theta}_{C}$, it is also attained at $(P^{\mathrm{3prod}},z)$, where Step~2(b) gives $R_{H},R_{L}>0$. For $\theta>\underline{\theta}_{C}$, the unique equilibrium outcome is therefore that of Proposition~\ref{prop:c-pricing} (CQ1) with $\theta$ in place of $q^{\textrm{base}}$: $D_{L}^{C}=0$, and $R_{L}>0$ except at $\theta=\underline{\theta}_{L}(\phi)$. At $\theta=\underline{\theta}_{C}$, which requires $\underline{\theta}_{C}\geq\underline{\theta}_{L}(\phi)$, the three-product outcome of part~2 is a second equilibrium outcome.

Let $-\tfrac{t}{2}<\theta<\underline{\theta}_{C}$. The bound is attained only at $(P^{\mathrm{3prod}},z)$, where Step~2(b) gives $R_{H},R_{L}>0$ whatever $\underline{\theta}_{L}(\phi)$ is. The three-product configuration is therefore the unique equilibrium outcome: the AI sells $D_{AI}=1+\theta>0$ at the price $z>0$, and $L$ regains the demand $D_{L}=-g-\tfrac{\theta}{2}>0$.

Let $\theta=\theta_C<\theta_L(\phi)$. The bound of Step~1 is attained at $(P^C,P_{AI}^C)$ and at $(P^{3\mathrm{prod}},z)$. At $(P^C,P_{AI}^C)$, Step~2(a) gives $R_L<0$, so $L$ would rather exit and no allowed play puts probability one on $(1,1)$. By Step~1, every allowed play yields less than $\bar{\pi}$ at every price pair not attaining the bound. At $(P^{3\mathrm{prod}},z)$, Step~2(b) gives
$R_H,R_L>0$, so $(1,1)$ is the only allowed play by Lemma~B.6(b), and it yields $\bar{\pi}$. Lemma~B.1(b), with $\mathcal{P}^*=\{(P^{3\mathrm{prod}},z)\}$ and $\pi^*=\bar{\pi}$, then shows that the three-product configuration is again the unique equilibrium outcome, with $D_L>0$ as $\theta_C<-2g$.

\noindent\textit{Step 4 (Part~3).} Let $\theta\leq-\tfrac{t}{2}$, so that $z\leq0$. Then no consumer obtains positive utility from the AI: $U_{AI}=z-P_{AI}\leq-P_{AI}$ under $(1,1)$, and $U_{AI}<0$ under every other profile. So AI revenue is zero at all prices and under every profile, and the AI sells nothing if $\theta<-\tfrac{t}{2}$. By Lemma~\ref{lemma:no-ai-revenue}, profit is at most $\pi^{B}$, with equality only if $P=P^{B}=\tfrac{t}{4}$ and both creators participate and sell $q_{j}-P^{B}$. Fix a pair $(P^{B},P_{AI})$. The AI gives no consumer positive utility, ties go to creators' content and $P^{B}>\tfrac{t-1}{2}$. Each participating creator thus sells $q_{j}-P^{B}$ whatever the other creator does (Lemma~\ref{lemma:no-ai}(a)), and contributions affect no payoff. Her revenue $V_{j}^{B}$ covers her cost, so both creators participate in every allowed play, $L$ by Assumption~\ref{assumption:weak-participation} if $\phi=\phi^{B}$. Lemma~\ref{lemma:stage2}(b) and~(c) with $\mathcal{P}^{*}=\{(P^{B},P_{AI}):P_{AI}\geq0\}$ and $\pi^{*}=\pi^{B}$ gives the benchmark outcome, with no AI sales if $\kappa_{0}>\kappa_{0}^{*}$.
\end{proof}

\subsection{Proof of Proposition~\ref{prop:heterogeneous-data}}
\label{apx:proof-heterogeneous-data}

\begin{proof}
Here $\kappa_{0}=0$, $\tau_{H}=\tfrac12$ and $0<\tau_{L}\leq\tfrac12$, so $z=q^{\textrm{base}}+\tfrac{q_{H}}{2}+\tau_{L}q_{L}$ is the $q_{AI}$ of the statement. CQ1 is the case $z<q_{H}$ and the price pairs of the statement are the displacement prices $\mathcal{P}(z)$. A lower transfer rate lowers $z$, so the platform sells a less capable AI. We show that it still sets the displacement prices, and that equal splitting still pays $L$ enough to participate.

\noindent\textit{Step 1 (Profit bound).} By Lemma~\ref{lemma:c-bound}, every profile yields at most $(1-\alpha)F(z)$ at any price pair. Equality requires $(1,1)$ and a pair in $\mathcal{P}(z)$, where $D_{L}=0$, $D_{H}=D_{H}^{C}(z)$ and $D_{AI}=1-D_{H}^{C}(z)$.

\noindent\textit{Step 2 (Creators' payoffs).} Under $(1,1)$ at a pair in $\mathcal{P}(z)$, $H$'s net payoff is positive and $L$ earns $V_{L}^{C}=\tfrac{\alpha}{2}G(z)$ (Lemma~\ref{lemma:c-payoffs}, which applies as $\theta>\underline{\theta}_{C}$ by Lemma~\ref{lemma:c-bound}). As $G$ increases and $z>\tfrac{q_{H}}{2}$, $V_{L}^{C}>\tfrac{\alpha}{2}G(\tfrac{q_{H}}{2})=\alpha q_{H}(4-q_{H})/16$, and
\[
\frac{16}{\alpha}\Bigl[\frac{\alpha q_{H}(4-q_{H})}{16}-\phi^{B}q_{L}\Bigr]=q_{H}(4-5q_{L})+3q_{L}(q_{H}-q_{L})>0,
\]
since $q_{L}\leq q_{H}/\sqrt{2}<\tfrac45$. Thus $V_{L}^{C}>\phi^{B}q_{L}\geq\phi q_{L}$ for every $\tau_{L}\in(0,\tfrac12]$: $L$ receives half of the creators' share of AI revenue, which stays large even when her own data adds little to AI quality.

\noindent\textit{Step 3 (Equilibrium).} At every pair in $\mathcal{P}(z)$, $P>\tfrac{q_{H}}{2}>\tfrac{t-1}{2}$. By Step~2 and Lemma~\ref{lemma:ai-at-quality}(b), $(1,1)$ is the only allowed play there, and it yields $(1-\alpha)F(z)$. By Step~1 and Lemma~\ref{lemma:stage2}(c), every allowed play yields strictly less at every other price pair. Lemma~\ref{lemma:stage2}(b) with $\mathcal{P}^{*}=\mathcal{P}(z)$ and $\pi^{*}=(1-\alpha)F(z)$ shows that every equilibrium has prices in $\mathcal{P}(z)$, play $(1,1)$ and the demands of Step~1, and that every pair in $\mathcal{P}(z)$ is an equilibrium price pair. This proves the characterization and item~3.

\noindent\textit{Step 4 (Comparative statics).} As $dz/d\tau_{L}=q_{L}>0$ and $F$ is strictly increasing, the AI price $z$ and the platform's profit $(1-\alpha)F(z)$ strictly increase in $\tau_{L}$ (item~1). In CQ1, the human price $P=\tfrac{q_{H}+z}{2}$ and AI demand $D_{AI}=1-D_{H}^{C}(z)$ strictly increase in $\tau_{L}$ and $D_{H}^{C}(z)=\tfrac{q_{H}-z}{2}$ strictly decreases in $\tau_{L}$; in CQ2, $D_{AI}=1$ (item~2). Both creators' gross payoffs, $\tfrac{\alpha}{2}G(z)$ and $V_{H}(z)$, also strictly increase in $\tau_{L}$ (Lemma~\ref{lemma:c-payoffs}), as claimed in the discussion that follows the proposition in Section~\ref{subsec:heterogeneous-data}.
\end{proof}

\subsection{Proof of Corollary~\ref{cor:reversal}}
\label{apx:proof-reversal}

\begin{proof}
Under NC, $L$'s participation constraint binds and $H$'s is slack (Proposition~\ref{prop:nc-pricing}), so $V_{L}^{NC}/q_{L}=\phi<V_{H}^{NC}/q_{H}$. Under C, we split the gap between the two ratios into two parts: $H$'s direct revenue per unit of quality, and the excess of $L$'s AI payment per unit of quality over $H$'s. We show that the second part is larger, because both creators receive the same AI payment, which is larger relative to $L$'s lower quality. By Proposition~\ref{prop:c-pricing}, $s_{H}=s_{L}=1$ and $D_{L}^{C}=0$, so \eqref{eq:creator-payoff} and $1/q_{L}-1/q_{H}=2g/(q_{H}q_{L})$ give
\[
\frac{V_{H}^{C}}{q_{H}}-\frac{V_{L}^{C}}{q_{L}}
=\frac{\alpha D_{H}^{C}P^{C}}{q_{H}}-\frac{\alpha D_{AI}^{C}P_{AI}^{C}}{2}\Bigl(\frac{1}{q_{L}}-\frac{1}{q_{H}}\Bigr)
=\frac{\alpha}{q_{H}q_{L}}\bigl(q_{L}D_{H}^{C}P^{C}-g\,D_{AI}^{C}P_{AI}^{C}\bigr).
\]
Also by Proposition~\ref{prop:c-pricing}, $D_{H}^{C}P^{C}\leq\tfrac{g}{2}q_{H}$ (in CQ1, $D_{H}^{C}=\tfrac{g-q^{\textrm{base}}}{2}\leq\tfrac{g}{2}$ and $P^{C}=q_{H}-D_{H}^{C}$; in CQ2, $D_{H}^{C}=0$), $D_{AI}^{C}=1-D_{H}^{C}\geq1-\tfrac{g}{2}$ and $P_{AI}^{C}=q^{\textrm{base}}+\tfrac{t}{2}\geq\tfrac{t}{2}$, so
\[
q_{L}D_{H}^{C}P^{C}-g\,D_{AI}^{C}P_{AI}^{C}
\leq\frac{g}{2}\Bigl[q_{H}q_{L}-\Bigl(1-\frac{g}{2}\Bigr)t\Bigr]<0,
\]
because $q_{H}q_{L}<q_{L}<\tfrac{t}{2}<(1-\tfrac{g}{2})t$. Hence $V_{H}^{C}/q_{H}<V_{L}^{C}/q_{L}$, as claimed.
\end{proof}

\subsection{Proof of Proposition~\ref{prop:quality-weighted}}
\label{apx:proof-quality-weighted}

\begin{proof}
Quality weighting changes how creators split their share of AI revenue, but not what consumers buy or what the platform earns under any profile. The platform's preferred prices are therefore unchanged, and the only question is whether $L$ still participates when part of her AI payment goes to $H$. Steps~1--4 answer it for equal transfer rates (the first part of the proposition) and Steps~5 and~6 for heterogeneous transfer rates (the second part).

\noindent\textit{Step 1 (Profit bound with equal transfer rates).} Let $\tau_{H}=\tau_{L}=\tfrac12$, so that $z=q^{\textrm{base}}+\tfrac{t}{2}=P_{AI}^{C}$, and we let $\mathcal{P}^{C}\equiv\mathcal{P}(z)$, $D_{H}^{C}\equiv D_{H}^{C}(z)$, $D_{AI}^{C}\equiv1-D_{H}^{C}$ and $\pi^{C}\equiv(1-\alpha)F(z)$ be the equilibrium prices, demands and profit of Proposition~\ref{prop:c-pricing}. Lemma~\ref{lemma:c-bound} holds under either sharing rule. So every profile yields at most $\pi^{C}$ at every price pair, with equality only under $(1,1)$ at the pairs in $\mathcal{P}^{C}$. There, demands are $D_{L}=0$, $D_{H}=D_{H}^{C}$ and $D_{AI}=D_{AI}^{C}$. Prices, demands, profit and consumer surplus are thus those of Proposition~\ref{prop:c-pricing} once we show that $(1,1)$ is the only allowed play on $\mathcal{P}^{C}$.

\noindent\textit{Step 2 (Redistribution).} Under $(1,1)$, creator $j$'s quality-weighted share is $q_{j}/t$: $H$'s share rises from $\tfrac12$ to $q_{H}/t$ and $L$'s falls to $q_{L}/t$. On $\mathcal{P}^{C}$,
\[
V_{H}^{C,\mathrm{QW}}-V_{H}^{C}=V_{L}^{C}-V_{L}^{C,\mathrm{QW}}=\frac{\alpha(q_{H}-q_{L})}{2t}\,D_{AI}^{C}P_{AI}^{C}>0,
\]
and creators' total net payoff is unchanged. Each creator now receives the AI payment $\alpha D_{AI}^{C}P_{AI}^{C}/t$ per unit of quality, and $D_{L}^{C}=0$, so
\begin{equation}
\frac{V_{H}^{C,\mathrm{QW}}}{q_{H}}-\frac{V_{L}^{C,\mathrm{QW}}}{q_{L}}=\frac{\alpha D_{H}^{C}P^{C}}{q_{H}}\;\geq\;0 ,
\label{eq:qw-ratio}
\end{equation}
with strict inequality in CQ1, where $D_{H}^{C}P^{C}>0$, and equality in CQ2, where $D_{H}^{C}=0$.

\noindent\textit{Step 3 (Positive net payoffs).} By \eqref{eq:qw-ratio}, it suffices to show $V_{L}^{C,\mathrm{QW}}>\phi q_{L}$. By Corollary~\ref{cor:data-supplier}, $V_{L}^{C}$ increases in $q^{\textrm{base}}$ and so does $V_{L}^{C,\mathrm{QW}}=(2q_{L}/t)V_{L}^{C}$. Hence $V_{L}^{C,\mathrm{QW}}$ is smallest at $q^{\textrm{base}}=0$ (CQ1), where $D_{AI}^{C}=1-\tfrac{g}{2}$ and $P_{AI}^{C}=\tfrac{t}{2}$. As $\phi\leq\phi^{B}$ (Assumption~\ref{assumption:benchmark-participation}) and $q_{L}<1$,
\begin{equation}
\frac{V_{L}^{C,\mathrm{QW}}}{q_{L}}-\phi\;\geq\;\frac{\alpha(4-q_{H}+q_{L})}{8}-\phi^{B}=\frac{\alpha\bigl[(q_{H}-2q_{L})^{2}+q_{L}(8-5q_{L})\bigr]}{16\,q_{L}}>0 .
\label{eq:qw-positive}
\end{equation}

\noindent\textit{Step 4 (Equilibrium with equal transfer rates).} At every pair in $\mathcal{P}^{C}$, $P>\tfrac{q_{H}}{2}>\tfrac{t-1}{2}$. By Step~3 and Lemma~\ref{lemma:ai-at-quality}(b), $(1,1)$ is the only allowed play there, and it yields $\pi^{C}$. By Step~1 and Lemma~\ref{lemma:stage2}(c), every allowed play yields strictly less elsewhere. Lemma~\ref{lemma:stage2}(b) with $\mathcal{P}^{*}=\mathcal{P}^{C}$ and $\pi^{*}=\pi^{C}$ then proves the first part of the proposition, the case of equal transfer rates.

\noindent\textit{Step 5 (The threshold $\tau_{L}^{\star}$).} Now let $q_{H}/q_{L}\geq\sqrt{2}$, $\tau_{H}=\tfrac12$ and $0<\tau_{L}\leq\tfrac12$, and we write $h\equiv\tfrac{q_{H}}{2}$ and $\ell\equiv\tau_{L}q_{L}$, so that $z=q^{\textrm{base}}+h+\ell$. By Lemma~\ref{lemma:c-bound}, the profit bound $(1-\alpha)F(z)$, the price pairs $\mathcal{P}(z)$ that attain it under $(1,1)$ and the demands there are those of Proposition~\ref{prop:heterogeneous-data}. Under $(1,1)$ on $\mathcal{P}(z)$, AI revenue is $G(z)$ (Lemma~\ref{lemma:c-payoffs}). $L$ sells nothing and receives the share $\ell/(h+\ell)$ of the creators' part of this revenue. Her payoff is thus $\alpha\eta(\tau_{L})$, where $\eta(\tau_{L})\equiv\tfrac{\ell}{h+\ell}G(z)$ does not depend on $\phi$ or $\alpha$. A lower transfer rate now reduces her payment twice, through lower AI revenue and through her smaller share. Formally, $\eta$ is continuous and strictly increasing, as $\ell/(h+\ell)$ and $G(z)$ are positive and strictly increase in $\ell$. It tends to zero as $\tau_{L}\downarrow0$. At $\tau_{L}=\tfrac12$, $\ell/(h+\ell)=q_{L}/t$ and $z\geq\tfrac{t}{2}$, so $\eta(\tfrac12)\geq(q_{L}/t)G(\tfrac{t}{2})=q_{L}(4-q_{H}+q_{L})/8$, which exceeds $(\phi^{B}/\alpha)q_{L}\geq(\phi/\alpha)q_{L}$ by \eqref{eq:qw-positive}. Hence a unique $\tau_{L}^{\star}\in(0,\tfrac12)$ solves $\eta(\tau_{L}^{\star})=(\phi/\alpha)q_{L}$. It depends on $\phi$ and $\alpha$ only through $\phi/\alpha$, and $L$'s net payoff is negative below $\tau_{L}^{\star}$, zero at $\tau_{L}^{\star}$ and positive above it.

\noindent\textit{Step 6 (Equilibrium with heterogeneous transfer rates).} $H$'s share $h/(h+\ell)$ is at least $\tfrac12$, so her payoff is at least her payoff under equal splitting, and her net payoff is positive (Lemma~\ref{lemma:c-payoffs}). If $\tau_{L}\geq\tau_{L}^{\star}$, $L$'s net payoff is nonnegative. As $P>\tfrac{q_{H}}{2}>\tfrac{t-1}{2}$ at every pair in $\mathcal{P}(z)$, Lemma~\ref{lemma:ai-at-quality}(b) shows that $(1,1)$ is the only allowed play there. Lemmas~\ref{lemma:c-bound} and~\ref{lemma:stage2}(b)--(c) then show that the outcome of Proposition~\ref{prop:heterogeneous-data} is the equilibrium outcome. If $\tau_{L}<\tau_{L}^{\star}$, $L$'s net payoff under $(1,1)$ is negative at every pair in $\mathcal{P}(z)$, so she would rather exit than contribute. Then $(1,1)$ is not a Nash equilibrium at any of these pairs, and no equilibrium under quality weighting has this outcome.
\end{proof}